\documentclass[aps,pra,amsmath,notitlepage,nofootinbib,twocolumn,superscriptaddress,10pt]{revtex4-2}
\pdfoutput=1
\usepackage[utf8]{inputenc}
\usepackage[english]{babel}
\usepackage{microtype}
\usepackage[normalem]{ulem}
\usepackage{graphicx}
\usepackage{dcolumn}
\usepackage{bm}
\usepackage{amsmath,amssymb,amsfonts}
\usepackage{booktabs}
\usepackage{array}
\usepackage{mathrsfs}
\usepackage{braket}
\usepackage{dsfont}
\usepackage{xcolor}
\usepackage{mathtools}
\usepackage{tikz}
\usepackage[breaklinks=true,colorlinks=true,linkcolor=teal,urlcolor=teal,citecolor=teal]{hyperref}
\usepackage{orcidlink}
\usepackage{cleveref}
\begin{document}

\title{Optimal Waveforms for Dipole Moment Estimation with Coherent States}

\begin{abstract}
We investigate quantum sensing for spectroscopy in a system consisting of a two-level atom coupled to a continuum of modes. We focus on optimizing the pulse shape of a coherent state to maximize the quantum Fisher information (QFI) of the emitted light with the aim of estimating the atom’s dipole moment, which is proportional to its spontaneous emission rate. To achieve this, we derive a set of coupled differential equations, which include the standard optical Bloch equations as a subset and whose solution directly yields the QFI of the emitted light without resorting to finite-difference methods. Furthermore, we analyze the factors that govern its optimization, provide analytic solutions in both the long and the short pulse width limits, and examine the role of the average photon number of the pulses. We then show that under the closed (periodic) boundary conditions, the harmonic (plane-wave) with frequency equal to half the spontaneous emission rate and a phase determined by detuning are optimal in the long pulse width limit.
\end{abstract}

\author{Karthik Chinni}
\email{karthik.chinni@polymtl.ca}
\affiliation{D\'epartement de g\'enie physique, \'Ecole polytechnique de Montr\'eal, Montr\'eal, QC, H3T 1J4, Canada}
\author{Nicol\'as Quesada}
\email{nicolas.quesada@polymtl.ca}
\affiliation{D\'epartement de g\'enie physique, \'Ecole polytechnique de Montr\'eal, Montr\'eal, QC, H3T 1J4, Canada}
\maketitle 
\date{\today}
\section{Introduction}
Quantum sensing provides a framework for extracting information about physical systems with high precision, and light-matter interactions play an important role in this field~\cite{dowling2015quantum, taylor2016quantum, tan2019nonclassical, lawrie2019quantum, berchera2019quantum, datta2025sensing, holdsworth2025saturation, polzik1992spectroscopy}. Quantum sensing has applications in various areas including interferometry~\cite{caves1981quantum, bondurant1984squeezed, dowling2008quantum}, microscopy~\cite{casacio2021quantum,datta2020quantum,de2020quantum} and spectroscopy~\cite{mukamel2020roadmap,kira2011quantum, dorfman2016nonlinear, albarelli2023fundamental, darsheshdar2024role, khan2024does, kalashnikov2014quantum, fujihashi2021achieving, raymer2013entangled, leibfried2004toward}. In this work, we focus on the paradigm of spectroscopy where the goal is to probe a parameter of the system of interest by analyzing the light that has interacted with it. The figure of merit used for the precision of estimating a parameter is the quantum Fisher information, which bounds the variance of an unbiased estimator via the well-known quantum Cramér-Rao bound~\cite{cramer1946mathematical,rao1945information,frechet1943extension}. However, computing this quantity is typically challenging since it requires the full knowledge of the state and its derivative when it is computed in a naive manner. Here, we focus on the setup in which a coherent state of the quantized electromagnetic field propagating in one dimension is coupled to a two-level atom with the goal of identifying the optimal pulse for the detection of the dipole moment of the atom as shown in Fig.~\eqref{fig:atom_waveguide}. In contrast to cavity-based approaches, which typically involve a single optical mode~\cite{bernad2019optimal, genoni2012optimal,montenegro2022probing}, we consider quantized light encompassing many temporal modes, where the interaction with the atom gives rise to temporal correlations in the scattered field as in \cite{albarelli2023fundamental, darsheshdar2024role}. This would naturally suggest evaluating the QFI via the matrix product operator (MPO) representation of the system and its derivatives. However, in the absence of losses, \textit{i.e.} assuming that we can measure all the scattered light, the system becomes bipartite (waveguide and matter) allowing the computation of the QFI through the double-sided master equation without having access to the full state of the light \cite{gammelmark2014fisher, yang2023efficient}. In particular, the MPO form along with the Schmidt decomposition of the bipartite systems permits this simplification. See recent works~\cite{yang2025quantum, khan2025tensor, midha2025metrology} for computing the QFI in the presence of loss. The problem of estimating the Rabi frequency has been investigated in \cite{kiilerich2014estimation, kiilerich2016bayesian} for the case of a two-level atom driven by classical light with constant amplitude. This setting differs from the estimation of the coupling parameter under arbitrary pulse driving, which is the focus of the present work.

In our work, we focus on the problem of electromagnetic field propagating in one dimension (which we refer to as the waveguide hereafter) interacting with an atom without losses and find optimal pulses that maximize the QFI through analytic tools and numerical simulations. According to \cite{yang2023efficient}, the QFI is given as the double derivative of a function of the generalized density operator with respect to the parameter that is being measured where the generalized density operator evolves under the double sided master equation. This is typically evaluated through finite-difference methods. Here we simplify this by presenting the QFI as the solution to a set of coupled ordinary differential equations (ODEs), which contain optical Bloch equations as a subset, and thereby allowing us to bypass instabilities associated with the finite-difference methodologies. We further solve these ODEs for various standard pulse shapes to illustrate the factors that play a role in maximizing the QFI. For instance, we show that the QFI is governed by factors beyond the atomic excitation probability consistent with single-photon pulse results in~\cite{albarelli2023fundamental, darsheshdar2024role}. We then derive analytic expressions for the QFI in two complementary regimes of the probe pulse: low- and high-photon-density limits of the pulse. In these cases, the QFI can be cast into a bilinear form, which enables us an analytic derivation of the optimal pulses. Finally, we show through numerical simulations that under the closed-boundary conditions, the harmonic at the frequency equal to half the spontaneous emission rate and an overall phase dependent on detuning achieves the optimal QFI per unit photon of $4$. Under the periodic boundary conditions, the same QFI is obtained by the plane wave with the frequency shifted by half the spontaneous emission rate from the detuning.

The rest of the manuscript is organized as follows. In Sec.~\ref{sec:Hamiltonian}, we provide general background material, starting from the Hamiltonian of the problem. In Sec.~\ref{sec:QFI_introduction}, we present standard results in quantum estimation theory and review the method for computing the QFI efficiently using the two-sided master equation formalism. Using this toolkit, we obtain our results in Sec.~\ref{sec:real_pulses} for the case of real pulses and zero detuning. More specifically, we present the set of coupled ODEs whose solution yields the QFI, examine the factors that play a role in optimizing the QFI through the analysis of standard pulses, study the role of average photon number, obtain the optimal pulse for the QFI in the short and the long pulse width limit analytically and present the optimal pulse in the general case using the results of numerical optimization. Following this, in Sec.~\ref{sec:complex_pulses}, we generalize the results of the previous section to the case of nonzero detuning and complex pulses.
\section{Atom driven by a coherent state}
\label{sec:Hamiltonian}
\begin{figure}[t]
\includegraphics[width=1.0\linewidth]{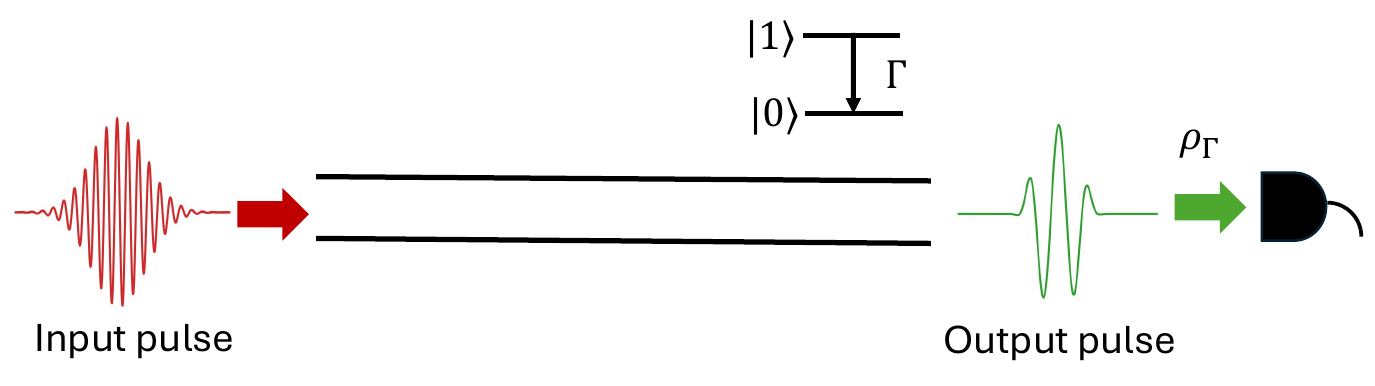}
\caption{Our setup consists of a two-level atom coupled to a coherent state of the quantized electromagnetic field propagating in one-dimension (represented by a waveguide). The light emitted by the atom is analyzed to extract the dipole moment of the atom. We denote the decay rate of the atom into the waveguide of interest by $\Gamma = \Gamma_{\parallel}$, and assume that we have access to all the scattered light, $\Gamma_{\perp} = 0$.}
\label{fig:atom_waveguide}
\end{figure}
The Hamiltonian of a two-level atom coupled to a waveguide, as illustrated in Fig.~\eqref{fig:atom_waveguide}, is given by \cite{quesada2022beyond}
\begin{align}\label{eq:Hami}
\tilde H = \tilde H_{\text{atom}}    + \tilde H_{\text{wg}} + \tilde V,
\end{align}
where
\begin{align}
\label{eq:atom_hami}
\tilde H_{\text{atom}} =\;& \omega_0 \sigma^\dagger \sigma = \omega_0 \ket{1} \bra{1}, \\
\label{eq:wg_hami}
\tilde H_{\text{wg}} =& \sum_{\eta}  \omega_1 \int d x \ \psi^\dagger_{\eta}(x) \psi_{\eta}(x)   \\
& +\sum_{\eta}\frac{i v}{2}  \int d x \left( \left[\frac{\partial}{\partial x} \psi_{\eta}^\dagger(x) \right] \psi_{\eta}(x) - \text{H.c.} \right), \nonumber\\
\label{eq:int_hami}
\tilde V =&  \sum_{\eta} i \sqrt{v \Gamma_{\eta}} \left(\sigma \psi_{\eta}^\dagger(0) - \text{H.c.}\right),
\end{align}
$\hbar=1$, $\omega_0/\omega_1$ are the frequencies of the atom/waveguide, $v$ is the group velocity of the waveguide mode, $\Gamma_{\eta} $ is the decay rate of the atom into the waveguide $\eta$, $\ket{0/1}$ are the ground/excited states of the atom and $\sigma = \ket{0}\bra{1}$, $\eta \in \{\parallel, \perp\}$, $\psi_{\parallel}$ and $\psi_{\perp}$ are the modes corresponding to the waveguide of interest (the scattered light which is accessible for measurement) and the perpendicular waveguide (accounts for loss) respectively. The canonical commutation relations of the field and atom are
\begin{align}
[\psi_{\eta}(x),\psi_{\eta'}^\dagger(x')]=\delta(x-x') \delta_{\eta,\eta'} \text{ and } \left\{ \sigma,\sigma^\dagger \right\} = 1.
\end{align}
In this work, we consider a simplified model by neglecting losses, $\Gamma_{\perp}=0$. This allows us to analyze the QFI of the emitted light without resorting to the matrix product operator (MPO) formalism. We also assume point coupling between the waveguide and the atom at position $x=0$ and ignore dispersion within each field in the waveguide. The total number of excitations in the system is conserved under this Hamiltonian,
\begin{align}
\label{eq:conservation_law}
    [\tilde{H},N]=0 \text{ where } N =  \sigma^{\dagger}\sigma + \int d x \ \psi^{\dagger}(x) \psi(x).
\end{align}
\subsection{Interaction Picture}
\label{subsection:interaction_picture}
We now move to the interaction picture, meaning that we will move some trivial time evolution into the operators, while the nontrivial part will be carried by the state vector. Rearranging the Hamiltonian in Eq.~\eqref{eq:Hami} and setting $\Gamma_{\perp}=0$, $\Gamma_{\parallel}=\Gamma$, we have
\begin{align}
\tilde{H} &= H_0 + H_1,
\end{align}
where 
\begin{align}
    H_{0} &= \omega_1 N + \frac{i}{2}  v \int d x \left( \left[\frac{\partial}{\partial x} \psi^\dagger(x) \right] \psi(x) - \text{H.c.} \right), \\
    H_{1} &=\delta \sigma^{\dagger}\sigma     +i  \sqrt{v \Gamma}\left[ \sigma \psi^\dagger(0) - \text{H.c.}  \right],
\end{align}
and $\delta = \omega_{0}-\omega_{1}$. Moving to the interaction picture with respect to $H_{0}$, we have
\begin{align}
\label{eq:interaction_Hamiltonian}
    H_{\text{int}}(t) &= \mathcal{U}_0^{\dagger}(t) H_{1} \mathcal{U}_0(t) \\
    &= \delta \sigma^{\dagger}\sigma   +i \sqrt{v \Gamma} \left( \sigma \psi^\dagger(-v t) - \text{H.c.} \right),
\end{align}
where $\mathcal{U}_0(t) = \exp\left(-it H_{0}  \right)$ and $t_{0}=0$. In this work, we focus only on the initial states where the light pulse is a coherent state in the waveguide and the atom is in its ground state
\begin{align}
\label{eq:init_state}
\ket{\phi(0)} = \mathcal{D}(\tilde{f}(x)) \ket{\text{vac}}  \otimes \ket{g}.
\end{align}
The displacement operator is $\mathcal{D}(\tilde{f}(x)) \equiv \exp(\int dx \tilde{f}(x) \psi^\dagger (x) -\text{H.c.} )$ and $\tilde{f}(x)$ is the shape of the pulse. Note that the pulse shape is normalized to average photon number, 
\begin{align}
    \int dx \bra{\phi(0)} \psi^{\dagger}(x)\psi(x) \ket{\phi(0)} &=\int dx |\tilde{f}(x)|^{2} = \alpha^{2}.
\end{align}
The state of the system at an arbitrary time can then be written as \cite{mollow1975pure, fischer2018scattering}
\begin{align}\label{eq:state_arbitrary_time}
\ket{\phi(t)} \equiv \Lambda(t)\mathcal{U}_0(t) \mathcal{T}\left[e^{-i \int_{0}^t d\tau H_{\mathcal{D}}(\tau)}\right] \ket{g} \otimes \ket{\text{vac}},
\end{align}
where 
\begin{align}\label{dynamics}
\Lambda(t) &=\mathcal{D}\left(\tilde{f}(x - v t) e^{ -i \omega_1 t}\right),
\end{align}
and 
\begin{align}\label{eq:effective_Hamiltonian}
    H_{\mathcal{D}}(t)  &=\delta  \sigma^\dagger \sigma  +i  \sqrt{v \Gamma} \left( \sigma \tilde{f}^*(-v t) - \text{H.c.} \right) \nonumber \\
    &\qquad +i   \sqrt{v \Gamma} \left( \sigma \psi^\dagger(-v t) - \text{H.c.} \right) .
\end{align}
For more details regarding Eq.~\eqref{eq:state_arbitrary_time}, see Appendix~\eqref{appendix:Mollow_transformation}. Note that the factors $\Lambda(t)$ and $\mathcal{U}_{0}(t)$ do not play any role in the computation of QFI since they are independent of $\Gamma$, which is the parameter whose precision is of interest to us. Therefore, these factors can be safely ignored. From the above, it can be seen that an atom interacting with a coherent state is the same as interacting with the vacuum at the cost of adding the term $i \sqrt{v \Gamma} \left(\sigma \tilde{f}^*(-vt) - \text{H.c.} \right)$ to the Hamiltonian of the atom. Following Fischer et al.~\cite{fischer2018derivation}, tracing out the waveguide we obtain the usual Lindblad master equation,
\begin{align}
\label{eq:master_eqn}
    \dot{\rho}(t) &= i [\rho, H] + \Gamma \sigma \rho \sigma^{\dagger} - \frac{\Gamma}{2} \left(\sigma^{\dagger} \sigma \rho + \rho \sigma^{\dagger} \sigma\right) ,\\
    \label{eq:driving_Hamiltonian}
    H & = \delta \sigma^{\dagger}\sigma +  i \sqrt{\Gamma} \left(f(t)^{*}\sigma - f(t) \sigma^{\dagger}\right),  
\end{align}
where $\rho = \text{Tr}_{\text{wg}}(\ket{\phi(t)} \bra{\phi(t)})$ and $f(t) \equiv \sqrt{v} \tilde{f}(x = -v t)$. Taking expectation values of the Pauli operators, we recover the usual optical Bloch equations. Note that the treatment above accounts for the electromagnetic field fully quantum mechanically.
\section{Quantum Fisher Information}
\label{sec:QFI_introduction}
The precision of an estimate of a parameter $\theta$ is quantified through the Fisher information. The variance of any unbiased estimator, for which the mean value of the estimator equals the true value, the mean-square error is bounded by the classical Fisher information (CFI) $\mathcal{C}(\rho_{\Gamma},\Pi)$ according to the Cramer-Rao bound~\cite{cramer1946mathematical,rao1945information,frechet1943extension}
\begin{align}
\label{eq:Cramer-Rao bound}
    \sigma^{2}(\theta) \geq \frac{1}{n \mathcal{C}(\rho_{\theta},\Pi)},
\end{align}
where $n$ is the number of repetitions of the experiment, $\rho_{\theta}$ is the state of the system and $\Pi$ represents the positive operator valued measure (POVM) elements corresponding to a measurement basis. The CFI quantifies the sensitivity of the probability distribution, $p(x,\theta)=\text{Tr}\left(\rho_{\theta}\Pi_{x}\right)$, to the parameter that we are trying to estimate. It is given by 
\begin{align}
    \mathcal{C}(\rho_{\theta}, \Pi) &= \sum_{x} \frac{1}{p(x,\theta)} \left(\frac{\partial p(x,\theta)}{\partial \theta}\right)^{2}.
\end{align}
Note that the CFI depends on the detection method, and we obtain the quantum Fisher information (QFI) by maximizing this quantity over all the measurement basis~\cite{paris2009quantum}. Hence, the QFI provides a more stringent bound on the Cramer-Rao bound in Eq.~\eqref{eq:Cramer-Rao bound} known as the quantum Cramer-Rao bound \cite{helstrom1967minimum,holevo2011probabilistic, braunstein1994statistical, paris2009quantum}. The QFI can be also defined using the symmetric logarithmic derivative, $\mathcal{L}_{\theta}$, which is a Hermitian operator defined implicitly through the equation \cite{helstrom1967minimum, helstrom1332quantum, holevo2011probabilistic}
\begin{align}
    \frac{\partial \rho_{\theta}}{\partial \theta} &= \frac{1}{2}\left(\rho_{\theta} \mathcal{L}_{\theta} + \mathcal{L}_{\theta} \rho_{\theta} \right).
\end{align}
Then,
\begin{align}
    \mathcal{F}(\rho_{\theta}) &= \text{max}_{\Pi} \mathcal{C}(\rho_{\theta},\Pi) = \text{Tr}\left(\rho_{\theta}\mathcal{L}_{\theta}^{2}\right).
\end{align}
The above reduces to the following expression for a mixed state with the spectral decomposition, ${{\rho }_{\theta }}=\sum\limits_{n}{}{{p}_{n}}\left| {{\psi }_{n}} \right\rangle \left\langle  {{\psi }_{n}} \right|$ \cite{liu2014quantum},
\begin{align}
    \label{eq:qfi_mixed_state}    \mathcal{F}\left(\rho_{\theta}\right)  &= \sum_{n}\,\frac{(\partial_{\theta} p_n)^2}{p_n} + \sum_n\,4p_n\,\langle\partial_{\theta}\psi_{n}|\partial_{\theta}\psi_{n}\rangle \nonumber\\
    & -\sum_{m,n}\frac{8p_m p_n}{p_m +p_n}|\langle\partial_{\theta}\psi_{m}|\psi_{n}\rangle|^2 .
\end{align}
In the case of a pure state $\ket{\psi_{\theta}}$, the above expression further simplifies to~\cite{paris2009quantum} 
\begin{align}
    \mathcal{F}(|\psi_{\theta}\rangle) &= 4 \left(\braket{\partial_{\theta}\psi_{\theta} |\partial_{\theta}\psi_{\theta}} - |\braket{\partial_{\theta}\psi_{\theta}|\psi_{\theta}}|^{2}\right).
\end{align}
Furthermore, note that the real part of $\braket{\partial_{\theta}\psi_{\theta}|\psi_{\theta}}$ is zero for real states, hence the QFI is just given by $ \mathcal{F}(|\psi_{\theta}\rangle) = 4\braket{\partial_{\theta}\psi_{\theta} |\partial_{\theta}\psi_{\theta}}$. The straightforward evaluation of this quantity is inefficient since it requires keeping track of the full state of the system and taking its derivative with respect to the parameter of interest. However, in our case, which involves light interacting with a two-level atom, due to the bipartite nature of the system, the computation of the QFI can be reduced to taking double derivative of quantity that is a function of the solution of four complex ODEs \cite{gammelmark2014fisher, yang2023efficient}.

More specifically, it was shown that the QFI associated with (i) the light emitted, $\mathcal{F}_{L}$,  or (ii) the atom and the light emitted (global QFI), $\mathcal{F}_{G}$, can be computed in a straight forward manner. The QFI can be expressed as \cite{yang2023efficient}
\begin{align}
\label{eq:QFI_evaluation}
    \mathcal{F}_{L(G)}(t) &= -4 \partial_{\zeta}^{2} I_{E(G)}(\theta, \theta + \zeta,t )|_{\zeta=0},
\end{align}
where $I_{E}(\theta_{1}, \theta_{2})$ is the fidelity between two states characterized by $\theta_{1}$ and $\theta_{2}$ of the emitted field and $I_{G}(\theta_{1}, \theta_{2})$ is the fidelity between two global states. These fidelities can be written as
\begin{align}
    I_{E}(\theta_{1},\theta_{2}, t) &= \text{Tr}\left(\sqrt{\mu_{\theta_{1},\theta_{2}}(t)\mu^{\dagger}_{\theta_{1},\theta_{2}}(t)}\right),
\end{align}
and 
\begin{align}
    I_{G}(\theta_{1},\theta_{2}, t) &= |\text{Tr}\left(\mu_{\theta_{1},\theta_{2}}(t)\right)|,
\end{align}
where $\mu_{\theta_{1},\theta_{2}}(t)$ is the generalized density operator. The generalized density operator has the same shape as the density operator associated with the atom, and evolves under the following differential equation
\begin{align}
\label{eq:two_sided}
\frac{d\mu(t)}{dt} &= -i H_{A}(\Gamma_{1},t) \mu(t)  +i \mu(t) H_{A}(\Gamma_{2},t) \nonumber \\
& + J_{A}(\Gamma_{1},t) \mu(t) J_{A}^{\dagger}(\Gamma_{2},t)  -\frac{1}{2}J_{A}^{\dagger}(\Gamma_{1},t)J_{A}(\Gamma_{1},t) \mu(t) \nonumber \\
& -\frac{1}{2} \mu(t) J_{A}^{\dagger}(\Gamma_{2},t) J_{A}(\Gamma_{2},t), 
\end{align}
where $\mu(0)=\rho_{A}(0)$ is the initial state of the atom, $J_{A}(t)$ and $H_{A}(t)$ are the jump operators of the atom and the Hamiltonian of the atom respectively in the interaction picture. Hence with this method, one only needs to evolve four complex ODEs and take the double derivative of a quantity dependent on the solution of these ODEs as opposed to keeping track of all the modes of the waveguide. Note that we do not show the explicit dependence of the operator $\mu(t) \equiv \mu_{\theta_{1}, \theta_{2}}(t)$ on $\theta_{1}$ and $\theta_{2}$ in the above ODEs for the sake of compactness. Also, in the above, all the operators that act on the generalized density operator from the left are $\Gamma_{1}$ dependent while the ones acting from the right are the $\Gamma_{2}$ dependent, and if $\Gamma_{1}=\Gamma_{2}$, we obtain the usual Lindblad master equation. Note that the QFI expression shown in Eq.~\eqref{eq:QFI_evaluation} is typically evaluated using the finite-difference methodologies which may introduce potential instabilities. To avoid this issue, we derive coupled ODEs whose solution directly provides us the QFI.
\section{Real pulses and zero detuning}
\label{sec:real_pulses}
In the section, we focus on real pulses and zero detuning, analyze the factors that play a role in maximizing the QFI and present an optimal pulse. We are interested exclusively in the QFI in the long-time limit, which is the time after the pulse leaves the waveguide and the atom spontaneously decays to the ground state. In this limit, the QFI of the emitted field $\mathcal{F}_{L}$ and the global QFI $\mathcal{F}_{G}$ coincide since the atom decays to the ground state. In this work, we consider only the global QFI in the long-time limit, as it is governed by fewer ODEs than $\mathcal{F}_{L}$ even though the coupled 
ordinary differential equations (ODEs) for both these quantities can be derived in the same manner. Hereafter, we omit the subscript on the global QFI in the remainder of the paper for compactness, $\mathcal{F}_{G}(t)=\mathcal{F}(t)$. We also denote the asymptotic value by $\mathcal{F^{\infty}} = \mathcal{F}(t\rightarrow \infty)$. We show in Appendix~\eqref{appendix:ODEs_derivation} that for a real pulse $f(t)$ and zero detuning, the global QFI is governed by the following four coupled ODEs
\begin{align}
    \dot{x}(t)   &= - \frac{\Gamma}{2}x(t) + 2\sqrt{\Gamma}   f(t) z(t), \label{eq:xdot} \\
    \dot{z}(t)   &= - \Gamma z(t) -2\sqrt{\Gamma} f(t) x(t) -\Gamma,  \label{eq:zdot} \\
    \dot{\xi}(t)  &= - \frac{\Gamma}{2}\xi(t) + \frac{x(t)}{4}+ \frac{f(t)}{2\sqrt{\Gamma}}, \label{eq:xidot} \\
    \dot{\mathcal{F}}(t) &= \frac{1}{2 \Gamma}\left(1+z(t)\right)+ \frac{4f(t)}{\sqrt{\Gamma}} \xi(t), \label{eq:Fdot}
\end{align}
with the initial conditions given by $x(0)=0$, $z(0)=-1$, $\xi(0)=0$ and $\mathcal{F}(0)=0$. The ODE~\eqref{eq:Fdot} is associated with the QFI and can be evaluated in this case without resorting to finite-difference methodologies. Here $x(t)$ and $z(t)$ are the usual optical Bloch equations evolving under the Hamiltonian in Eq.~\eqref{eq:driving_Hamiltonian} with $\delta=0$ and a real pulse $f^{*}(t)=f(t)$, $H  =  \sqrt{\Gamma} f(t) \sigma_{y}$. The solution to the ODE~\eqref{eq:Fdot} can be written as sum of three contributions
\begin{align}
\label{eq:QFI_expression}
    \mathcal{F}(t) &= \mathcal{F}_{p}(t) +\mathcal{F}_{z}(t) +\mathcal{F}_{x}(t),
 \end{align}
 where 
 \begin{align}
 \label{eq:Fp_definition}
     \mathcal{F}_{p}(t) &= \frac{2}{\Gamma} \int_{0}^{t} d\tau\, f(\tau) \left(\int_0^\tau d\tau' \, e^{-\frac{\Gamma(\tau - \tau')}{2}} f(\tau') \right), \\
     \label{eq:Fz_definition}
    \mathcal{F}_{z}(t) &=  \int_{0}^{t} d\tau \left(\frac{1+z(\tau)}{2\Gamma}\right),  \\
    \label{eq:Fx_definition}
    \mathcal{F}_{x}(t) &= \frac{1}{\sqrt{\Gamma}}\int_0^t d\tau\, f(\tau) \left( \int_0^\tau d\tau'\, e^{-\frac{\Gamma(\tau - \tau')}{2}} x(\tau') \right). 
 \end{align}
We focus here on the QFI per unit photon, $\mathcal{F}^{\infty}/\alpha^2$, since this quantity remains bounded by a constant independent of $\alpha$, allowing for meaningful comparisons across different pulse amplitudes. Note that the factor $\mathcal{F}_{p}(t)$ is only dependent on the pulse function and hence we denote it with subscript ``$p$". It can be shown that $\mathcal{F}_{p}(t)/\alpha^{2}$ is always positive and independent of $\alpha^{2}$, so its properties do not change with $\alpha^{2}$ (see Appendix~\eqref{appendix:Fp_properties_real1}). Furthermore, it can be re-casted into bilinear form in $f(\tau)$ that allows us to prove that $\mathcal{F}_{p}^{\infty}/\alpha^{2} \leq 4 $ (See Appendix~\eqref{appendix:Fp_properties_real2}). The second term, which is dependent on $z(t)$, is the cumulative probability of the atom as described in Eq.~\eqref{eq:master_eqn} to be in the excited state, and this contribution is also always positive. This term implies that more the input light is able to excite the atom, the more information we have about the atom's dipole moment. The third term $\mathcal{F}_{x}(t)$ is the average value of $x(t)$ with an exponential memory kernel weighted with the pulse function, and this is the only term that could have a negative contribution, so identifying the pulse that would reduce this contribution is essential to maximizing the QFI. The fact that this term can be negative is evident from the observation that the Hamiltonian generates a rotation around the $+y$-axis when the pulse $f(t)$ is positive, and around the $-y$-axis when $f(t)$ is negative. It should be noted that the second term and the third compete because as the atom is excited strongly, the magnitude of $x(t)$ also increases.
\subsection{Standard pulses}
\begin{table*}[t]
    \begin{ruledtabular}
        \renewcommand{\arraystretch}{2.8} 
        \begin{tabular}{ccccc}
            Pulse type & $f(t)$ & $T_\sigma$ & $\Gamma^{2}\mathcal{F}_{\text{long}}^{\infty}/\alpha^{2}$ & $\Gamma^{2}\mathcal{F}_{\text{short}}^{\infty}/\alpha^{2}$ \\
            \hline
            Rectangular &
            $\dfrac{\alpha}{\sqrt{T}} \bigl(\Theta(t)-\Theta(t-T)\bigr)$ &
            $\dfrac{T}{\sqrt{12}}$ &
            $\dfrac{8}{\Gamma T}\Bigl[2\bigl(1 - e^{-\tfrac{\Gamma T}{2}}\bigr) - \Gamma T e^{-\tfrac{\Gamma T}{2}}\Bigr]$ &
            $\begin{aligned}[t]
              &4 \Bigl[1 - \dfrac{2}{\Gamma T}\Bigl(1-e^{-\tfrac{\Gamma T}{2}}\Bigr)\Bigr] \\
              &+\; \sin^{2}\!\bigl(\alpha \sqrt{T}\bigr)/\alpha^2
            \end{aligned}$ \\
            \hline
            Gaussian &
            $\dfrac{\alpha}{\pi^{1/4}\sqrt{T}} \exp\!\Bigl[-\dfrac{(t-t_{0})^{2}}{2T^{2}}\Bigr]$ &
            $\dfrac{T}{\sqrt{2}}$ &
            $\begin{aligned}[t] &2\Gamma T \biggl(\sqrt{\pi}\,(\Gamma^{2}T^{2}+2)\,e^{\tfrac{\Gamma^{2}T^{2}}{4}}  \\
            &\qquad\quad\mathrm{erfc}\!\Bigl(\tfrac{\Gamma T}{2}\Bigr)- 2\Gamma T \biggr)   \end{aligned}$ &
            $\begin{aligned}[t]
              &2\sqrt{\pi}\,\Gamma T\, e^{\tfrac{\Gamma^{2}T^{2}}{4}} 
                 \,\mathrm{erfc}\!\Bigl(\tfrac{\Gamma T}{2}\Bigr) \\
              &+\;\sin^{2}\!\bigl(\pi^{1/4}\alpha\sqrt{2T}\bigr)/\alpha^2
            \end{aligned}$ \\
            \hline
            Decreasing Exp. &
            $\dfrac{\alpha}{\sqrt{T}} e^{-\tfrac{t}{2T}} \,\Theta(t)$ &
            $T$ &
            $\dfrac{8\Gamma T}{(1 + \Gamma T)^2}$ &
            $\dfrac{4 \Gamma T}{\Gamma T+1} \;+\; \sin^{2}\!\bigl(2 \alpha \sqrt{T}\bigr)/\alpha^{2}$ \\
            \hline
            Rising Exp. &
            $\dfrac{\alpha}{\sqrt{T}} e^{\tfrac{t-t_{0}}{2 T}} \,\Theta(t_{0}-t)$ &
            $T$ &
            $\dfrac{8\Gamma T}{(1 + \Gamma T)^2}$ &
            $\dfrac{4 \Gamma T}{\Gamma T+1} \;+\; \sin^{2}\!\bigl(2 \alpha \sqrt{T}\bigr)/\alpha^{2}$ \\
            \hline
            Symmetric Exp. &
            $\dfrac{\alpha}{\sqrt{T}} e^{-\tfrac{|t-t_{0}|}{T}}$ &
            $\dfrac{T}{\sqrt{2}}$ &
            $\dfrac{64 \Gamma T}{(2 + \Gamma T)^3}$ &
            $\dfrac{4\Gamma T(\Gamma T+4)}{(\Gamma T+2)^{2}} \;+\; \sin^{2}\!\bigl(2 \alpha \sqrt{T}\bigr)/\alpha^{2}$ \\
        \end{tabular}
    \end{ruledtabular}
\caption{\label{table:pulses} 
    Analytic expressions for standard pulse shapes, their variance, and quantum Fisher information (QFI) in the long and short pulse width limits. For Gaussian, rising-exponential, and symmetric-exponential pulses, $t_{0}$ is chosen sufficiently large so that the pulse is effectively zero at $t=0$.}
\end{table*}

\begin{figure}[t]
\includegraphics[width=0.9\linewidth]{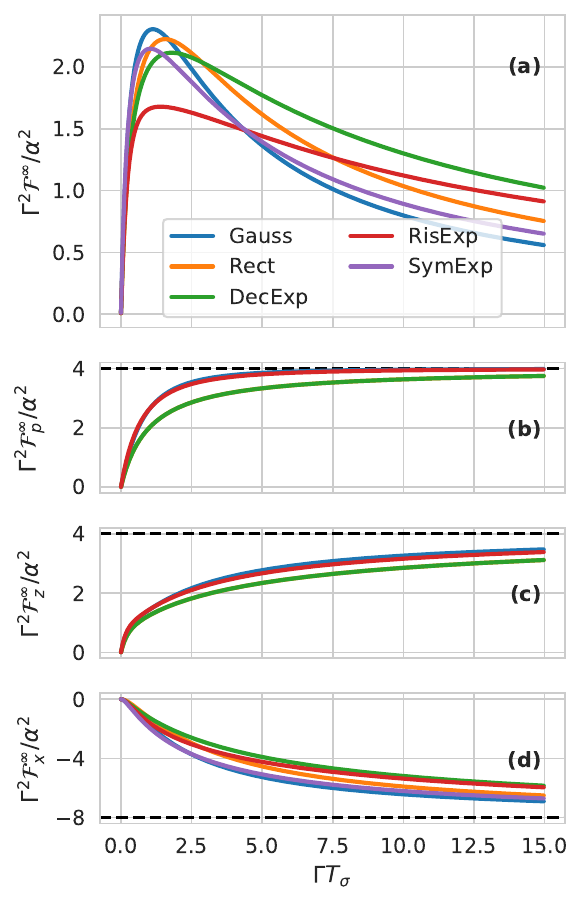}
\caption{\textbf{(a)} The QFI per unit photon, $\mathcal{F}^{\infty}/\alpha^{2}=(\mathcal{F}_{p}^{\infty}+\mathcal{F}_{z}^{\infty}+\mathcal{F}_{x}^{\infty})/\alpha^{2}$, of standard pulses plotted as a function of the variance of the pulse $T_{\sigma}$ for $\alpha^{2}=1$. See Table \eqref{table:pulses} for more details about the pulses. All these pulses show similar behavior with $T_{\sigma}$: the QFI initially increases with the pulse width, reaches a maximum, and then decreases toward zero as the width becomes larger. \textbf{(b, c)} Both $\mathcal{F}_{p}^{\infty}/\alpha^2$ and $\mathcal{F}_{z}^{\infty}/\alpha^2$  are always positive and approach $4$ as $T_{\sigma} \rightarrow \infty$. \textbf{(d)} $\mathcal{F}_{x}^{\infty}/\alpha^2$ can be the only negative term among the three terms that contribute to QFI and it approaches $-8$ as $T_{\sigma}\rightarrow \infty$. See also Fig.~\eqref{fig:qfi_system_size} for the QFI plotted at other values of $\alpha^{2}$. }
\label{fig:qfi_standard}
\end{figure}
\begin{figure*}[t!]
\includegraphics[width=\textwidth]{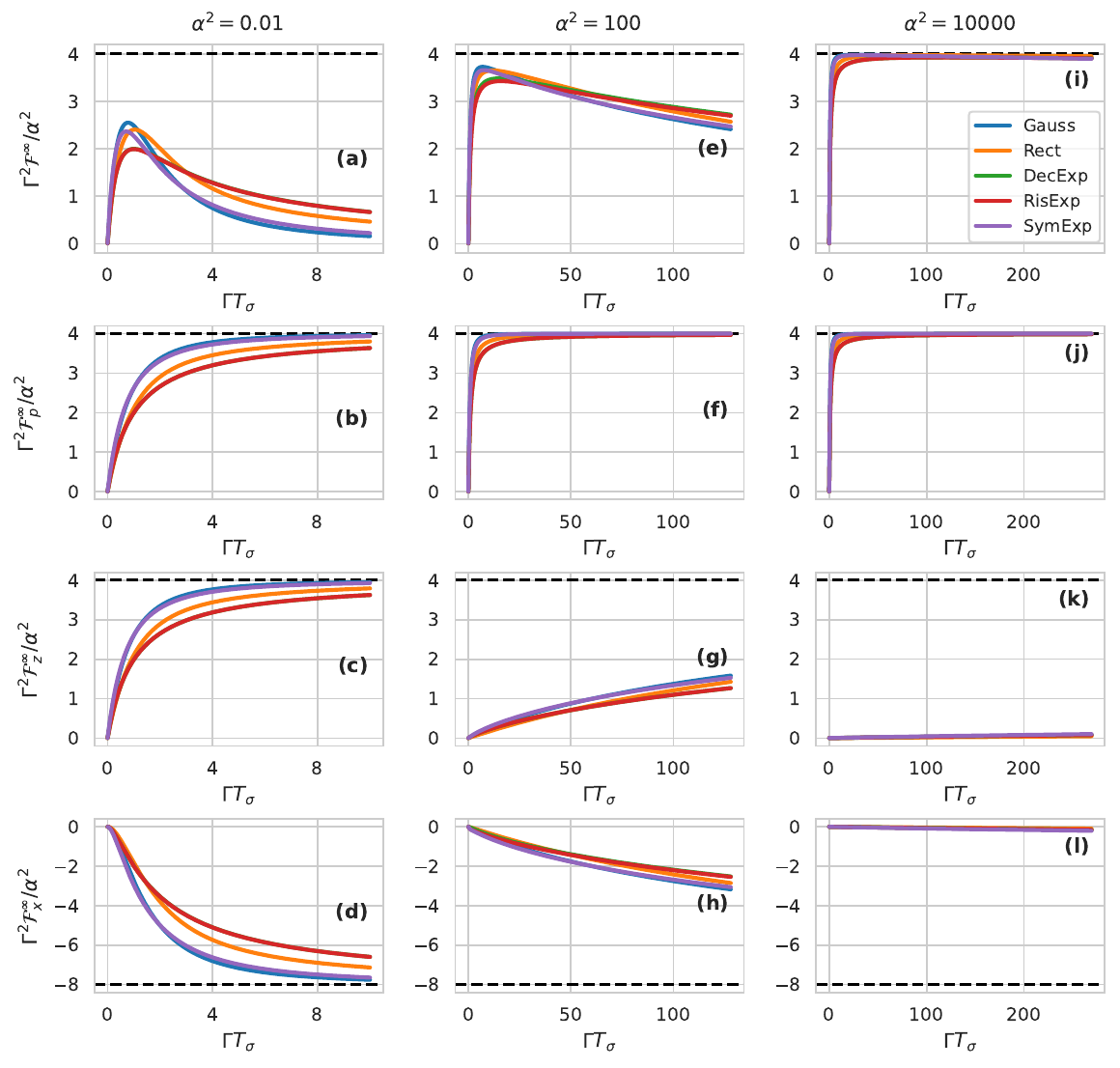}
\caption{The QFI per unit photon of $\alpha^{2}=10^{-2}$ for the standard pulses is plotted in (a) while the contributions $\mathcal{F}_{p}^{\infty}/\alpha^{2}$, $\mathcal{F}_{z}^{\infty}/\alpha^{2}$ and $\mathcal{F}_{x}^{\infty}/\alpha^{2}$ are plotted in \textbf{(b)}, \textbf{(c)} and \textbf{(d)} respectively. These quantities are all defined below Eq.~\eqref{eq:QFI_expression}, and satisfy $\mathcal{F}^{\infty}=\mathcal{F}_{p}^{\infty}+\mathcal{F}_{z}^{\infty}+\mathcal{F}_{x}^{\infty}$, with $\mathcal{F}_{x}^{\infty}$ being the only term that can have negative contribution. \textbf{(e-h):} Same as (a-d) for $\alpha^{2}=10^{2}$.\textbf{(i-l):} Same as (a-d) for $\alpha^{2}=10^{4}$. The QFI per unit photon exhibits similar behavior as $\alpha^{2}$ varies: it rises sharply for small $\Gamma T_{\sigma}$, reaches a maximum, and then decreases to zero as $\Gamma T_{\sigma}$ increases further. However, for larger $\alpha^{2}$, the decay toward zero is slower. It is evident that $\mathcal{F}_{p}^{\infty}/\alpha^{2}$ exhibits the same behavior across all $\alpha^{2}$ values, as expected. Therefore, any variation in the behavior of QFI per unit photon can be attributed to changes in $\mathcal{F}_{z}^{\infty}/\alpha^{2}$ and $\mathcal{F}_{x}^{\infty}/\alpha^{2}$. As $\alpha^{2}$ increases, the negative contribution from $\mathcal{F}_{x}^{\infty}/\alpha^{2}$ decreases at smaller $T_{\sigma}$ enabling all the standard pulses to become optimal for these smaller widths and large $\alpha^{2}$. The legend for all the plots is provided in (i). Note that the behavior of QFI per unit photon in \textbf{(a)} is identical to the single-photon pulse behavior shown in the work by Albarelli et al.~\cite{albarelli2023fundamental} for the reasons explained in the text.}
\label{fig:qfi_system_size}
\end{figure*}

In this subsection, we analyze the QFI,  $\mathcal{F}^{\infty}/\alpha^2$, associated with various standard pulses using the coupled ODEs presented above. These standard pulse shapes were studied in the context of single-photon pulses interacting with a two-level atom~\cite{albarelli2023fundamental, wang2011efficient}. The analytic expressions for the pulse shapes $f(t)$ analyzed here are given in Table~\ref{table:pulses}. The QFI of these pulses is plotted in Fig.~\eqref{fig:qfi_standard}  along with the contributions from $\mathcal{F}_{p}^{\infty}$, $\mathcal{F}_{z}^{\infty}$ and $\mathcal{F}_{x}^{\infty}$. It can be seen here that all the standard pulses have similar behavior as a function of variance of the pulses, $T_{\sigma}$. That is, sharp increase at small $T_{\sigma}$ until the QFI reaches maximum and then slow decrease to zero for large values of $T_{\sigma}$. The behavior of the QFI in the short pulse width limit can be attributed to $\mathcal{F}_{p}^{\infty}$ since the atom does not have time to respond to short pulses and both $x(t)$ and $z(t)$ remain closer to the initial values in this regime as corroborated by Fig.~\eqref{fig:qfi_system_size} (also see section~\eqref{sec:short_pulses} for more details).
As shown in Fig.~\eqref{fig:qfi_standard}(a), $\mathcal{F}_{p}^{\infty}/\alpha^2$ is a monotonically increasing function with $T_{\sigma}$ for all the pulses: it starts from zero and approaches $4$ in the limit $T_{\sigma} \rightarrow \infty$, which is the maximum possible value of this term. Focusing on the large $T_{\sigma}$ region, the influence of the second and the third terms can be understood using the quasi-steady state approximation, where we assume that the pulse changes slowly compared to the spontaneous emission rate. Hence, the atom has time to approach its steady state temporarily in time (quasi-steady state), which allows us to compute the steady state values, $x_{s.s.}(t)$ and $z_{s.s.}(t)$, by setting the time derivatives of $x(t)$ and $z(t)$ to zero. Using this, we obtain the following expression for $\mathcal{F}_{z}^{\infty}$ for the rectangular pulse
 \begin{align}
    \mathcal{F}_{z}^{\infty} &\approx \int_{0}^{\infty} dt \frac{4 f^{2}(t)}{1+8 f^{2}(t)} =  4 \alpha^{2}\frac{T_{\sigma}+1 }{T_{\sigma}+8 \alpha^{2}}, \\
 \mathcal{F}_{x}^{\infty} &\approx- 4\int_{0}^{\infty}\,d\tau f(\tau) \int_{0}^{\tau}\,d\tau'\frac{e^{\frac{1}{2}(\tau'-\tau)}f(\tau')} {1 + 8f(\tau')^{2}} \\
 &\approx -8 \alpha^2\frac{\left(T_{\sigma}+2(e^{-T_{\sigma}/2}-1) \right)}{T_{\sigma} + 8 \alpha^2 },
 \end{align}
 which approaches $-8 \alpha^{2}$ in the limit $T_{\sigma}/\alpha^{2} \gg 1$. As a result, $\mathcal{F}^{\infty}= 4\alpha^{2} + 4 \alpha^{2}-8\alpha^{2} =0$ in this limit. Note that all the other pulses considered have similar shape in the limit of large $T_{\sigma}$. Hence, for these pulses, the QFI is zero both in the limit $T_{\sigma} \rightarrow 0$ and $T_{\sigma} \rightarrow \infty$, and the optimal width to obtain best the QFI is determined by competition between the terms with positive contribution, $\mathcal{F}_{p}^{\infty}$ and $\mathcal{F}_{z}^{\infty}$, and the negative contribution resulting from $\mathcal{F}_{x}^{\infty}$. 
\subsection{Short-pulse-width regime}
\label{sec:short_pulses}
 Let $s \equiv t/T$ be a dimensionless parameter whose value changes between $0$ and $1$ as $t$ varies between $0$ and the width of the pulse, $T$. Assuming the shape of the pulse is scale invariant, we can factor out the width of the pulse from the pulse function as $f(t) \equiv \left( \alpha/\sqrt{T}\right) \phi(t/T)$ where $\phi(s)$ is a dimensionless function with unit norm, $\int ds \; \phi(s)^{2} = 1$. Note that all the pulses that are of interest to us in this work can be expressed in this manner. That is, the standard pulses considered in this work, the harmonic basis functions and the Hermite Gaussian basis functions are all scale invariant. For the definition of the harmonic basis functions and the Hermite Gaussian basis functions, see Appendix \eqref{appendix:basis}. Using this parametrization, the dominant terms in the ODEs become evident in the limit $\sqrt{\Gamma T} \ll \alpha$, which then allows us to obtain the QFI analytically in this limit. Expressing ODEs in Eqs.~\eqref{eq:xdot} to \eqref{eq:Fdot} in terms of derivatives of $s$ and keeping the leading order terms, we have
\begin{align}
\label{eq:short_time_limit}
\frac{d}{ds}x(s)  &= 2 \alpha \sqrt{\Gamma T}\phi(s) z(s), \\
\frac{d}{ds}z(s) &= -2 \alpha\sqrt{\Gamma T} \phi(s) x(s), \\
\label{eq:ODE_xi(s)}
\frac{d}{ds}\tilde{\xi}(s) &= -\frac{\Gamma T}{2} \tilde{\xi}(s)+\frac{\alpha \sqrt{\Gamma T} }{2}\phi(s), \\
\frac{d}{ds} \tilde{\mathcal{F}}(s) &= 4\alpha\sqrt{\Gamma T} \phi(s)\tilde{\xi}(s),
\end{align}
where we define unitless quantities, $\tilde{\xi}(s) \equiv \Gamma \xi(s)$ and $\tilde{\mathcal{F}} \equiv \Gamma^{2} \mathcal{F}(s)$. Note that the term $(\Gamma T/2) \tilde{\xi}(s)$ cannot be ignored in Eq.~\eqref{eq:ODE_xi(s)} because $\tilde{\xi}(s)$ can be on the order of $\alpha$. The QFI in this limit can be therefore expressed in terms of time as
\begin{align}
    \mathcal{F}_{\text{short}}(t) &=  
    \begin{cases}
    \mathcal{F}_{p}(t)  & t \leq T \\
    \mathcal{F}_{p}(T) + \sin^{2}\left(A\right) \left(1-e^{-\Gamma(t-T)}\right) & t> T,
    \end{cases}
\end{align}
where $A$ is the area of the pulse, $A = \int_{0}^{T} d\tau f(\tau) = \alpha \sqrt{T} \int_{0}^{1} ds \phi(s)$, and $\mathcal{F}_{p}(t)$ is defined in Eq.~\eqref{eq:Fp_definition}. In the long-time limit, $t \rightarrow \infty$, which is of interest to us, the above reduces to
\begin{align}
\label{eq:QFI_short_pulse}
    \mathcal{F}_{\text{short}}^{\infty} &= \mathcal{F}_{p}(T) + \sin^{2}\left(A\right), 
\end{align}
where $\mathcal{F}_{p}(T)$ is the QFI gained during the pulse duration and $\sin^{2}\left(A\right)$ quantifies the information obtained during spontaneous emission. For the QFI analytic expressions associated with the standard pulses in this limit, see Table~\ref{table:pulses}.

Focusing on the optimal pulse in this limit for large $\alpha^{2} \gg 1$, the problem is equivalent to maximizing $\mathcal{F}_{p}(T)/\alpha^{2}$ since the factor $\sin^{2}\left(A\right)/\alpha^{2}$ becomes negligible as $\alpha^{2}$ increases. The optimization can be done by first expressing QFI in the bilinear form, $\mathcal{F}_{p}(T)/\alpha^{2} = \sum_{m,n}c_{n} K_{m,n}c_{n}$ where $c_{m}$ and $c_{n}$ are the coefficients of the pulse $f(t)$ in some basis, and diagonalizing the $K$ matrix. The optimal QFI is the maximum eigenvalue and the optimal pulse is the corresponding eigenvector. For pulses that satisfy the condition $f(t=0)=f(t=T)=0$, the eigenvector corresponding to the maximum eigenvalue is a superposition of first few odd harmonics with the maximum support on the first harmonic, $\sqrt{2/T} \sin(\pi t/T)$. As $\Gamma T$ becomes larger, the support over the first harmonic increases and it is the eigenvector corresponding to maximum eigenvalue in the limit $\Gamma T \rightarrow \infty$ (see Appendix~\eqref{appendix:Fp_properties_real2} for more details). Also, note that when the pulse is very narrow, $\mathcal{F}_{\text{short}}^{(\infty)} = 2A^{2}$ since $\mathcal{F}_{p}(T) \approx A^{2}$ and $\sin^{2}\left(A\right) \approx A^{2}$. The rectangular pulse is optimal in this case, as this is the shape that maximizes the area under the pulse for a given finite width $T$ and the normalization, $\int dt f(t)^{2}=\alpha^2$. This can be shown in a straightforward manner using the Cauchy-Schwarz inequality,
\begin{align}
    A^{2} = \left(\int_{0}^{T} dt f(t) \right)^2 \leq \left(\int_{0}^{T} dt\right) \left(\int_{0}^{T} dt f(t)^2 \right),
\end{align}
where the bound is saturated when $f(t)$ is a constant pulse.  
\subsection{QFI behavior as a function of average photon number}
\label{sec:QFI_alpha2}
We analyze how the behavior of the QFI changes for the standard pulses as a function of $\alpha^2$ in this subsection. As illustrated in Fig.~\eqref{fig:qfi_system_size}, the QFI has a similar behavior for all $\alpha^2$: the value increases with the variance until it reaches the optimal width and then decreases to zero for larger variances. However, as the value of $\alpha^2$ increases, the maximum value of the QFI increases for all the pulses, and the non-zero width where it goes to zero also increases. It should be noted particularly that for large $\alpha^2$, as seen in Fig.~\eqref{fig:qfi_system_size},  the QFI per unit photon of all the standard pulses approaches $4$, which is the optimal value as shown in subsection \ref{sec:optimal_pulses}.
This can be understood by analyzing the behavior of three components of $\mathcal{F}^{\infty}/\alpha^2$ as defined in \eqref{eq:QFI_expression}. The behavior of $\mathcal{F}_{p}^{\infty}/\alpha^{2}$ is independent of $\alpha$, and it approaches $4$ as the variance of the pulse $T_{\sigma}$ is increased for all the standard pulses since they all resemble a square pulse in the $T_{\sigma}\rightarrow \infty$. Hence, the system-size effects arise only from $\mathcal{F}_{z}^{\infty}/\alpha^2$ and $\mathcal{F}_{x}^{\infty}/\alpha^2$. As the value of $\alpha$ is increased, the contribution both from $\mathcal{F}_{z}^{\infty}/\alpha^2$ and $\mathcal{F}_{x}^{\infty}/\alpha^2$ remains closer to zero for a wider range of $T_{\sigma}$ as opposed to dominating negative contribution from $\mathcal{F}_{x}^{\infty}/\alpha^2$ even at smaller $T_{\sigma}$ when $\alpha$ is small. This is because the range of the short-width limit defined by $\sqrt{\Gamma T} \ll \alpha$ increases as the value of $\alpha$ is increased, where $\mathcal{F}_{x}^{\infty}/\alpha^2$ and $\mathcal{F}_{z}^{\infty}/\alpha^2$ do not contribute as already shown in Eq.~\eqref{eq:QFI_short_pulse}. Hence, all the standard pulses reach global maximum of QFI for large $\alpha^2$ at the optimal width. For small $\alpha^2$ (Fig.~\eqref{fig:qfi_system_size}(a)), the QFI behavior of the pulses is identical to single-photon pulses studied in \cite{albarelli2023fundamental} for the reasons explained in the next subsection.
\subsection{Long pulse width regime}
As the width of a pulse increases, its amplitude decreases because its normalization is fixed, which is proportional to its energy. Hence, we can resort to perturbation theory to derive analytic expressions for the system of interest. Using the parametrization from the previous subsection $f(t) \equiv \left( \alpha/\sqrt{T}\right) \phi(t/T) \equiv \varepsilon \phi(t/T)$, we see that the parameter $\varepsilon \ll 1$ in this limit. The perturbation theory expression up to second order in $\varepsilon$ is then given by 
\begin{align}
\label{eq:QFI_long}
&\mathcal{F}_{\text{long}}(t) =\mathcal{F}_{p}(t) + \frac{1}{\Gamma}\int_0^t d\tau\, p(\tau)^2 \nonumber\\
&- \frac{2}{\sqrt{\Gamma}} \int_{0}^t d\tau\, f(\tau) \left( \int_{0}^\tau d\tau'\, e^{-\frac{(\tau - \tau')}{2}} p(\tau') \right),
\end{align}
where 
\begin{align}
\label{eq:pt_definition}
    p(t) &\equiv  \sqrt{\Gamma} \int_{0}^{t} d\tau \; e^{\frac{\Gamma (\tau-t)}{2}}f(\tau),
\end{align}
and $\mathcal{F}_{p}$ is defined in Eq.~\eqref{eq:Fp_definition}. For more details on the perturbation theory result, see Appendix~\eqref{appendix:perturbation_theory_real}. Note that $|p(t)|^{2}=(1+z(t))/2$ is the probability of the atom to be in the excited state up to second order in $\varepsilon$. In the long-time limit, which is the time after the atom spontaneously decayed to the ground state, the above expression for the QFI reduces to 
\begin{align}
    \label{eq:qfi_long_time_limit}
    \mathcal{F}_{\text{long}}^{\infty} &= 16\int_{0}^{\infty} dt \left(\dot{q}(t)\right)^{2},
\end{align}
where
\begin{align}
\label{eq:qt_definition}
    q(t) &\equiv \frac{\partial}{\partial \Gamma} \left(\frac{p(t)}{\sqrt{\Gamma}}\right) =\int_{0}^{t} d\tau \left(\frac{\tau -t}{2}\right) e^{\Gamma\left(\frac{\tau-t}{2}\right)} f(\tau)\\
    &= -\frac{1}{2\sqrt{\Gamma}}  \int_{0}^{t} d\tau' \;  e^{\frac{\Gamma (\tau'-t)}{2}} p(\tau').
\end{align}
 The expression for the QFI in this limit is quadratic in pulse shape $f(t)$, and therefore it can be written in the bilinear form. More specifically, evaluating the QFI in a set of basis functions, the above expression can be written as $\mathcal{F}^{\infty}_{\text{long}} = \sum_{m,n}c_{m}K_{mn}c_{n}$, where $c_{n}$ is the $n^{th}$ coefficient of the pulse $f(t)$ in a given basis set and the dimension of the matrix $K$ is the number of basis functions considered. Here, we choose the basis of harmonic functions (see Appendix~\eqref{appendix:basis}) that satisfy the boundary condition $f(0)=f(T)=0$, and take the symmetric part of the matrix $K^{S} = (K+K^{T})/2$ since the antisymmetric part does not contribute to the QFI, $\sum_{m,n} c_{m}K^{A}_{mn}c_{n}=-\sum_{m,n} c_{m}K^{A}_{mn}c_{n} =0$. where $K^{A} = ((K-K^{T})/2)$. Therefore, the maximum QFI can then be found by diagonalizing the symmetric part of the matrix $K^{S}_{m,n}$ where the eigenvector corresponding to the maximum eigenvalue is the optimal pulse and eigenvalue is the optimized QFI. It can be shown that the matrix $K^S$ becomes diagonal in the $\Gamma T \gg 1$ limit with the harmonic $\sqrt{(2\alpha^{2}/T)}\sin(\pi t/2T)$ having the largest eigenvalue of $4 \alpha^{2}$. In terms of contributions of the individual terms in Eq.~\eqref{eq:QFI_expression}, this pulse ensures that the contribution of $\mathcal{F}_{x}=0$, which is the only term with potential negative contribution, and the other two terms provide an equal contribution of $2\alpha^{2}$. The details associated with the diagonalisation are explicilty shown in Appendix \eqref{appendix:diagonalizing_qfi_long_time_real}.

In Appendix~\eqref{appendix:single_photon_pulses}, we have shown that the QFI up to second order in $\varepsilon$ in the long pulse width limit, given in Eq.\eqref{eq:qfi_long_time_limit}, is identical to the QFI of a single-photon pulse, whose expression is provided in \cite{albarelli2023fundamental}. In the single-photon pulse case, $p(t)$ corresponds to the probability amplitude of the atom to be in the excited state up to an overall sign. Also, the analytic expressions for the QFI in this limit for the standard pulses, which are evidently identical to the expressions for the single-photon pulses, are provided in Table \eqref{table:pulses}. In summary, for real pulses, the maximum QFI per unit photon for both the coherent state in the long time limit and the single photon pulse is $4$ with the optimal pulse given by $\sqrt{(2\alpha^{2}/T)}\sin(\pi t/2T)$. Also, see \cite{sourav2025} for more information on the optimal QFI of a single-photon pulse. 
\subsection{Optimal Pulses}
\label{sec:optimal_pulses}
\begin{figure*}[t]
\includegraphics[width = \textwidth]{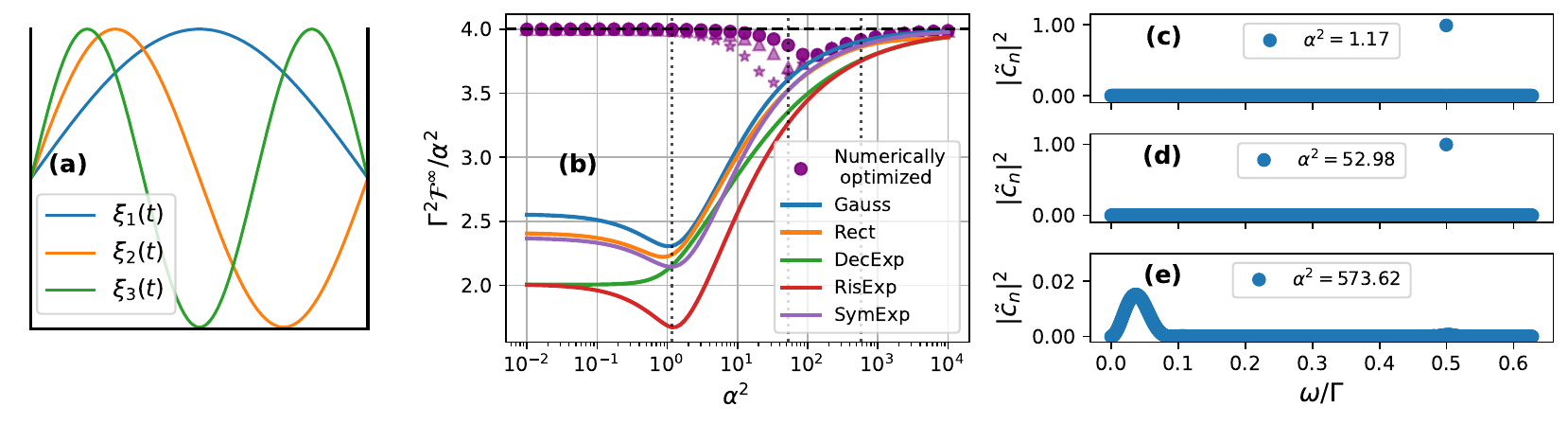}
\caption{\textbf{(a)} The first three harmonics $\xi_{n} = \sin(n\pi t/T)$ are shown here. \textbf{(b)} The QFI of the numerically optimized pulse, expressed in the basis of harmonics, is shown for $\Gamma T=700$ (stars), $\Gamma T=2000$ (triangles) and $\Gamma T=5000$ (dots) for $n_{\text{max}}=1000$, which is the maximum number of basis functions considered. We consider $10$ random initial seeds and plot the maximum QFI value obtained from these seeds. As seen, the numerically optimal QFI per unit photon gets closer to $4$ as the length of pulse $T$ is increased. In other words, the optimal pulse is $\sqrt{2/T}\sin(\Gamma t/2)$ in the regime $\alpha/\sqrt{\Gamma T} \ll 1$ as confirmed by the plots (c) and (d). However, outside this long pulse width regime, the Gaussian pulse (along with other standard pulses as seen in Fig.~\eqref{fig:qfi_system_size}) becomes optimal for larger values of $\alpha^{2}$.  For reference, the optimal QFI (maximized over the pulse width) of standard pulses is plotted as a function of $\alpha^{2}$. \textbf{(c-e)} The normalized population $|\tilde{c}_{n}|^2=|c_{n}|^{2}/\alpha^{2}$ of the optimal pulse in the harmonic basis functions is shown here corresponding to three different values of $\alpha^{2}$ identified by the vertical dotted lines in (b). For smaller values of $\alpha^{2}$ ((c) and (d) for $\Gamma T=5000$), it can be seen that all the population of the optimized pulse is in the harmonic $\omega=\Gamma/2$ where  $\omega=n\pi/T$.}
\label{fig:qfi_optimization}
\end{figure*}
In this section, we present the results of the numerical optimization for identifying the real pulse that maximizes the QFI. For this we consider the basis of harmonics with the boundary condition $f(0)=f(T)=0$, and the first few harmonics are shown in Fig.~\eqref{fig:qfi_optimization}(a). In Fig.~\eqref{fig:qfi_optimization}(b), different colored curves correspond to the optimal QFI of various standard pulses obtained by optimizing the width $T_{\sigma}$ for each $\alpha^{2}$. For $\alpha^{2} \gtrsim 10$, the optimal QFI per unit photon of these pulses grow according to a power law reaching values closer to $4$ for the reasons explained in Sec.~\ref{sec:QFI_alpha2}. For smaller values of $\alpha^{2}$, the optimal QFI of these pulses has a different behavior. The data points in purple (circles, triangles and stars) correspond to the optimal values of the QFI from the numerical optimization for $\Gamma T=700$, $\Gamma T=2000$ and $\Gamma T=5000$ respectively and $\Gamma T$ is the maximum width of the pulse. As can be noted, the maximum QFI per unit photon is at most $4$. For each $\alpha$ and $\Gamma T$, we present the maximum value of the QFI obtained from 10 initial seeds among which 9 initial seeds are chosen randomly and one seed initialized according to $c_{n}=\alpha \delta_{n, [T/2\pi]}$, which corresponds to the optimal pulse obtained analytically in the long pulse width limit. The optimization is performed using LBFGSB optimizer in the  JAXopt package~\cite{jaxopt_implicit_diff} and an ODE solver from Diffrax package~\cite{kidger2021on} compatible with JAX. As can be seen from comparing data points (circles, stars and triangles) in Fig.~\eqref{fig:qfi_optimization}(b), the optimal value of QFI per unit photon is less than $4$ is due to limited width of the pulse. The normalized population of the optimal pulse in the basis is shown in Fig.~\eqref{fig:qfi_optimization}(c) for three different values of $\alpha^{2}$. For smaller $\alpha^{2}$, we see that $\sqrt{2/T}\sin(\Gamma t/2)$ is the optimal pulse and the optimal pulse approaches a Gaussian pulse for larger values of $\alpha^{2}$. All of this can be understood in the following manner: the pulse $\sin(\Gamma t/2)$ in the long pulse width limit, which is defined as $\alpha/\sqrt{\Gamma T} \ll 1$ (same as the single-photon limit for real pulses), is the optimal pulse with the QFI per unit photon $4$. As we increase $\alpha$, the width of the pulse needs to be longer to remain in this long pulse width regime. However, for very large $\alpha^{2}$, all the standard pulses also have optimal QFI of $4$ as explained in Sec.~\ref{sec:short_pulses}, and the numerical optimization provides us the Gaussian pulse for large $\alpha^{2}$ since it reaches $4$ faster compared to other pulses.

In summary, for real pulses, $\sqrt{2/T}\sin(\Gamma t/2)$ is optimal in the long pulse width limit. However, when the width of the pulse is limited, all the standard pulses reach the optimal value for large $\alpha^{2}$ and the rectangular pulse becomes optimal for very short widths.
\section{Complex Pulses}
\label{sec:complex_pulses}
In this section, we analyze the general case where the pulse can be complex $f(t)$ with nonzero detuning in the system. In this case, the relevant equations of motion for the global QFI are given by
\begin{align}
    x_1'(t) &= -(\Gamma/2) x_1(t) + 2\sqrt{\Gamma} f_{r}(t) z_1(t) - \delta y_1(t), \label{eq:complex_ODE_x1_noncompact}\\
    y_1'(t) &= -(\Gamma/2)y_1(t) - 2\sqrt{\Gamma} f_{i}(t) z_1(t) + \delta x_1(t), \label{eq:complex_ODE_y1_noncompact}\\
    z_1'(t) &= -\Gamma z_1(t) - 2\sqrt{\Gamma} f_{r}(t) x_1(t) + 2 \sqrt{\Gamma} f_{i}(t) y_1(t)\nonumber \\
    &\qquad -\Gamma, \label{eq:complex_ODE_z1_noncompact}\\
    w_{1}'(t) &= \left( f_{i}(t) x_1(t) + f_{r}(t) y_1(t) \right)/2 \sqrt{\Gamma}, \label{eq:complex_ODE_w1_noncompact} \\
    x_2'(t) &= -(\Gamma/2)x_2(t) + 2 \sqrt{\Gamma}f_{r}(t) z_2(t) - \delta y_2(t) \nonumber \\
    &\qquad - (y_1(t)/4) + \left(f_{i}(t)/2 \sqrt{\Gamma} \right), \label{eq:complex_ODE_x2_noncompact} \\
    y_2'(t) &= -(\Gamma/2)y_2(t) - 2\sqrt{\Gamma} f_{i}(t) z_2(t)  + \delta x_2(t) \nonumber  \\
    & \qquad + (x_1(t)/4) + \left(f_{r}(t)/2\sqrt{\Gamma} \right), \label{eq:complex_ODE_y2_noncompact}
\end{align}
\begin{align}
    z_2'(t) &= -\Gamma z_2(t) - 2\sqrt{\Gamma} f_{r}(t) x_2(t) + 2\sqrt{\Gamma} f_{i}(t) y_2(t) \nonumber\\
    &\qquad - \Gamma w_{1}(t), \label{eq:complex_ODE_z2_noncompact}\\
    w_{2}'(t) &= \left( f_{i}(t) x_2(t) + f_{r}(t) y_2(t) \right)/2 \sqrt{\Gamma}, \label{eq:complex_ODE_w2_noncompact}\\
    \mathcal{F}'(t) &= (1/2\Gamma) \left( 1 + z_1(t) \right)  +4 \frac{d}{dt}\left(2 w_{2}(t)-w_{1}(t)^{2} \right), \label{eq:complex_ODE_F_noncompact}
\end{align}
where the first three equations of motion associated with $x_{1}(t)$, $y_{1}(t)$ and $z_{1}(t)$ describe the evolution of atom's $\langle \sigma_{i} \rangle$ for $i=\{x,y,z\}$ under the Lindblad master equation shown in Eq.~\eqref{eq:master_eqn}.

The variable $w_{1}(t)$ has been introduced for compactness,  and note that $\Gamma \dot{w}_{1}(t) = i \sqrt{\Gamma} \langle  (f(t)^{*}\sigma - f(t) \sigma^{\dagger}) \rangle$ is the work done on the atom under the drive shown in Eq.~\eqref{eq:master_eqn}. Note that the variables $x_{2}(t)$, $y_{2}(t)$ and $z_{2}(t)$  are not associated with $\langle \sigma_{i}\rangle$ of a different atom, but result from the derivatives of $x_{1}(t)$, $y_{1}(t)$, $z_{1}(t)$ with respect to $\zeta$, which is an auxiliary variable used in obtaining the QFI (see Appendix \eqref{appendix:ODEs_derivation} for more details). These variables with subscript $2$ have the units of time, and they have been labeled in this manner because the equations of motion of these variables are similar to the those of $x_{1}(t)$, $y_{1}(t)$ and $z_{1}(t)$ with additional driving terms. Likewise, $w_{2}(t)$ is the analog of $w_{1}(t)$. The QFI can then be compactly written as
\begin{align}
\label{eq:QFI_expression_complex}
    \mathcal{F}(t) &=  \int_{0}^{t} d\tau \left(\frac{1+z_{1}(\tau)}{2\Gamma}\right) + 4 \left(2 w_{2}(t)-w_{1}(t)^{2}\right).
 \end{align}
 Note that alternatively we can move to a different frame of reference defined by $H_{0}= \delta \sigma^{\dagger}\sigma$ where the interaction Hamiltonian is given by $H_{\text{int}}=  i \sqrt{\Gamma} \left(f(t)^{*}e^{-i\delta t}\sigma - f(t) e^{i\delta t}\sigma^{\dagger}\right)$. In this frame of reference, the effect of detuning is replaced by adding a phase factor to the complex pulse $e^{i\delta t} f(t)$. Since this unitary transformation is independent of the parameter we are trying to estimate, the QFI is invariant under this transformation.
  
As before, we can analyze the optimal pulses in the long pulse width limit by resorting to the perturbation theory. The QFI in this limit up to second order in $\varepsilon$ is given by 
\begin{align}
    \label{eq:qfi_complex_long_time_limit}
    \mathcal{F}_{\text{long}}^{\infty} &= 16\int_{0}^{\infty} dt \left(\dot{Q}(t)\right)^{2},
\end{align}
where
\begin{align}
    Q(t) &\equiv \frac{\partial}{\partial \Gamma} \left(\frac{P(t)}{\sqrt{\Gamma}}\right) = -\frac{1}{2\sqrt{\Gamma}}\int_{0}^{t} d\tau \; e^{\frac{\Gamma}{2}(\tau-t)}P(\tau), \\
    P(t) &\equiv \sqrt{\Gamma} \int_{0}^{t} d\tau \; e^{\frac{\Gamma}{2} (\tau-t)} e^{-i\delta \tau} f^{*}(\tau),
\end{align}
which is the generalization of variables $p(t)$ and $q(t)$ defined in the real case with zero detuning as shown in Eq.~\eqref{eq:pt_definition} and Eq.~\eqref{eq:qt_definition}. As before, $|P(t)|^{2} = (1+z(t))/2$ is the probability of the atom to be in the excited state. Also, since $Q(t)$ is quadratic in pulse shape $f(t)$, we follow similar procedure for optimizing QFI, and through diagonalization we find that the optimal pulse in this limit is given by $f(t) = \sqrt{2/T}e^{-i\delta t} \sin(\Gamma t/2)$ with $\Gamma^{2}\mathcal{F}_{\text{opt}}^{\infty} =4$ under the boundary conditions $f(0)=f(T)=0$. For more details, see Appendix~\eqref{appendix:diagonalize_open_boundary_conditions}. Hence, the optimal pulse in this case is identical to the optimal pulse in the real case since the extra phase factor can be associated with moving to a different frame of reference where the effect of detuning is replaced by a phase in the pulse. If we optimize under the periodic boundary conditions $f(0)=f(T)$, we obtain the optimal pulse $\sqrt{\alpha/T}e^{i\omega t}$ where $\omega = \delta \pm \Gamma/2$ with $\Gamma^{2}\mathcal{F}_{\text{opt}}^{\infty} =4$ as shown in Appendix \eqref{appendix:diagonalize_periodic_boundary_conditions}.  We also find through numerical optimization that the maximum QFI per unit photon is $4$ even when the pulses are allowed to be complex and arbitrary width (as opposed to long pulse width regime), and we do not gain any QFI when allowing the pulses to complex. In other words, the optimal pulse is given by $f(t) = \sqrt{2/T}e^{-i\delta t} \sin(\Gamma t/2)$ or $\sqrt{\alpha/T}e^{i \left(\delta \pm \Gamma/2\right) t}$ with the maximum QFI,  $\mathcal{F}^{\infty}/\alpha^{2} = 4$.

Also, unlike the case of real pulses, the QFI in this limit for complex pulses is not equal to that of the single-photon pulses. This can be understood by the fact for single photon pulses, the QFI is given by
\begin{align}
    \mathcal{F}_{\text{single}}^{\infty} &= 16 \int_{0}^{\infty} dt \; |\partial_{t}Q(t)|^{2}  \nonumber\\
    &- \frac{4}{\Gamma^2} \left|\int_{0}^{\infty} dt \; \left(\partial_{t}P(t)\right) P(t)^{*} \right|^{2},
\end{align}
which is fourth order in the pulse shape $f(t)$ due to the presence of second term, which is zero for real pulses. See Appendix~\eqref{appendix:single_photon_complex_pulses} for more details. In the long pulse width limit for a coherent state, $\mathcal{F}_{\text{long}}^{\infty}/\alpha^2 = (\varepsilon^{2}/\alpha^2) \mathcal{F}^{(2)} + (\varepsilon^{4}/\alpha^2) \mathcal{F}^{(4)} + \mathcal{O}(\varepsilon^{6})$ where $\varepsilon = \alpha/\sqrt{\Gamma T}$, so the fourth-order term per unit photon goes to zero as $\alpha$ is decreased, so we expect it to be different from the single-photon QFI. Note that the second order contribution from the coherent state still matches with $ 16 \int_{0}^{\infty} dt \; |\partial_{t}Q(t)|^{2}$.
\section{Conclusion and Outlook}
We obtained an explicit solution of the double-sided master equation for the setup in which a coherent state is sent through a waveguide coupled to a simple matter system, a two-level atom. This allowed us to identify the physical factors that influence the value of the QFI and obtain the optimal pulse that maximizes it. Through this analysis, we also made connections with the results obtained for the single-photon pulse. It would be interesting to identify how the presence of losses modifies the QFI, and in particular, the optimal pulse. While some work has been performed with Gaussian pulses along this direction \cite{khan2025tensor}, a comprehensive analysis remains to be developed. It is also worthwhile to explore the broader class of systems in which the QFI can be cast in this form, \textit{i.e.}, as the explicit solution to a set of coupled ODEs.

The code that has been used to produce results in this manuscript is available at \cite{OptimalWaveformsForDipoleEstimation}.

\section{Acknowledgments}
The authors kindly thank Francesco Albarelli, Animesh Datta, Aiman Khan and Sourav Das for insightful discussions on quantum estimation theory. The authors are also grateful to Rodrigo A Vargas-Hernández and Alexandre de Camargo for help with the code. This work was supported by the Ministère de l’Économie et de l’Innovation du Québec and the Natural Sciences and Engineering Research Council of Canada. This work has been funded by the European Union’s Horizon Europe Research and Innovation Programme under agreement 101070700 project MIRAQLS. This research was enabled in part by support provided by the Digital Research Alliance of Canada.
\bibliography{References.bib}

\begin{thebibliography}{53}%
\makeatletter
\providecommand \@ifxundefined [1]{%
 \@ifx{#1\undefined}
}%
\providecommand \@ifnum [1]{%
 \ifnum #1\expandafter \@firstoftwo
 \else \expandafter \@secondoftwo
 \fi
}%
\providecommand \@ifx [1]{%
 \ifx #1\expandafter \@firstoftwo
 \else \expandafter \@secondoftwo
 \fi
}%
\providecommand \natexlab [1]{#1}%
\providecommand \enquote  [1]{``#1''}%
\providecommand \bibnamefont  [1]{#1}%
\providecommand \bibfnamefont [1]{#1}%
\providecommand \citenamefont [1]{#1}%
\providecommand \href@noop [0]{\@secondoftwo}%
\providecommand \href [0]{\begingroup \@sanitize@url \@href}%
\providecommand \@href[1]{\@@startlink{#1}\@@href}%
\providecommand \@@href[1]{\endgroup#1\@@endlink}%
\providecommand \@sanitize@url [0]{\catcode `\\12\catcode `\$12\catcode `\&12\catcode `\#12\catcode `\^12\catcode `\_12\catcode `\%12\relax}%
\providecommand \@@startlink[1]{}%
\providecommand \@@endlink[0]{}%
\providecommand \url  [0]{\begingroup\@sanitize@url \@url }%
\providecommand \@url [1]{\endgroup\@href {#1}{\urlprefix }}%
\providecommand \urlprefix  [0]{URL }%
\providecommand \Eprint [0]{\href }%
\providecommand \doibase [0]{https://doi.org/}%
\providecommand \selectlanguage [0]{\@gobble}%
\providecommand \bibinfo  [0]{\@secondoftwo}%
\providecommand \bibfield  [0]{\@secondoftwo}%
\providecommand \translation [1]{[#1]}%
\providecommand \BibitemOpen [0]{}%
\providecommand \bibitemStop [0]{}%
\providecommand \bibitemNoStop [0]{.\EOS\space}%
\providecommand \EOS [0]{\spacefactor3000\relax}%
\providecommand \BibitemShut  [1]{\csname bibitem#1\endcsname}%
\let\auto@bib@innerbib\@empty
\bibitem [{\citenamefont {Dowling}\ and\ \citenamefont {Seshadreesan}(2015)}]{dowling2015quantum}%
  \BibitemOpen
  \bibfield  {author} {\bibinfo {author} {\bibfnamefont {J.~P.}\ \bibnamefont {Dowling}}\ and\ \bibinfo {author} {\bibfnamefont {K.~P.}\ \bibnamefont {Seshadreesan}},\ }\bibfield  {title} {\bibinfo {title} {Quantum optical technologies for metrology, sensing, and imaging},\ }\href@noop {} {\bibfield  {journal} {\bibinfo  {journal} {Journal of Lightwave Technology}\ }\textbf {\bibinfo {volume} {33}},\ \bibinfo {pages} {2359} (\bibinfo {year} {2015})}\BibitemShut {NoStop}%
\bibitem [{\citenamefont {Taylor}\ and\ \citenamefont {Bowen}(2016)}]{taylor2016quantum}%
  \BibitemOpen
  \bibfield  {author} {\bibinfo {author} {\bibfnamefont {M.~A.}\ \bibnamefont {Taylor}}\ and\ \bibinfo {author} {\bibfnamefont {W.~P.}\ \bibnamefont {Bowen}},\ }\bibfield  {title} {\bibinfo {title} {Quantum metrology and its application in biology},\ }\href@noop {} {\bibfield  {journal} {\bibinfo  {journal} {Physics Reports}\ }\textbf {\bibinfo {volume} {615}},\ \bibinfo {pages} {1} (\bibinfo {year} {2016})}\BibitemShut {NoStop}%
\bibitem [{\citenamefont {Tan}\ and\ \citenamefont {Jeong}(2019)}]{tan2019nonclassical}%
  \BibitemOpen
  \bibfield  {author} {\bibinfo {author} {\bibfnamefont {K.~C.}\ \bibnamefont {Tan}}\ and\ \bibinfo {author} {\bibfnamefont {H.}~\bibnamefont {Jeong}},\ }\bibfield  {title} {\bibinfo {title} {Nonclassical light and metrological power: An introductory review},\ }\href@noop {} {\bibfield  {journal} {\bibinfo  {journal} {AVS Quantum Science}\ }\textbf {\bibinfo {volume} {1}} (\bibinfo {year} {2019})}\BibitemShut {NoStop}%
\bibitem [{\citenamefont {Lawrie}\ \emph {et~al.}(2019)\citenamefont {Lawrie}, \citenamefont {Lett}, \citenamefont {Marino},\ and\ \citenamefont {Pooser}}]{lawrie2019quantum}%
  \BibitemOpen
  \bibfield  {author} {\bibinfo {author} {\bibfnamefont {B.~J.}\ \bibnamefont {Lawrie}}, \bibinfo {author} {\bibfnamefont {P.~D.}\ \bibnamefont {Lett}}, \bibinfo {author} {\bibfnamefont {A.~M.}\ \bibnamefont {Marino}},\ and\ \bibinfo {author} {\bibfnamefont {R.~C.}\ \bibnamefont {Pooser}},\ }\bibfield  {title} {\bibinfo {title} {Quantum sensing with squeezed light},\ }\href@noop {} {\bibfield  {journal} {\bibinfo  {journal} {Acs Photonics}\ }\textbf {\bibinfo {volume} {6}},\ \bibinfo {pages} {1307} (\bibinfo {year} {2019})}\BibitemShut {NoStop}%
\bibitem [{\citenamefont {Berchera}\ and\ \citenamefont {Degiovanni}(2019)}]{berchera2019quantum}%
  \BibitemOpen
  \bibfield  {author} {\bibinfo {author} {\bibfnamefont {I.~R.}\ \bibnamefont {Berchera}}\ and\ \bibinfo {author} {\bibfnamefont {I.~P.}\ \bibnamefont {Degiovanni}},\ }\bibfield  {title} {\bibinfo {title} {Quantum imaging with sub-poissonian light: challenges and perspectives in optical metrology},\ }\href@noop {} {\bibfield  {journal} {\bibinfo  {journal} {Metrologia}\ }\textbf {\bibinfo {volume} {56}},\ \bibinfo {pages} {024001} (\bibinfo {year} {2019})}\BibitemShut {NoStop}%
\bibitem [{\citenamefont {Datta}(2025)}]{datta2025sensing}%
  \BibitemOpen
  \bibfield  {author} {\bibinfo {author} {\bibfnamefont {A.}~\bibnamefont {Datta}},\ }\bibfield  {title} {\bibinfo {title} {Sensing with quantum light: a perspective},\ }\href@noop {} {\bibfield  {journal} {\bibinfo  {journal} {Nanophotonics}\ }\textbf {\bibinfo {volume} {14}},\ \bibinfo {pages} {1993} (\bibinfo {year} {2025})}\BibitemShut {NoStop}%
\bibitem [{\citenamefont {Holdsworth}\ \emph {et~al.}(2025)\citenamefont {Holdsworth}, \citenamefont {Adamczyk},\ and\ \citenamefont {Agarwal}}]{holdsworth2025saturation}%
  \BibitemOpen
  \bibfield  {author} {\bibinfo {author} {\bibfnamefont {T.}~\bibnamefont {Holdsworth}}, \bibinfo {author} {\bibfnamefont {J.}~\bibnamefont {Adamczyk}},\ and\ \bibinfo {author} {\bibfnamefont {G.~S.}\ \bibnamefont {Agarwal}},\ }\bibfield  {title} {\bibinfo {title} {Saturation of the cram$\backslash$'er-rao bound for the atomic resonance frequency with phased array of hyperbolic secant pulses},\ }\href@noop {} {\bibfield  {journal} {\bibinfo  {journal} {arXiv preprint arXiv:2505.08192}\ } (\bibinfo {year} {2025})}\BibitemShut {NoStop}%
\bibitem [{\citenamefont {Polzik}\ \emph {et~al.}(1992)\citenamefont {Polzik}, \citenamefont {Carri},\ and\ \citenamefont {Kimble}}]{polzik1992spectroscopy}%
  \BibitemOpen
  \bibfield  {author} {\bibinfo {author} {\bibfnamefont {E.}~\bibnamefont {Polzik}}, \bibinfo {author} {\bibfnamefont {J.}~\bibnamefont {Carri}},\ and\ \bibinfo {author} {\bibfnamefont {H.}~\bibnamefont {Kimble}},\ }\bibfield  {title} {\bibinfo {title} {Spectroscopy with squeezed light},\ }\href@noop {} {\bibfield  {journal} {\bibinfo  {journal} {Physical review letters}\ }\textbf {\bibinfo {volume} {68}},\ \bibinfo {pages} {3020} (\bibinfo {year} {1992})}\BibitemShut {NoStop}%
\bibitem [{\citenamefont {Caves}(1981)}]{caves1981quantum}%
  \BibitemOpen
  \bibfield  {author} {\bibinfo {author} {\bibfnamefont {C.~M.}\ \bibnamefont {Caves}},\ }\bibfield  {title} {\bibinfo {title} {Quantum-mechanical noise in an interferometer},\ }\href@noop {} {\bibfield  {journal} {\bibinfo  {journal} {Physical Review D}\ }\textbf {\bibinfo {volume} {23}},\ \bibinfo {pages} {1693} (\bibinfo {year} {1981})}\BibitemShut {NoStop}%
\bibitem [{\citenamefont {Bondurant}\ and\ \citenamefont {Shapiro}(1984)}]{bondurant1984squeezed}%
  \BibitemOpen
  \bibfield  {author} {\bibinfo {author} {\bibfnamefont {R.~S.}\ \bibnamefont {Bondurant}}\ and\ \bibinfo {author} {\bibfnamefont {J.~H.}\ \bibnamefont {Shapiro}},\ }\bibfield  {title} {\bibinfo {title} {Squeezed states in phase-sensing interferometers},\ }\href@noop {} {\bibfield  {journal} {\bibinfo  {journal} {Physical Review D}\ }\textbf {\bibinfo {volume} {30}},\ \bibinfo {pages} {2548} (\bibinfo {year} {1984})}\BibitemShut {NoStop}%
\bibitem [{\citenamefont {Dowling}(2008)}]{dowling2008quantum}%
  \BibitemOpen
  \bibfield  {author} {\bibinfo {author} {\bibfnamefont {J.~P.}\ \bibnamefont {Dowling}},\ }\bibfield  {title} {\bibinfo {title} {Quantum optical metrology--the lowdown on high-n00n states},\ }\href@noop {} {\bibfield  {journal} {\bibinfo  {journal} {Contemporary physics}\ }\textbf {\bibinfo {volume} {49}},\ \bibinfo {pages} {125} (\bibinfo {year} {2008})}\BibitemShut {NoStop}%
\bibitem [{\citenamefont {Casacio}\ \emph {et~al.}(2021)\citenamefont {Casacio}, \citenamefont {Madsen}, \citenamefont {Terrasson}, \citenamefont {Waleed}, \citenamefont {Barnscheidt}, \citenamefont {Hage}, \citenamefont {Taylor},\ and\ \citenamefont {Bowen}}]{casacio2021quantum}%
  \BibitemOpen
  \bibfield  {author} {\bibinfo {author} {\bibfnamefont {C.~A.}\ \bibnamefont {Casacio}}, \bibinfo {author} {\bibfnamefont {L.~S.}\ \bibnamefont {Madsen}}, \bibinfo {author} {\bibfnamefont {A.}~\bibnamefont {Terrasson}}, \bibinfo {author} {\bibfnamefont {M.}~\bibnamefont {Waleed}}, \bibinfo {author} {\bibfnamefont {K.}~\bibnamefont {Barnscheidt}}, \bibinfo {author} {\bibfnamefont {B.}~\bibnamefont {Hage}}, \bibinfo {author} {\bibfnamefont {M.~A.}\ \bibnamefont {Taylor}},\ and\ \bibinfo {author} {\bibfnamefont {W.~P.}\ \bibnamefont {Bowen}},\ }\bibfield  {title} {\bibinfo {title} {Quantum-enhanced nonlinear microscopy},\ }\href@noop {} {\bibfield  {journal} {\bibinfo  {journal} {Nature}\ }\textbf {\bibinfo {volume} {594}},\ \bibinfo {pages} {201} (\bibinfo {year} {2021})}\BibitemShut {NoStop}%
\bibitem [{\citenamefont {Datta}(2020)}]{datta2020quantum}%
  \BibitemOpen
  \bibfield  {author} {\bibinfo {author} {\bibfnamefont {A.}~\bibnamefont {Datta}},\ }\bibfield  {title} {\bibinfo {title} {Quantum-enhanced stimulated emission microscopy},\ }in\ \href@noop {} {\emph {\bibinfo {booktitle} {Emerging Imaging and Sensing Technologies for Security and Defence V; and Advanced Manufacturing Technologies for Micro-and Nanosystems in Security and Defence III}}},\ Vol.\ \bibinfo {volume} {11540}\ (\bibinfo {organization} {SPIE},\ \bibinfo {year} {2020})\ p.\ \bibinfo {pages} {115400A}\BibitemShut {NoStop}%
\bibitem [{\citenamefont {de~Andrade}\ \emph {et~al.}(2020)\citenamefont {de~Andrade}, \citenamefont {Kerdoncuff}, \citenamefont {Berg-S{\o}rensen}, \citenamefont {Gehring}, \citenamefont {Lassen},\ and\ \citenamefont {Andersen}}]{de2020quantum}%
  \BibitemOpen
  \bibfield  {author} {\bibinfo {author} {\bibfnamefont {R.~B.}\ \bibnamefont {de~Andrade}}, \bibinfo {author} {\bibfnamefont {H.}~\bibnamefont {Kerdoncuff}}, \bibinfo {author} {\bibfnamefont {K.}~\bibnamefont {Berg-S{\o}rensen}}, \bibinfo {author} {\bibfnamefont {T.}~\bibnamefont {Gehring}}, \bibinfo {author} {\bibfnamefont {M.}~\bibnamefont {Lassen}},\ and\ \bibinfo {author} {\bibfnamefont {U.~L.}\ \bibnamefont {Andersen}},\ }\bibfield  {title} {\bibinfo {title} {Quantum-enhanced continuous-wave stimulated raman scattering spectroscopy},\ }\href@noop {} {\bibfield  {journal} {\bibinfo  {journal} {Optica}\ }\textbf {\bibinfo {volume} {7}},\ \bibinfo {pages} {470} (\bibinfo {year} {2020})}\BibitemShut {NoStop}%
\bibitem [{\citenamefont {Mukamel}\ \emph {et~al.}(2020)\citenamefont {Mukamel}, \citenamefont {Freyberger}, \citenamefont {Schleich}, \citenamefont {Bellini}, \citenamefont {Zavatta}, \citenamefont {Leuchs}, \citenamefont {Silberhorn}, \citenamefont {Boyd}, \citenamefont {S{\'a}nchez-Soto}, \citenamefont {Stefanov} \emph {et~al.}}]{mukamel2020roadmap}%
  \BibitemOpen
  \bibfield  {author} {\bibinfo {author} {\bibfnamefont {S.}~\bibnamefont {Mukamel}}, \bibinfo {author} {\bibfnamefont {M.}~\bibnamefont {Freyberger}}, \bibinfo {author} {\bibfnamefont {W.}~\bibnamefont {Schleich}}, \bibinfo {author} {\bibfnamefont {M.}~\bibnamefont {Bellini}}, \bibinfo {author} {\bibfnamefont {A.}~\bibnamefont {Zavatta}}, \bibinfo {author} {\bibfnamefont {G.}~\bibnamefont {Leuchs}}, \bibinfo {author} {\bibfnamefont {C.}~\bibnamefont {Silberhorn}}, \bibinfo {author} {\bibfnamefont {R.~W.}\ \bibnamefont {Boyd}}, \bibinfo {author} {\bibfnamefont {L.~L.}\ \bibnamefont {S{\'a}nchez-Soto}}, \bibinfo {author} {\bibfnamefont {A.}~\bibnamefont {Stefanov}}, \emph {et~al.},\ }\bibfield  {title} {\bibinfo {title} {Roadmap on quantum light spectroscopy},\ }\href@noop {} {\bibfield  {journal} {\bibinfo  {journal} {Journal of physics B: Atomic, molecular and optical physics}\ }\textbf {\bibinfo {volume} {53}},\ \bibinfo {pages} {072002} (\bibinfo {year} {2020})}\BibitemShut {NoStop}%
\bibitem [{\citenamefont {Kira}\ \emph {et~al.}(2011)\citenamefont {Kira}, \citenamefont {Koch}, \citenamefont {Smith}, \citenamefont {Hunter},\ and\ \citenamefont {Cundiff}}]{kira2011quantum}%
  \BibitemOpen
  \bibfield  {author} {\bibinfo {author} {\bibfnamefont {M.}~\bibnamefont {Kira}}, \bibinfo {author} {\bibfnamefont {S.~W.}\ \bibnamefont {Koch}}, \bibinfo {author} {\bibfnamefont {R.~P.}\ \bibnamefont {Smith}}, \bibinfo {author} {\bibfnamefont {A.~E.}\ \bibnamefont {Hunter}},\ and\ \bibinfo {author} {\bibfnamefont {S.~T.}\ \bibnamefont {Cundiff}},\ }\bibfield  {title} {\bibinfo {title} {Quantum spectroscopy with schr{\"o}dinger-cat states},\ }\href@noop {} {\bibfield  {journal} {\bibinfo  {journal} {Nature Physics}\ }\textbf {\bibinfo {volume} {7}},\ \bibinfo {pages} {799} (\bibinfo {year} {2011})}\BibitemShut {NoStop}%
\bibitem [{\citenamefont {Dorfman}\ \emph {et~al.}(2016)\citenamefont {Dorfman}, \citenamefont {Schlawin},\ and\ \citenamefont {Mukamel}}]{dorfman2016nonlinear}%
  \BibitemOpen
  \bibfield  {author} {\bibinfo {author} {\bibfnamefont {K.~E.}\ \bibnamefont {Dorfman}}, \bibinfo {author} {\bibfnamefont {F.}~\bibnamefont {Schlawin}},\ and\ \bibinfo {author} {\bibfnamefont {S.}~\bibnamefont {Mukamel}},\ }\bibfield  {title} {\bibinfo {title} {Nonlinear optical signals and spectroscopy with quantum light},\ }\href@noop {} {\bibfield  {journal} {\bibinfo  {journal} {Reviews of Modern Physics}\ }\textbf {\bibinfo {volume} {88}},\ \bibinfo {pages} {045008} (\bibinfo {year} {2016})}\BibitemShut {NoStop}%
\bibitem [{\citenamefont {Albarelli}\ \emph {et~al.}(2023)\citenamefont {Albarelli}, \citenamefont {Bisketzi}, \citenamefont {Khan},\ and\ \citenamefont {Datta}}]{albarelli2023fundamental}%
  \BibitemOpen
  \bibfield  {author} {\bibinfo {author} {\bibfnamefont {F.}~\bibnamefont {Albarelli}}, \bibinfo {author} {\bibfnamefont {E.}~\bibnamefont {Bisketzi}}, \bibinfo {author} {\bibfnamefont {A.}~\bibnamefont {Khan}},\ and\ \bibinfo {author} {\bibfnamefont {A.}~\bibnamefont {Datta}},\ }\bibfield  {title} {\bibinfo {title} {Fundamental limits of pulsed quantum light spectroscopy: Dipole moment estimation},\ }\href@noop {} {\bibfield  {journal} {\bibinfo  {journal} {Physical Review A}\ }\textbf {\bibinfo {volume} {107}},\ \bibinfo {pages} {062601} (\bibinfo {year} {2023})}\BibitemShut {NoStop}%
\bibitem [{\citenamefont {Darsheshdar}\ \emph {et~al.}(2024)\citenamefont {Darsheshdar}, \citenamefont {Khan}, \citenamefont {Albarelli},\ and\ \citenamefont {Datta}}]{darsheshdar2024role}%
  \BibitemOpen
  \bibfield  {author} {\bibinfo {author} {\bibfnamefont {E.}~\bibnamefont {Darsheshdar}}, \bibinfo {author} {\bibfnamefont {A.}~\bibnamefont {Khan}}, \bibinfo {author} {\bibfnamefont {F.}~\bibnamefont {Albarelli}},\ and\ \bibinfo {author} {\bibfnamefont {A.}~\bibnamefont {Datta}},\ }\bibfield  {title} {\bibinfo {title} {Role of chirping in pulsed single-photon spectroscopy},\ }\href@noop {} {\bibfield  {journal} {\bibinfo  {journal} {Physical Review A}\ }\textbf {\bibinfo {volume} {110}},\ \bibinfo {pages} {043710} (\bibinfo {year} {2024})}\BibitemShut {NoStop}%
\bibitem [{\citenamefont {Khan}\ \emph {et~al.}(2024)\citenamefont {Khan}, \citenamefont {Albarelli},\ and\ \citenamefont {Datta}}]{khan2024does}%
  \BibitemOpen
  \bibfield  {author} {\bibinfo {author} {\bibfnamefont {A.}~\bibnamefont {Khan}}, \bibinfo {author} {\bibfnamefont {F.}~\bibnamefont {Albarelli}},\ and\ \bibinfo {author} {\bibfnamefont {A.}~\bibnamefont {Datta}},\ }\bibfield  {title} {\bibinfo {title} {Does entanglement enhance single-molecule pulsed biphoton spectroscopy?},\ }\href@noop {} {\bibfield  {journal} {\bibinfo  {journal} {Quantum Science and Technology}\ }\textbf {\bibinfo {volume} {9}},\ \bibinfo {pages} {035004} (\bibinfo {year} {2024})}\BibitemShut {NoStop}%
\bibitem [{\citenamefont {Kalashnikov}\ \emph {et~al.}(2014)\citenamefont {Kalashnikov}, \citenamefont {Pan}, \citenamefont {Kuznetsov},\ and\ \citenamefont {Krivitsky}}]{kalashnikov2014quantum}%
  \BibitemOpen
  \bibfield  {author} {\bibinfo {author} {\bibfnamefont {D.~A.}\ \bibnamefont {Kalashnikov}}, \bibinfo {author} {\bibfnamefont {Z.}~\bibnamefont {Pan}}, \bibinfo {author} {\bibfnamefont {A.~I.}\ \bibnamefont {Kuznetsov}},\ and\ \bibinfo {author} {\bibfnamefont {L.~A.}\ \bibnamefont {Krivitsky}},\ }\bibfield  {title} {\bibinfo {title} {Quantum spectroscopy of plasmonic nanostructures},\ }\href@noop {} {\bibfield  {journal} {\bibinfo  {journal} {Physical Review X}\ }\textbf {\bibinfo {volume} {4}},\ \bibinfo {pages} {011049} (\bibinfo {year} {2014})}\BibitemShut {NoStop}%
\bibitem [{\citenamefont {Fujihashi}\ and\ \citenamefont {Ishizaki}(2021)}]{fujihashi2021achieving}%
  \BibitemOpen
  \bibfield  {author} {\bibinfo {author} {\bibfnamefont {Y.}~\bibnamefont {Fujihashi}}\ and\ \bibinfo {author} {\bibfnamefont {A.}~\bibnamefont {Ishizaki}},\ }\bibfield  {title} {\bibinfo {title} {Achieving two-dimensional optical spectroscopy with temporal and spectral resolution using quantum entangled three photons},\ }\href@noop {} {\bibfield  {journal} {\bibinfo  {journal} {The Journal of Chemical Physics}\ }\textbf {\bibinfo {volume} {155}} (\bibinfo {year} {2021})}\BibitemShut {NoStop}%
\bibitem [{\citenamefont {Raymer}\ \emph {et~al.}(2013)\citenamefont {Raymer}, \citenamefont {Marcus}, \citenamefont {Widom},\ and\ \citenamefont {Vitullo}}]{raymer2013entangled}%
  \BibitemOpen
  \bibfield  {author} {\bibinfo {author} {\bibfnamefont {M.~G.}\ \bibnamefont {Raymer}}, \bibinfo {author} {\bibfnamefont {A.~H.}\ \bibnamefont {Marcus}}, \bibinfo {author} {\bibfnamefont {J.~R.}\ \bibnamefont {Widom}},\ and\ \bibinfo {author} {\bibfnamefont {D.~L.}\ \bibnamefont {Vitullo}},\ }\bibfield  {title} {\bibinfo {title} {Entangled photon-pair two-dimensional fluorescence spectroscopy (epp-2dfs)},\ }\href@noop {} {\bibfield  {journal} {\bibinfo  {journal} {The Journal of Physical Chemistry B}\ }\textbf {\bibinfo {volume} {117}},\ \bibinfo {pages} {15559} (\bibinfo {year} {2013})}\BibitemShut {NoStop}%
\bibitem [{\citenamefont {Leibfried}\ \emph {et~al.}(2004)\citenamefont {Leibfried}, \citenamefont {Barrett}, \citenamefont {Schaetz}, \citenamefont {Britton}, \citenamefont {Chiaverini}, \citenamefont {Itano}, \citenamefont {Jost}, \citenamefont {Langer},\ and\ \citenamefont {Wineland}}]{leibfried2004toward}%
  \BibitemOpen
  \bibfield  {author} {\bibinfo {author} {\bibfnamefont {D.}~\bibnamefont {Leibfried}}, \bibinfo {author} {\bibfnamefont {M.~D.}\ \bibnamefont {Barrett}}, \bibinfo {author} {\bibfnamefont {T.}~\bibnamefont {Schaetz}}, \bibinfo {author} {\bibfnamefont {J.}~\bibnamefont {Britton}}, \bibinfo {author} {\bibfnamefont {J.}~\bibnamefont {Chiaverini}}, \bibinfo {author} {\bibfnamefont {W.~M.}\ \bibnamefont {Itano}}, \bibinfo {author} {\bibfnamefont {J.~D.}\ \bibnamefont {Jost}}, \bibinfo {author} {\bibfnamefont {C.}~\bibnamefont {Langer}},\ and\ \bibinfo {author} {\bibfnamefont {D.~J.}\ \bibnamefont {Wineland}},\ }\bibfield  {title} {\bibinfo {title} {Toward heisenberg-limited spectroscopy with multiparticle entangled states},\ }\href@noop {} {\bibfield  {journal} {\bibinfo  {journal} {Science}\ }\textbf {\bibinfo {volume} {304}},\ \bibinfo {pages} {1476} (\bibinfo {year} {2004})}\BibitemShut {NoStop}%
\bibitem [{\citenamefont {Cram{\'e}r}(1946)}]{cramer1946mathematical}%
  \BibitemOpen
  \bibfield  {author} {\bibinfo {author} {\bibfnamefont {H.}~\bibnamefont {Cram{\'e}r}},\ }\href@noop {} {\emph {\bibinfo {title} {Mathematical Methods of Statistics}}},\ \bibinfo {series} {Princeton Mathematical Series}, Vol.~\bibinfo {volume} {9}\ (\bibinfo  {publisher} {Princeton University Press},\ \bibinfo {address} {Princeton, NJ},\ \bibinfo {year} {1946})\BibitemShut {NoStop}%
\bibitem [{\citenamefont {Rao}\ \emph {et~al.}(1945)\citenamefont {Rao} \emph {et~al.}}]{rao1945information}%
  \BibitemOpen
  \bibfield  {author} {\bibinfo {author} {\bibfnamefont {C.~R.}\ \bibnamefont {Rao}} \emph {et~al.},\ }\bibfield  {title} {\bibinfo {title} {Information and the accuracy attainable in the estimation of statistical parameters},\ }\href@noop {} {\bibfield  {journal} {\bibinfo  {journal} {Bull. Calcutta Math. Soc}\ }\textbf {\bibinfo {volume} {37}},\ \bibinfo {pages} {81} (\bibinfo {year} {1945})}\BibitemShut {NoStop}%
\bibitem [{\citenamefont {Fr{\'e}chet}(1943)}]{frechet1943extension}%
  \BibitemOpen
  \bibfield  {author} {\bibinfo {author} {\bibfnamefont {M.}~\bibnamefont {Fr{\'e}chet}},\ }\bibfield  {title} {\bibinfo {title} {Sur l'extension de certaines evaluations statistiques au cas de petits echantillons},\ }\href@noop {} {\bibfield  {journal} {\bibinfo  {journal} {Revue de l'Institut International de Statistique}\ ,\ \bibinfo {pages} {182}} (\bibinfo {year} {1943})}\BibitemShut {NoStop}%
\bibitem [{\citenamefont {Bern{\'a}d}\ \emph {et~al.}(2019)\citenamefont {Bern{\'a}d}, \citenamefont {Sanavio},\ and\ \citenamefont {Xuereb}}]{bernad2019optimal}%
  \BibitemOpen
  \bibfield  {author} {\bibinfo {author} {\bibfnamefont {J.~Z.}\ \bibnamefont {Bern{\'a}d}}, \bibinfo {author} {\bibfnamefont {C.}~\bibnamefont {Sanavio}},\ and\ \bibinfo {author} {\bibfnamefont {A.}~\bibnamefont {Xuereb}},\ }\bibfield  {title} {\bibinfo {title} {Optimal estimation of matter-field coupling strength in the dipole approximation},\ }\href@noop {} {\bibfield  {journal} {\bibinfo  {journal} {Physical Review A}\ }\textbf {\bibinfo {volume} {99}},\ \bibinfo {pages} {062106} (\bibinfo {year} {2019})}\BibitemShut {NoStop}%
\bibitem [{\citenamefont {Genoni}\ and\ \citenamefont {Invernizzi}(2012)}]{genoni2012optimal}%
  \BibitemOpen
  \bibfield  {author} {\bibinfo {author} {\bibfnamefont {M.~G.}\ \bibnamefont {Genoni}}\ and\ \bibinfo {author} {\bibfnamefont {C.}~\bibnamefont {Invernizzi}},\ }\bibfield  {title} {\bibinfo {title} {Optimal quantum estimation of the coupling constant of jaynes-cummings interaction},\ }\href@noop {} {\bibfield  {journal} {\bibinfo  {journal} {The European Physical Journal Special Topics}\ }\textbf {\bibinfo {volume} {203}},\ \bibinfo {pages} {49} (\bibinfo {year} {2012})}\BibitemShut {NoStop}%
\bibitem [{\citenamefont {Montenegro}\ \emph {et~al.}(2022)\citenamefont {Montenegro}, \citenamefont {Genoni}, \citenamefont {Bayat},\ and\ \citenamefont {Paris}}]{montenegro2022probing}%
  \BibitemOpen
  \bibfield  {author} {\bibinfo {author} {\bibfnamefont {V.}~\bibnamefont {Montenegro}}, \bibinfo {author} {\bibfnamefont {M.}~\bibnamefont {Genoni}}, \bibinfo {author} {\bibfnamefont {A.}~\bibnamefont {Bayat}},\ and\ \bibinfo {author} {\bibfnamefont {M.}~\bibnamefont {Paris}},\ }\bibfield  {title} {\bibinfo {title} {Probing of nonlinear hybrid optomechanical systems via partial accessibility},\ }\href@noop {} {\bibfield  {journal} {\bibinfo  {journal} {Physical Review Research}\ }\textbf {\bibinfo {volume} {4}},\ \bibinfo {pages} {033036} (\bibinfo {year} {2022})}\BibitemShut {NoStop}%
\bibitem [{\citenamefont {Gammelmark}\ and\ \citenamefont {M{\o}lmer}(2014)}]{gammelmark2014fisher}%
  \BibitemOpen
  \bibfield  {author} {\bibinfo {author} {\bibfnamefont {S.}~\bibnamefont {Gammelmark}}\ and\ \bibinfo {author} {\bibfnamefont {K.}~\bibnamefont {M{\o}lmer}},\ }\bibfield  {title} {\bibinfo {title} {Fisher information and the quantum cram{\'e}r-rao sensitivity limit of continuous measurements},\ }\href@noop {} {\bibfield  {journal} {\bibinfo  {journal} {Physical review letters}\ }\textbf {\bibinfo {volume} {112}},\ \bibinfo {pages} {170401} (\bibinfo {year} {2014})}\BibitemShut {NoStop}%
\bibitem [{\citenamefont {Yang}\ \emph {et~al.}(2023)\citenamefont {Yang}, \citenamefont {Huelga},\ and\ \citenamefont {Plenio}}]{yang2023efficient}%
  \BibitemOpen
  \bibfield  {author} {\bibinfo {author} {\bibfnamefont {D.}~\bibnamefont {Yang}}, \bibinfo {author} {\bibfnamefont {S.~F.}\ \bibnamefont {Huelga}},\ and\ \bibinfo {author} {\bibfnamefont {M.~B.}\ \bibnamefont {Plenio}},\ }\bibfield  {title} {\bibinfo {title} {Efficient information retrieval for sensing via continuous measurement},\ }\href@noop {} {\bibfield  {journal} {\bibinfo  {journal} {Physical Review X}\ }\textbf {\bibinfo {volume} {13}},\ \bibinfo {pages} {031012} (\bibinfo {year} {2023})}\BibitemShut {NoStop}%
\bibitem [{\citenamefont {Yang}\ \emph {et~al.}(2025)\citenamefont {Yang}, \citenamefont {Ketkar}, \citenamefont {Audenaert}, \citenamefont {Huelga},\ and\ \citenamefont {Plenio}}]{yang2025quantum}%
  \BibitemOpen
  \bibfield  {author} {\bibinfo {author} {\bibfnamefont {D.}~\bibnamefont {Yang}}, \bibinfo {author} {\bibfnamefont {M.}~\bibnamefont {Ketkar}}, \bibinfo {author} {\bibfnamefont {K.}~\bibnamefont {Audenaert}}, \bibinfo {author} {\bibfnamefont {S.~F.}\ \bibnamefont {Huelga}},\ and\ \bibinfo {author} {\bibfnamefont {M.~B.}\ \bibnamefont {Plenio}},\ }\bibfield  {title} {\bibinfo {title} {Quantum cramer-rao precision limit of noisy continuous sensing},\ }\href@noop {} {\bibfield  {journal} {\bibinfo  {journal} {arXiv preprint arXiv:2504.12400}\ } (\bibinfo {year} {2025})}\BibitemShut {NoStop}%
\bibitem [{\citenamefont {Khan}\ \emph {et~al.}(2025)\citenamefont {Khan}, \citenamefont {Albarelli},\ and\ \citenamefont {Datta}}]{khan2025tensor}%
  \BibitemOpen
  \bibfield  {author} {\bibinfo {author} {\bibfnamefont {A.}~\bibnamefont {Khan}}, \bibinfo {author} {\bibfnamefont {F.}~\bibnamefont {Albarelli}},\ and\ \bibinfo {author} {\bibfnamefont {A.}~\bibnamefont {Datta}},\ }\bibfield  {title} {\bibinfo {title} {A tensor network approach to sensing quantum light-matter interactions},\ }\href@noop {} {\bibfield  {journal} {\bibinfo  {journal} {arXiv preprint arXiv:2504.12399}\ } (\bibinfo {year} {2025})}\BibitemShut {NoStop}%
\bibitem [{\citenamefont {Midha}\ and\ \citenamefont {Gopalakrishnan}(2025)}]{midha2025metrology}%
  \BibitemOpen
  \bibfield  {author} {\bibinfo {author} {\bibfnamefont {S.}~\bibnamefont {Midha}}\ and\ \bibinfo {author} {\bibfnamefont {S.}~\bibnamefont {Gopalakrishnan}},\ }\bibfield  {title} {\bibinfo {title} {Metrology of open quantum systems from emitted radiation},\ }\href@noop {} {\bibfield  {journal} {\bibinfo  {journal} {arXiv preprint arXiv:2504.13815}\ } (\bibinfo {year} {2025})}\BibitemShut {NoStop}%
\bibitem [{\citenamefont {Kiilerich}\ and\ \citenamefont {M{\o}lmer}(2014)}]{kiilerich2014estimation}%
  \BibitemOpen
  \bibfield  {author} {\bibinfo {author} {\bibfnamefont {A.~H.}\ \bibnamefont {Kiilerich}}\ and\ \bibinfo {author} {\bibfnamefont {K.}~\bibnamefont {M{\o}lmer}},\ }\bibfield  {title} {\bibinfo {title} {Estimation of atomic interaction parameters by photon counting},\ }\href@noop {} {\bibfield  {journal} {\bibinfo  {journal} {Physical Review A}\ }\textbf {\bibinfo {volume} {89}},\ \bibinfo {pages} {052110} (\bibinfo {year} {2014})}\BibitemShut {NoStop}%
\bibitem [{\citenamefont {Kiilerich}\ and\ \citenamefont {M{\o}lmer}(2016)}]{kiilerich2016bayesian}%
  \BibitemOpen
  \bibfield  {author} {\bibinfo {author} {\bibfnamefont {A.~H.}\ \bibnamefont {Kiilerich}}\ and\ \bibinfo {author} {\bibfnamefont {K.}~\bibnamefont {M{\o}lmer}},\ }\bibfield  {title} {\bibinfo {title} {Bayesian parameter estimation by continuous homodyne detection},\ }\href@noop {} {\bibfield  {journal} {\bibinfo  {journal} {Physical Review A}\ }\textbf {\bibinfo {volume} {94}},\ \bibinfo {pages} {032103} (\bibinfo {year} {2016})}\BibitemShut {NoStop}%
\bibitem [{\citenamefont {Quesada}\ \emph {et~al.}(2022)\citenamefont {Quesada}, \citenamefont {Helt}, \citenamefont {Menotti}, \citenamefont {Liscidini},\ and\ \citenamefont {Sipe}}]{quesada2022beyond}%
  \BibitemOpen
  \bibfield  {author} {\bibinfo {author} {\bibfnamefont {N.}~\bibnamefont {Quesada}}, \bibinfo {author} {\bibfnamefont {L.}~\bibnamefont {Helt}}, \bibinfo {author} {\bibfnamefont {M.}~\bibnamefont {Menotti}}, \bibinfo {author} {\bibfnamefont {M.}~\bibnamefont {Liscidini}},\ and\ \bibinfo {author} {\bibfnamefont {J.}~\bibnamefont {Sipe}},\ }\bibfield  {title} {\bibinfo {title} {Beyond photon pairs—nonlinear quantum photonics in the high-gain regime: a tutorial},\ }\href@noop {} {\bibfield  {journal} {\bibinfo  {journal} {Advances in Optics and Photonics}\ }\textbf {\bibinfo {volume} {14}},\ \bibinfo {pages} {291} (\bibinfo {year} {2022})}\BibitemShut {NoStop}%
\bibitem [{\citenamefont {Mollow}(1975)}]{mollow1975pure}%
  \BibitemOpen
  \bibfield  {author} {\bibinfo {author} {\bibfnamefont {B.}~\bibnamefont {Mollow}},\ }\bibfield  {title} {\bibinfo {title} {Pure-state analysis of resonant light scattering: Radiative damping, saturation, and multiphoton effects},\ }\href@noop {} {\bibfield  {journal} {\bibinfo  {journal} {Physical Review A}\ }\textbf {\bibinfo {volume} {12}},\ \bibinfo {pages} {1919} (\bibinfo {year} {1975})}\BibitemShut {NoStop}%
\bibitem [{\citenamefont {Fischer}\ \emph {et~al.}(2018)\citenamefont {Fischer}, \citenamefont {Trivedi}, \citenamefont {Ramasesh}, \citenamefont {Siddiqi},\ and\ \citenamefont {Vu{\v{c}}kovi{\'c}}}]{fischer2018scattering}%
  \BibitemOpen
  \bibfield  {author} {\bibinfo {author} {\bibfnamefont {K.~A.}\ \bibnamefont {Fischer}}, \bibinfo {author} {\bibfnamefont {R.}~\bibnamefont {Trivedi}}, \bibinfo {author} {\bibfnamefont {V.}~\bibnamefont {Ramasesh}}, \bibinfo {author} {\bibfnamefont {I.}~\bibnamefont {Siddiqi}},\ and\ \bibinfo {author} {\bibfnamefont {J.}~\bibnamefont {Vu{\v{c}}kovi{\'c}}},\ }\bibfield  {title} {\bibinfo {title} {Scattering into one-dimensional waveguides from a coherently-driven quantum-optical system},\ }\href@noop {} {\bibfield  {journal} {\bibinfo  {journal} {Quantum}\ }\textbf {\bibinfo {volume} {2}},\ \bibinfo {pages} {69} (\bibinfo {year} {2018})}\BibitemShut {NoStop}%
\bibitem [{\citenamefont {Fischer}(2018)}]{fischer2018derivation}%
  \BibitemOpen
  \bibfield  {author} {\bibinfo {author} {\bibfnamefont {K.}~\bibnamefont {Fischer}},\ }\bibfield  {title} {\bibinfo {title} {Derivation of the quantum-optical master equation based on coarse-graining of time},\ }\href@noop {} {\bibfield  {journal} {\bibinfo  {journal} {Journal of Physics Communications}\ }\textbf {\bibinfo {volume} {2}},\ \bibinfo {pages} {091001} (\bibinfo {year} {2018})}\BibitemShut {NoStop}%
\bibitem [{\citenamefont {Paris}(2009)}]{paris2009quantum}%
  \BibitemOpen
  \bibfield  {author} {\bibinfo {author} {\bibfnamefont {M.~G.}\ \bibnamefont {Paris}},\ }\bibfield  {title} {\bibinfo {title} {Quantum estimation for quantum technology},\ }\href@noop {} {\bibfield  {journal} {\bibinfo  {journal} {International Journal of Quantum Information}\ }\textbf {\bibinfo {volume} {7}},\ \bibinfo {pages} {125} (\bibinfo {year} {2009})}\BibitemShut {NoStop}%
\bibitem [{\citenamefont {Helstrom}(1967)}]{helstrom1967minimum}%
  \BibitemOpen
  \bibfield  {author} {\bibinfo {author} {\bibfnamefont {C.~W.}\ \bibnamefont {Helstrom}},\ }\bibfield  {title} {\bibinfo {title} {Minimum mean-squared error of estimates in quantum statistics},\ }\href@noop {} {\bibfield  {journal} {\bibinfo  {journal} {Physics letters A}\ }\textbf {\bibinfo {volume} {25}},\ \bibinfo {pages} {101} (\bibinfo {year} {1967})}\BibitemShut {NoStop}%
\bibitem [{\citenamefont {Holevo}(2011)}]{holevo2011probabilistic}%
  \BibitemOpen
  \bibfield  {author} {\bibinfo {author} {\bibfnamefont {A.~S.}\ \bibnamefont {Holevo}},\ }\href@noop {} {\emph {\bibinfo {title} {Probabilistic and statistical aspects of quantum theory}}},\ Vol.~\bibinfo {volume} {1}\ (\bibinfo  {publisher} {Springer Science \& Business Media},\ \bibinfo {year} {2011})\BibitemShut {NoStop}%
\bibitem [{\citenamefont {Braunstein}\ and\ \citenamefont {Caves}(1994)}]{braunstein1994statistical}%
  \BibitemOpen
  \bibfield  {author} {\bibinfo {author} {\bibfnamefont {S.~L.}\ \bibnamefont {Braunstein}}\ and\ \bibinfo {author} {\bibfnamefont {C.~M.}\ \bibnamefont {Caves}},\ }\bibfield  {title} {\bibinfo {title} {Statistical distance and the geometry of quantum states},\ }\href@noop {} {\bibfield  {journal} {\bibinfo  {journal} {Physical Review Letters}\ }\textbf {\bibinfo {volume} {72}},\ \bibinfo {pages} {3439} (\bibinfo {year} {1994})}\BibitemShut {NoStop}%
\bibitem [{\citenamefont {Helstrom}(1976)}]{helstrom1332quantum}%
  \BibitemOpen
  \bibfield  {author} {\bibinfo {author} {\bibfnamefont {C.}~\bibnamefont {Helstrom}},\ }\bibfield  {title} {\bibinfo {title} {Quantum detection and estimation theory},\ }\href@noop {} {\bibfield  {journal} {\bibinfo  {journal} {New York, Academic Press}\ } (\bibinfo {year} {1976})}\BibitemShut {NoStop}%
\bibitem [{\citenamefont {Liu}\ \emph {et~al.}(2014)\citenamefont {Liu}, \citenamefont {Jing}, \citenamefont {Zhong},\ and\ \citenamefont {Wang}}]{liu2014quantum}%
  \BibitemOpen
  \bibfield  {author} {\bibinfo {author} {\bibfnamefont {J.}~\bibnamefont {Liu}}, \bibinfo {author} {\bibfnamefont {X.-X.}\ \bibnamefont {Jing}}, \bibinfo {author} {\bibfnamefont {W.}~\bibnamefont {Zhong}},\ and\ \bibinfo {author} {\bibfnamefont {X.-G.}\ \bibnamefont {Wang}},\ }\bibfield  {title} {\bibinfo {title} {Quantum fisher information for density matrices with arbitrary ranks},\ }\href@noop {} {\bibfield  {journal} {\bibinfo  {journal} {Communications in Theoretical Physics}\ }\textbf {\bibinfo {volume} {61}},\ \bibinfo {pages} {45} (\bibinfo {year} {2014})}\BibitemShut {NoStop}%
\bibitem [{\citenamefont {Wang}\ \emph {et~al.}(2011)\citenamefont {Wang}, \citenamefont {Min{\'a}{\v{r}}}, \citenamefont {Sheridan},\ and\ \citenamefont {Scarani}}]{wang2011efficient}%
  \BibitemOpen
  \bibfield  {author} {\bibinfo {author} {\bibfnamefont {Y.}~\bibnamefont {Wang}}, \bibinfo {author} {\bibfnamefont {J.}~\bibnamefont {Min{\'a}{\v{r}}}}, \bibinfo {author} {\bibfnamefont {L.}~\bibnamefont {Sheridan}},\ and\ \bibinfo {author} {\bibfnamefont {V.}~\bibnamefont {Scarani}},\ }\bibfield  {title} {\bibinfo {title} {Efficient excitation of a two-level atom by a single photon in a propagating mode},\ }\href@noop {} {\bibfield  {journal} {\bibinfo  {journal} {Physical Review A—Atomic, Molecular, and Optical Physics}\ }\textbf {\bibinfo {volume} {83}},\ \bibinfo {pages} {063842} (\bibinfo {year} {2011})}\BibitemShut {NoStop}%
\bibitem [{\citenamefont {Das}\ \emph {et~al.}(2025)\citenamefont {Das}, \citenamefont {Khan}, \citenamefont {Albarelli},\ and\ \citenamefont {Datta}}]{sourav2025}%
  \BibitemOpen
  \bibfield  {author} {\bibinfo {author} {\bibfnamefont {S.}~\bibnamefont {Das}}, \bibinfo {author} {\bibfnamefont {A.}~\bibnamefont {Khan}}, \bibinfo {author} {\bibfnamefont {F.}~\bibnamefont {Albarelli}},\ and\ \bibinfo {author} {\bibfnamefont {A.}~\bibnamefont {Datta}},\ }\href@noop {} {} (\bibinfo {year} {2025}),\ \bibinfo {note} {in preparation}\BibitemShut {NoStop}%
\bibitem [{\citenamefont {Blondel}\ \emph {et~al.}(2021)\citenamefont {Blondel}, \citenamefont {Berthet}, \citenamefont {Cuturi}, \citenamefont {Frostig}, \citenamefont {Hoyer}, \citenamefont {Llinares-L{\'o}pez}, \citenamefont {Pedregosa},\ and\ \citenamefont {Vert}}]{jaxopt_implicit_diff}%
  \BibitemOpen
  \bibfield  {author} {\bibinfo {author} {\bibfnamefont {M.}~\bibnamefont {Blondel}}, \bibinfo {author} {\bibfnamefont {Q.}~\bibnamefont {Berthet}}, \bibinfo {author} {\bibfnamefont {M.}~\bibnamefont {Cuturi}}, \bibinfo {author} {\bibfnamefont {R.}~\bibnamefont {Frostig}}, \bibinfo {author} {\bibfnamefont {S.}~\bibnamefont {Hoyer}}, \bibinfo {author} {\bibfnamefont {F.}~\bibnamefont {Llinares-L{\'o}pez}}, \bibinfo {author} {\bibfnamefont {F.}~\bibnamefont {Pedregosa}},\ and\ \bibinfo {author} {\bibfnamefont {J.-P.}\ \bibnamefont {Vert}},\ }\bibfield  {title} {\bibinfo {title} {Efficient and modular implicit differentiation},\ }\href@noop {} {\bibfield  {journal} {\bibinfo  {journal} {arXiv preprint arXiv:2105.15183}\ } (\bibinfo {year} {2021})}\BibitemShut {NoStop}%
\bibitem [{\citenamefont {Kidger}(2021)}]{kidger2021on}%
  \BibitemOpen
  \bibfield  {author} {\bibinfo {author} {\bibfnamefont {P.}~\bibnamefont {Kidger}},\ }\emph {\bibinfo {title} {{O}n {N}eural {D}ifferential {E}quations}},\ \href@noop {} {Ph.D. thesis},\ \bibinfo  {school} {University of Oxford} (\bibinfo {year} {2021})\BibitemShut {NoStop}%
\bibitem [{Opt()}]{OptimalWaveformsForDipoleEstimation}%
  \BibitemOpen
  \href@noop {} {}\bibinfo {note} {For code, see \url{https://github.com/polyquantique/OptimalWaveformsForDipoleEstimation}}\BibitemShut {NoStop}%
\bibitem [{\citenamefont {Griffiths}\ and\ \citenamefont {Schroeter}(2018)}]{griffiths2018introduction}%
  \BibitemOpen
  \bibfield  {author} {\bibinfo {author} {\bibfnamefont {D.~J.}\ \bibnamefont {Griffiths}}\ and\ \bibinfo {author} {\bibfnamefont {D.~F.}\ \bibnamefont {Schroeter}},\ }\href@noop {} {\emph {\bibinfo {title} {Introduction to quantum mechanics}}}\ (\bibinfo  {publisher} {Cambridge university press},\ \bibinfo {year} {2018})\BibitemShut {NoStop}%
\end{thebibliography}%
\begin{widetext}
\appendix
\section{Mollow transformation}
\label{appendix:Mollow_transformation}
In this section, we show that the state of the system at an arbitrary time can be written as the expression given in Eq.\eqref{eq:state_arbitrary_time} if the initial state of system is a coherent state in the waveguide and the atom in its ground state
\begin{align}
\ket{\phi(t_0)} = \mathcal{D}(\tilde{f}(x)) \ket{\text{vac}}  \otimes \ket{g},
\end{align}
where the displacement operator is $\mathcal{D}(\tilde{f}(x)) = \exp\left(\int dx \tilde{f}(x) \psi^\dagger (x) -\text{H.c.} \right)$. This is the so-called Mollow transformation \cite{mollow1975pure,fischer2018scattering}. Using Eq.~\eqref{eq:interaction_Hamiltonian}, we can write the time evolution operator for the original Schr\"odinger picture Hamiltonian as follows
\begin{align}
\mathcal{U}(t,t_0) = \mathcal{U}_0(t,t_0) \mathcal{T} \exp\left(- i \int_{0}^t dt' H_{\text{int}}(t') \right), 
\end{align}
where 
\begin{align}
H_{\text{int}} &= \delta \sigma^{\dagger}\sigma   +i  \sqrt{v \Gamma_{\parallel}} \left( \sigma \psi^\dagger(-v t) - \text{H.c.} \right),
\end{align}
and $\delta = \omega_{0}-\omega_{1}$. We can now write
\begin{align}
\ket{\phi(t)} &= \mathcal{U}_0(t) \mathcal{T} \exp\left(- i \int_{0}^t dt' H_{\text{int}}(t') \right)  \mathcal{D}(\tilde{f}(x)) \ket{\text{vac}} \otimes \ket{g} \\
&=\mathcal{U}_0(t) \mathcal{D}(\tilde{f}(x)) \left[\mathcal{D}^\dagger (\tilde{f}(x))\mathcal{T} \exp\left(- i \int_{0}^t dt' H_{\text{int}}(t') \right)  \mathcal{D}(\tilde{f}(x)) \right]\ket{\text{vac}} \otimes \ket{g} \\
&=\left[\mathcal{U}_0(t) \mathcal{D}(\tilde{f}(x)) \mathcal{U}^\dagger _0(t)\right] \mathcal{U}_0(t) \left[\mathcal{D}^\dagger (\tilde{f}(x))\mathcal{T} \exp\left(-i \int_{0}^t dt' H_{\text{int}}(t') \right)  \mathcal{D}(\tilde{f}(x))\right] \ket{\text{vac}} \otimes \ket{g} .
\end{align}
Some of the terms in the two lines above can be simplified to
\begin{align}
    \mathcal{U}_0(t)\mathcal{D}(\tilde{f}(x)) \mathcal{U}^\dagger_0(t) &=  \mathcal{D}\left(\tilde{f}(x - vt) e^{-i \omega_1 t}\right), \\
    \mathcal{D}^\dagger (\tilde{f}(x))\mathcal{T} \exp\left(- i \int_{0}^t dt' H_{\text{int}}(t') \right)  \mathcal{D}(\tilde{f}(x)) &= 
    \mathcal{T} \exp\left(- i \int_{0}^t dt' \mathcal{D}^\dagger (\tilde{f}(x)) H_{\text{int}}(t') \mathcal{D}(\tilde{f}(x)) \right) ,
\end{align}
and in turn we can calculate
\begin{align}
H_{\mathcal{D}}(t) \equiv \mathcal{D}^\dagger (\tilde{f}(x)) H_{\text{int}}(t) \mathcal{D}(\tilde{f}(x)) = \underbrace{ \delta  \sigma^\dagger \sigma  +i \sqrt{v \Gamma} \left( \sigma \tilde{f}^*(-v t) - \text{H.c.} \right) }_{H_{\text{atom}}(t)} +i  \sqrt{v\Gamma} \left( \sigma \psi^\dagger(-v t) - \text{H.c.} \right),
\end{align}
which means that interacting with a coherent state is the same as interacting with the vacuum at the cost of adding the term $i  \sqrt{v \Gamma} \left( \tilde{f}^*(-v t)\sigma - \text{H.c.} \right)$ to the Hamiltonian of the atom.
The net ket is thus
\begin{align}
\ket{\phi(t)} &= \mathcal{D}\left(\tilde{f}(x - v t) e^{ -i \omega_1 t}\right) \mathcal{U}_0(t) \mathcal{T}\exp\left( -i \int_{0}^t dt' H_{\mathcal{D}}(t') \right) \ket{g} \otimes \ket{\text{vac}} \\
&\equiv \Lambda(t) \mathcal{U}_0(t) \ket{\varphi(t)},
\end{align}
where 
\begin{align}\Lambda(t) &=\mathcal{D}\left(\tilde{f}(x - v t) e^{ -i \omega_1 t}\right),
\end{align}
which is Eq.~\eqref{eq:state_arbitrary_time} in the main text.
\section{Derivation of ODEs}
\label{appendix:ODEs_derivation}
As mentioned in the paper, the generalized density operator evolves under the following differential equation,
\begin{align}
\frac{d\mu(t)}{dt} &= i \mu(t) H_{A}(\Gamma_{2},t) -i H_{A}(\Gamma_{1},t) \mu(t)  + J_{A}(\Gamma_{1},t) \mu(t) J_{A}^{\dagger}(\Gamma_{2},t)  \nonumber \\
& -\frac{1}{2}J_{A}^{\dagger}(\Gamma_{1},t)J_{A}(\Gamma_{1},t) \mu(t) -\frac{1}{2} \mu(t) J_{A}^{\dagger}(\Gamma_{2},t) J_{A}(\Gamma_{2},t), 
\end{align}
where $\mu(0)=\rho_{A}(0)=|0\rangle\langle0|$ is the initial state of the atom, $J_{A}(t)=  \sqrt{\Gamma}e^{-i\omega_{1} t}\sigma$ and $H_{A}(t)$ are the jump operator of the atom and the Hamiltonian of the atom respectively in the interaction picture. The Hamiltonian of the atom in the interaction picture is given by
\begin{align}
    H_{A}(\Gamma, t) =  \delta  \sigma^\dagger \sigma  +i \sqrt{\Gamma} \left( \sigma f^*(t) - \text{H.c.} \right).
\end{align}
Note that $J_{A}(\Gamma,t)$ always appears with its conjugate in the above differential equation for the generalized master equation, so that the effect of phase disappears. The generalized density operator can be decomposed in the Pauli basis as $\mu(t) =\frac{1}{2} \sum_{j=0}^{3}c_{j}\sigma_{j}$ where $c_{j}$ can be complex since $\mu(t)$ is not necessarily Hermitian. Defining $c_{j} = d_{j} + i d_{j+4}$ and $f(t) = f_{r}(t) + i f_{i}(t)$, the real differential equations associated with these variables can be expressed as 
\begin{align}\label{eq:ODE_d0}
    \dot{d}_0(t) &= -\frac{1}{4} \left(\Gamma_{1,2}^{-} \right)^2 d_0(t) - \frac{1}{4} \left(\Gamma_{1,2}^{-}\right)^2 d_3(t) + \Gamma_{1,2}^{-} d_5(t) f_i(t) + \left(\Gamma_{1,2}^{-}\right) d_6(t) f_r(t), \\
    \label{eq:ODE_d1}
    \dot{d}_1(t) &= -\frac{1}{4} \left(\Gamma_1 + \Gamma_2\right) d_1(t) - \delta d_2(t) - \frac{1}{4} \left(\Gamma_1 - \Gamma_2\right) d_6(t) + d_4(t) \Gamma_{1,2}^{-} f_i(t) + d_3(t) \Gamma_{1,2}^{+} f_r(t), \\
    \label{eq:ODE_d2}
    \dot{d}_2(t) &= \delta \, d_1(t) - \frac{1}{4} \left(\Gamma_1 + \Gamma_2\right) d_2(t) + \frac{1}{4} \left(\Gamma_1 - \Gamma_2\right) d_5(t) -d_3(t) \Gamma_{1,2}^{+} f_i(t) + d_4(t) \Gamma_{1,2}^{-} f_r(t), \\
    \label{eq:ODE_d3}
    \dot{d}_3(t) &= -\frac{1}{4} \left(\Gamma_{1,2}^{+}\right)^2 d_0(t) - \frac{1}{4} \left(\Gamma_{1,2}^{+}\right)^2 d_3(t) + \Gamma_{1,2}^{+} d_2(t) f_i(t) - \Gamma_{1,2}^{+} d_1(t) f_r(t), \\
    \label{eq:ODE_d4}
    \dot{d}_4(t) &= -\frac{1}{4} \left(\Gamma_{1,2}^{-}\right)^2 d_4(t) - \frac{1}{4} \left(\Gamma_{1,2}^{-}\right)^2 d_7(t) - \Gamma_{1,2}^{-} d_1(t) f_i(t) - \Gamma_{1,2}^{(-)} d_2(t) f_r(t), \\
    \label{eq:ODE_d5}
    \dot{d}_5(t) &= \frac{1}{4} \left(\Gamma_1 - \Gamma_2\right) d_2(t) - \frac{1}{4} \left(\Gamma_1 + \Gamma_2\right) d_5(t) - \delta d_6(t) - \Gamma_{1,2}^{-} d_0(t) f_i(t) + \Gamma_{1,2}^{+}d_7(t) f_r(t), \\
    \label{eq:ODE_d6}
    \dot{d}_6(t) &= -\frac{1}{4} \left(\Gamma_1 - \Gamma_2 \right) d_1(t) + \delta \, d_5(t) - \frac{1}{4} \left(\Gamma_1 + \Gamma_2 \right) d_6(t) - d_7(t) \Gamma_{1,2}^{+}  f_i(t) - d_0(t) \Gamma_{1,2}^{-} f_r(t), \\
    \label{eq:ODE_d7}
    \dot{d}_7(t) &= -\frac{1}{4} \left(\Gamma_{1,2}^{+} \right)^2 d_4(t) - \frac{1}{4} \left(\Gamma_{1,2}^{+} \right)^2 d_7(t) + \Gamma_{1,2}^{+} d_6(t) f_i(t) - \Gamma_{1,2}^{+}d_5(t) f_r(t),
\end{align}
where $\Gamma_{1,2}^{\pm} \equiv \sqrt{\Gamma_1} \pm \sqrt{\Gamma_2}$. We omit the dependence of $d_{j}(t)$ on $\Gamma_{1}$ and $\Gamma_{2}$ is not shown explicitly above for compactness, $d_{j}(t) \equiv d_{j}(\Gamma_{1}, \Gamma_{2},t)$. The global quantum Fisher information is given by
\begin{align}
    \mathcal{F}\left(\ket{\psi_{\Gamma}(t)}\right) &=-4\frac{d^{2}}{d\zeta^{2}} \left[|\text{Tr}\left(\mu(\Gamma, \Gamma+\zeta, t)\right)|\right]_{\zeta=0}=-4\frac{d^{2}}{d\zeta^{2}} \left[|c_{0}(\Gamma, \Gamma+\zeta,t)\right]|_{\zeta=0}\\
    &= -4\frac{d^{2}}{d\zeta^{2}} \left[ \sqrt{d_{0}^{2}(\Gamma, \Gamma+\zeta,t)+d_{4}^{2}(\Gamma, \Gamma+\zeta, t)} \right]_{\zeta=0}.
\end{align}
As can be seen from the above, only the difference between $\Gamma_{1}$ and $\Gamma_{2}$ plays a role, so we use define new variables $g_{j}(\delta, t) \equiv d_{j}(\Gamma_{1}=\Gamma,\Gamma_{2}=\Gamma_{1}+\zeta, t)$ for the sake of compactness. Evaluating the derivatives with respect to $\zeta$, the QFI is given by 
\begin{align}
    \frac{d}{dt}\mathcal{F}\left(\ket{\psi_{\Gamma}(t)}\right) &= -4 \frac{d}{d t}\left(\frac{d^{2} g_{0}(0, t)}{d \zeta^{2}}\right)-8 \left(\frac{d g_{4}(0, t)}{d \zeta}\right)\frac{d}{d t}\left(\frac{\partial g_{4}(0, t)}{d \zeta} \right).
\end{align}
Then, taking the $\zeta$ derivatives of the above ODEs  \cref{eq:ODE_d0,eq:ODE_d4,eq:ODE_d5,eq:ODE_d6,eq:ODE_d7}, we have
\begin{align}
    \frac{d}{d t}\left(\frac{d^{2} g_{0}(0, t)}{d\zeta^{2}} \right)&= -\frac{1}{4}\biggl(\frac{1}{2 \Gamma } g_{0}(0,t) +\frac{1}{2 \Gamma} g_{3}(0,t)-\frac{1}{\Gamma^{3/2}}f_{i}(t) g_{5}(0,t)-\frac{1}{\Gamma ^{3/2}}f_{r}(t) g_{6}(0,t) +\frac{4}{\sqrt{\Gamma }}f_{i}(t)\left(\frac{\partial g_{5}(0,t)}{\partial \zeta} \right) \nonumber \\
    &+\frac{4}{\sqrt{\Gamma}} f_{r}(t) \left(\frac{\partial g_{6}(0,t)}{\partial \zeta} \right)\biggr),\\
    \frac{d}{d t}\left(\frac{d^{2} g_{4}(0, t)}{d\zeta^{2}} \right)&= \frac{1}{2 \sqrt{\Gamma }}\left(f_{i}(t) g_{1}(0,t)+f_{r}(t) g_{2}(0,t) \right), \\
    \frac{d}{d t}\left(\frac{d^{2} g_{5}(0, t)}{d\zeta^{2}} \right)&= 2 \sqrt{\Gamma } f_{r}(t)
   \left(\frac{\partial g_{7}(0,t)}{\partial \zeta} \right)-\frac{\Gamma}{2}   \left(\frac{\partial g_{5}(0, t)}{\partial \zeta}\right)-\delta\left(\frac{\partial g_{6}(0,t)}{\partial \zeta}\right) -\frac{1}{4}
   g_{2}(0,t)-\frac{1}{4}
   g_{5}(0, t)\nonumber \\
   & +\frac{1}{2 \sqrt{\Gamma }}\left(f_{r}(t) g_{7}(0,t) + f_{i}(t) g_{0}(0,t)\right), \\
   \frac{d}{d t}\left(\frac{d^{2} g_{6}(0, t)}{d\zeta^{2}} \right) &= \delta\left(\frac{\partial g_{5}(0,t)}{\partial \zeta}\right)-\frac{\Gamma}{2} \left(  \frac{\partial g_{6}(0,t)}{\partial \zeta}\right)-2 \sqrt{\Gamma } f_{i}(t)
   \left(  \frac{\partial g_{7}(0,t)}{\partial \zeta}\right) + \frac{1}{4} g_{1}(0,t)-\frac{1}{4}
   g_{6}(0,t) \nonumber \\
   &+\frac{1}{2 \sqrt{\Gamma }} \left(f_{r}(t) g_{0}(0,t) - f_{i}(t) g_{7}(0,t)\right), \\
   \frac{d}{d t}\left(\frac{d^{2} g_{7}(0, t)}{d\zeta^{2}} \right) &= -\Gamma  \left(\frac{\partial g_{4}(0,t)}{\partial \zeta}\right)-2 \sqrt{\Gamma } f_{r}(t)
   \left(\frac{\partial g_{5}(0,t)}{\partial \zeta}\right)+2
   \sqrt{\Gamma } f_{i}(t) \left(\frac{\partial g_{6}(0,t)}{\partial \zeta}\right)-\Gamma  \left(\frac{\partial g_{7}(0,t)}{\partial \zeta}\right) \nonumber \\
   &-\frac{1}{2}
   g_{7}(0,t) -\frac{1}{2} g_{4}(0,t) +\frac{f_{i}(t)
   g_{6}(0,t)}{2 \sqrt{\Gamma }} -\frac{f_{r}(t) g_{5}(0,t)}{2 \sqrt{\Gamma }}.
\end{align}
As can be seen from the above, these ODEs further couple to the variables $g_{j}(0,t)$ for $j=\{0,1,...,7\}$,  whose ODEs are given by
\begin{align}
\frac{d}{dt}g_{0}(0,t)&=0, \\
\frac{d}{dt}g_{1}(0,t)&=-\frac{1}{2} \Gamma  g_{1}(0,t)-\delta  g_{2}(0,t)+2 \sqrt{\Gamma } f_{r}(t) g_{3}(0,t), \\
\frac{d}{dt}g_{2}(0,t)&= \delta  g_{1}(0,t)-\frac{1}{2} \Gamma  g_{2}(0,t)-2 \sqrt{\Gamma} f_{i}(t) g_{3}(0,t), \\
\frac{d}{dt}g_{3}(0,t)&=-\Gamma  g_{0}(0,t)-2 \sqrt{\Gamma } f_{r}(t)
   g_{1}(0,t)+2 \sqrt{\Gamma } f_{i}(t) g_{2}(0,t)-\Gamma  g_{3}(0,t), \\
\frac{d}{dt}g_{4}(0,t)&=0, \\
\frac{d}{dt}g_{5}(0,t)&=-\frac{1}{2} \Gamma  g_{5}(0,t)-\delta  g_{6}(0,t)+2 \sqrt{\Gamma } f_{r}(t) g_{7}(0,t), \\
\frac{d}{dt}g_{6}(0,t)&=\delta  g_{5}(0,t)-\frac{1}{2} \Gamma  g_{6}(0,t)-2 \sqrt{\Gamma } f_{i}(t) g_{7}(0,t), \\
\frac{d}{dt}g_{7}(0,t)&=-\Gamma g_{4}(0,t)-2 \sqrt{\Gamma } f_{r}(t) g_{5}(0,t)+2 \sqrt{\Gamma } f_{i}(t) g_{6}(0,t)-\Gamma g_{7}(0,t).
\end{align}
It should be first noted that $g_{0}(0, t)=1$, the equations associated with $g_{1}(0,t),g_{2}(0,t),g_{3}(0,t)$ are the optical Bloch equations and the variables $g_{4}(0,t),g_{5}(0,t),g_{6}(0,t),g_{7}(0,t)$ do not evolve in time since they have their initial values zero. Hence, we have the following ODEs
\begin{align}
    \frac{d}{dt}g_{1}(0,t)&=-\frac{1}{2} \Gamma  g_{1}(0,t)-\delta  g_{2}(0,t)+2 \sqrt{\Gamma } f_{r}(t) g_{3}(0,t) ,\\
    \frac{d}{dt}g_{2}(0,t)&= -\frac{1}{2} \Gamma  g_{2}(0,t) + \delta  g_{1}(0,t)-2 \sqrt{\Gamma} f_{i}(t) g_{3}(0,t), \\
    \frac{d}{dt}g_{3}(0,t)&=-\Gamma  g_{3}(0,t)-\Gamma -2 \sqrt{\Gamma } f_{r}(t)
   g_{1}(0,t)+2 \sqrt{\Gamma } f_{i}(t) g_{2}(0,t), \\
    \frac{d}{dt}\left(\frac{d g_{4}(0, t)}{d \zeta} \right)&= \frac{1}{2 \sqrt{\Gamma }}\left(f_{i}(t) g_{1}(0,t)+f_{r}(t) g_{2}(0,t) \right), \\
    \frac{d}{dt}\left(\frac{d g_{5}(0, t)}{d \zeta} \right)&= -\frac{\Gamma}{2}   \left(\frac{d g_{5}(0, t)}{d \zeta}\right) + 2 \sqrt{\Gamma } f_{r}(t)
   \left(\frac{d g_{7}(0,t)}{d \zeta} \right)-\delta\left(\frac{d g_{6}(0,t)}{d \zeta}\right) -\frac{1}{4}
   g_{2}(0,t) +\frac{1}{2 \sqrt{\Gamma }}f_{i}(t),  \\
   \frac{d}{dt}\left(\frac{d g_{6}(0, t)}{d \zeta} \right) &= -\frac{\Gamma}{2} \left(  \frac{dg_{6}(0,t)}{d \zeta}\right) + \delta\left(\frac{d g_{5}(0,t)}{d \zeta}\right)-2 \sqrt{\Gamma } f_{i}(t)
   \left(  \frac{d g_{7}(0,t)}{d \zeta}\right) + \frac{1}{4} g_{1}(0,t) +\frac{1}{2 \sqrt{\Gamma }} f_{r}(t),  \\
   \frac{d}{dt}\left(\frac{d g_{7}(0, t)}{d \zeta} \right) &= -\Gamma  \left(\frac{d g_{7}(0,t)}{d \zeta}\right)  -\Gamma  \left(\frac{d g_{4}(0,t)}{d \zeta}\right)-2 \sqrt{\Gamma } f_{r}(t)
   \left(\frac{d g_{5}(0,t)}{d \zeta}\right)+2
   \sqrt{\Gamma } f_{i}(t) \left(\frac{d g_{6}(0,t)}{d \zeta}\right),\\
   \frac{d}{dt}\mathcal{F}(t) &=  \frac{1}{2\Gamma } \left(1 +g_{3}(0,t)\right)+\frac{4}{\sqrt{\Gamma }}\left(f_{i}(t)\left(\frac{d g_{5}(0,t)}{d \zeta} \right) + f_{r}(t) \left(\frac{d g_{6}(0,t)}{d \zeta} \right)\right) \nonumber\\
   &+2 \left(\frac{d g_{4}(0, t)}{d \zeta}\right)\frac{d}{d t}\left(\frac{d g_{4}(0, t)}{d \zeta} \right).
\end{align}
Relabelling the above variables as $x_{1}(t) \equiv g_{1}(0,t), y_{1}(t) \equiv g_{2}(0,t), z_{1}(t) \equiv g_{3}(0,t), w_{1}(t) \equiv \frac{d g_{4}(0,t)}{d\zeta}, x_{2}(t) \equiv \frac{d g_{5}(0,t)}{d\zeta}, y_{2}(t) \equiv \frac{d g_{6}(0,t)}{d\zeta}, z_{2}(t) \equiv \frac{d g_{7}(0,t)}{d\zeta}$ and introducing the variable $ w_{2}(t) \equiv \frac{1}{2 \sqrt{\Gamma }}\left(f_{i}(t) x_{2}(t)+f_{r}(t) y_{2}(t) \right) $, we have the ODEs given by 
\begin{align}
x_1'(t) &= -(\Gamma/2) x_1(t) + 2\sqrt{\Gamma} f_{r}(t) z_1(t) - \delta y_1(t), \\
y_1'(t) &= -(\Gamma/2)y_1(t) - 2\sqrt{\Gamma} f_{i}(t) z_1(t) + \delta x_1(t), \\
z_1'(t) &= -\Gamma z_1(t) - 2\sqrt{\Gamma} f_{r}(t) x_1(t) + 2 \sqrt{\Gamma} f_{i}(t) y_1(t)  -\Gamma, \\
w_{1}'(t) &= \left( f_{i}(t) x_1(t) + f_{r}(t) y_1(t) \right)/2 \sqrt{\Gamma}, \\
x_2'(t) &= -(\Gamma/2)x_2(t) + 2 \sqrt{\Gamma}f_{r}(t) z_2(t) - \delta y_2(t) - (y_1(t)/4) + \left(f_{i}(t)/2 \sqrt{\Gamma} \right), \\
y_2'(t) &= -(\Gamma/2)y_2(t) - 2\sqrt{\Gamma} f_{i}(t) z_2(t)  + \delta x_2(t)  + (x_1(t)/4) + \left(f_{r}(t)/2\sqrt{\Gamma} \right),  \\
z_2'(t) &= -\Gamma z_2(t) - 2\sqrt{\Gamma} f_{r}(t) x_2(t) + 2\sqrt{\Gamma} f_{i}(t) y_2(t) - \Gamma w_{1}(t), \\
w_{2}'(t) &= \left( f_{i}(t) x_2(t) + f_{r}(t) y_2(t) \right)/2 \sqrt{\Gamma}, \\
\mathcal{F}'(t) &= (1/2\Gamma) \left( 1 + z_1(t) \right)  +4 \frac{d}{dt}\left(2 w_{2}(t)-w_{1}(t)^{2} \right),
\end{align}
which are the equations shown in Sec.~\ref{sec:complex_pulses}. Note that when the pulse is real with zero detuning, the variables $y_{1}(t), w_{1}(t), x_{2}(t), z_{2}(t)$ do not evolve in time and the number of ODEs reduce to four. In that case, the relevant variables needed for computing QFI are $x_{1}(t)$, $z_{1}(t)$, $y_{2}(t)$, which are labelled as $x(t)$, $z(t)$ and $\xi(t)$ in Sec.~\ref{sec:real_pulses}
\section{Short and long pulse width limit}
We provide analytic expressions for QFI for the standard pulses analyzed in this work both in the short pulse width and the long pulse width limit. The QFI expression in the short pulse width limit is straightforward, given $p(t)$ since $\mathcal{F}_{p}$ is given by
\begin{align}
    \mathcal{F}_{p}(t) &= \frac{2}{\Gamma^{3/2}} \int_{0}^{t} d\tau\, f(\tau) p(\tau).
\end{align}
Likewise, $\mathcal{F}_{\text{long}}$ can be computed using the expressions for $q(t)$. We provide expressions for $p(t)$ along with $q(t)$ here for all the standard pulses and the resulting QFI in Table \eqref{table:pulses}. 
For a rectangular pulse, 
\begin{align}
    p_{\text{Rect}}(t) &=
    \begin{cases}
        \frac{2 \alpha}{\sqrt{\Gamma T}}\left(1 -  e^{-\frac{\Gamma t}{2}}\right) & t \leq T \\
        \frac{2\alpha \bigl(e^{\frac{\Gamma T}{2}} - 1\bigr)}{\sqrt{\Gamma T}}e^{-\frac{\Gamma t}{2}} & t \geq T 
    \end{cases},
\end{align}
and 
\begin{align}
        q_{\text{Rect}}(t) &= \begin{cases}
        \frac{\alpha}{\Gamma^{2}\sqrt{T}}\left(2e^{-\tfrac{t\Gamma}{2}} + \Gamma t e^{-\tfrac{t\Gamma}{2}} -2 \right) & t \leq T \\
        \frac{\alpha}{\Gamma^{2} \sqrt{T}} e^{-\tfrac{t\Gamma}{2}}\left(\Gamma t +2  + e^{\tfrac{T\Gamma}{2}}\left(\Gamma T - \Gamma t -2\right)\right) & t\geq T
    \end{cases},
\end{align}
where $T$ is the width of the pulse. 
For the decreasing exponential function and $t \geq 0$, 
\begin{align}
    p_{\text{DecExp}}(t) &=\frac{2 \alpha \sqrt{\Gamma T}}{\Gamma T - 1} \left(e^{\frac{t( \Gamma T -1)}{2T}} - 1\right) e^{-\frac{\Gamma t}{2}},\\
    q_{\text{DecExp}}(t) &= \frac{\alpha \sqrt{T}}{(\Gamma T - 1)^{2}}\bigl[2T\bigl(1 - e^{\frac{t(\Gamma T - 1)}{2T}}\bigr) + t(\Gamma T - 1)\bigr]e^{-\frac{\Gamma t}{2}}.
\end{align}
For the ease of computation with the exponentially increasing pulse, the symmetric exponential pulse and the Gaussian pulse, we set $t_0=0$ in the computations below and set the lower limit in the definition of $p(t)$, $q(t)$ and $\mathcal{F}_{p}$ to $-\infty$ instead of zero. Since this corresponds to just shifting the pulse on the time axis, the QFI should be invariant. We then have for the rising exponential function, 
\begin{align}
p_{\text{RisExp}}(t) &=
\begin{cases}
    \frac{ 2 \alpha \sqrt{\Gamma T} }{ 1 + \Gamma T} e^{\frac{t}{2T}} & t \leq 0 \\
    \frac{ 2 \alpha \sqrt{T \Gamma} }{ 1 + T \Gamma }e^{ -\frac{\Gamma t}{2} }
 & t \geq 0
\end{cases},\\
q_{\text{RisExp}}(t) &= \begin{cases}
    -\frac{ 2 \alpha T^{3/2}}{\left( 1 + T \Gamma \right)^{2}} e^{\frac{t}{2T}} & t \leq 0 \\
    -\frac{\alpha \sqrt{T}}{\left( 1 + T \Gamma \right)^{2} }\left( t + \Gamma T t  + 2T\right)e^{-\frac{\Gamma t}{2} } & t \geq 0
\end{cases}.
\end{align}
For the symmetric exponential pulse, we have
\begin{align}
    p_{\text{SymExp}}(t) &=
    \begin{cases}
        \frac{2 \alpha \sqrt{\Gamma T} }{\Gamma T +2} e^{\frac{t}{T}} & t \leq 0 \\
        \frac{ 2 \alpha \sqrt{\Gamma T}}{\Gamma^{2} T^{2} -4} \left((\Gamma T + 2)e^{-\frac{t}{T}}  -4 e^{ -\frac{\Gamma t}{2}} \right)   & t \geq 0
    \end{cases}, \\
    q_{\text{SymExp}}(t) &= \begin{cases}
        -\frac{ 2\alpha T^{3/2}}{\left( \Gamma T  + 2\right)^{2}}  e^{\frac{t}{T}}  & t \leq 0 \\
        -\frac{ 2 \alpha \sqrt{T} }{ \left(\Gamma^{2} T^{2} -4\right)^{2}} \left[e^{-\frac{t}{T}} T \left(\Gamma T + 2 \right)^{2} - 2 e^{-\frac{\Gamma t}{2} }  \left( 4 \Gamma T^{2}  + t \Gamma^{2} T^{2}t - 4 t \right) \right] & t \geq 0
    \end{cases}.
\end{align}
The QFI evaluation of a Gaussian pulse is not straightforward. For the short pulse width limit, we then have
\begin{align}
    \mathcal{F}_{p}^{\infty} &= \frac{2}{\Gamma} \int_{-\infty}^{\infty} dt \int_{-\infty}^{t} d\tau f(t) f(\tau) e^{\frac{\Gamma}{2}(\tau-t)} \\
    &= \frac{1}{\Gamma} \int_{-\infty}^{\infty} dt \int_{-\infty}^{\infty} d\tau f(t) f(\tau) e^{\frac{\Gamma}{2}|\tau-t|} .
\end{align}
Let $s=\tau -t $ and $r = t$, then
\begin{align}
    \frac{\Gamma^{2} \mathcal{F}_{p}^{\infty}}{\alpha^{2}} &=\frac{\Gamma}{\alpha^{2}} \int_{-\infty}^{\infty} ds \int_{-\infty}^{\infty} dr \; f(r) f(s+r) e^{-\frac{\Gamma}{2}|s|} \\
    &= \Gamma \int_{-\infty}^{\infty} ds \; e^{-\frac{s^{2}}{4 T^{2}}} e^{-\frac{\Gamma}{2}|s|} = 2 \Gamma \int_{0}^{\infty} ds \; e^{-\frac{s^{2}}{4 T^{2}}} e^{-\frac{\Gamma}{2}s} \\
    &= 2 \Gamma T \sqrt{\pi}  e^{\frac{\Gamma^{2}T^{2}}{4}} \text{erfc}\left( \frac{\Gamma T}{2} \right) .
\end{align}
In the long pulse width limit, it is easier to evaluate the QFI in the frequency domain, 
\begin{align}
    \mathcal{F}_{\text{long}}^{\infty} &=16 \int_{-\infty}^{\infty} d\omega \; \omega^{2} |\tilde{q}(\omega)|^{2} ,
\end{align}
where $p(t) = \sqrt{\Gamma} (u*f) (t)$ and $u(t) = \sqrt{\Gamma}\Theta(t) e^{-\frac{\Gamma t}{2}}$. Hence, 
\begin{align}
    \tilde{p}(\omega) &= \left(\frac{2\sqrt{\Gamma}}{\Gamma - 2i \omega}\right)\frac{\alpha \sqrt{T}e^{-\tfrac{1}{2} T^{2}\omega^{2}}}{\pi^{1/4}}, \\
    \tilde{q}(\omega) &= \frac{2\alpha \sqrt{T}\,e^{-\tfrac{1}{2} T^{2}\omega^{2}}}{\pi^{1/4}\,\bigl(2\omega + i\Gamma \bigr)^{2}}, \\
    \frac{\Gamma^{2}\mathcal{F}_{\text{long}}^{\infty}}{\alpha^{2}} &= \frac{64\Gamma^{2}T}{\sqrt{\pi}}\int_{-\infty}^{\infty} d\omega \;\omega^{2}\frac{e^{-\omega^{2}T^{2}}}{\bigl(\Gamma^{2}+4\omega^{2}\bigr)^{2}}=2 \Gamma T \left( \sqrt{\pi}(\Gamma^{2}T^{2} + 2)e^{\frac{\Gamma^{2} T^{2}}{4}}\text{erfc}\left(\frac{\Gamma T}{2}\right) -2 \Gamma T \right).
\end{align}
\section{Basis functions}
\label{appendix:basis}
\subsection{Harmonics}
The harmonics are defined as \cite{griffiths2018introduction}
\begin{align}
    \xi_{n}(t) &= \sqrt{\frac{2}{T}}\sin\left(\frac{n\pi t}{T}\right),
\end{align}
where $n=\{1,2,...\}$. These functions go to zero at the boundaries and form an orthonormal basis for square integrable functions that are defined between $t=0$ and $t=T$. The orthonormality condition is given by 
\begin{align}
    \int_{0}^{T} dt \; \xi_{n}(t) \xi_{m}(t) = \delta_{m,n}.
\end{align}
The mean $\mu_{n}$ and the variance $\sigma_{n}^{2}$ of an arbitrary function in this basis set is given by
\begin{align}
    \mu_{n} &= \int_{0}^{T} dt\; t \xi_{n}(t)^{2} =\frac{T}{2},\\    
    \sigma_{n}^2 &= \int_{0}^{T} dt\; t^{2} \xi_{n}(t)^{2} - \left(\int_{0}^{T} dt\; t \xi_{n}(t)^{2}\right)^{2} = \frac{T^{2}}{12}\left( 1 - \frac{6}{n^{2}\pi^{2}}\right).
\end{align}
Note that if $n$ is reasonably large, the standard deviation of all the basis functions is at most $\sigma_{m} \leq T/\sqrt{12}$. Any real or complex function can then be expressed with the boundary condition $f(0)=f(T)=0$ as 
\begin{align}
\label{eq:fp_basis}
    f(t) &= \sqrt{\frac{2}{T}}\sum_{n=1}^{\infty} c_{n} \sin\left(\frac{n\pi t}{T}\right),
\end{align}
where $c_{n}$ is complex or real depending on whether the function $f(t)$ is real or complex, and the coefficients are normalized to the average photon number of the pulse
\begin{align}
    \sum_{n=1}^{\infty} |c_{n}|^{2}=|\alpha|^{2}.
\end{align}
\subsection{Hermite Gaussian basis}
\label{app:Hermite_basis}
\begin{figure*}[t]
\includegraphics[width = \textwidth]{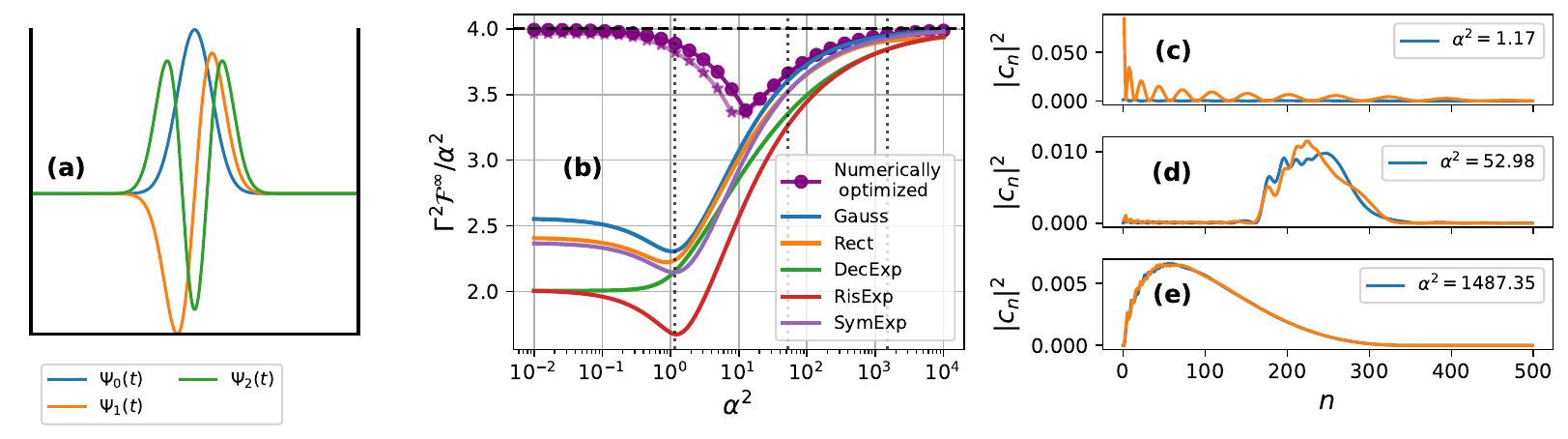}
\caption{\textbf{(a)} The first three Hermite-Gaussian functions are shown here. \textbf{(b)} The QFI of the numerically optimized pulse, expressed in the Hermite–Gaussian basis, is shown for $n_{\text{max}}=250$ (stars) and $n_{\text{max}}=500$ (dots) where $n_{\text{max}}$ is the maximum number of basis functions considered. We consider $10$ and $20$ random initial seeds for $n_{\text{max}}=250$ and $n_{\text{max}}=500$ respectively, and we plot the maximum QFI value obtained from these seeds. As in the case of harmonics, the optimal QFI per unit photon gets closer to $4$ as the length of pulse is increased. Note that the variance of a basis function is given by $\sigma_{n}  \sim \sqrt{n} T$ where $T$ is set to the optimal width of the Gaussian pulse, which is the first basis function. For reference, the optimal QFI (maximized over the pulse width) of standard pulses is plotted as a function of $\alpha^{2}$. \textbf{(c-e)} The population of the optimal pulse in the Hermite-Gaussian basis functions where blue line shows the population in the even basis functions and the orange curve is the population in the odd basis functions corresponding to three different values of $\alpha^{2}$ identified by the vertical dotted lines in (b).}
\label{fig:qfi_optimization_Hermite}
\end{figure*}
The Hermite-Gaussian functions are defined as \cite{griffiths2018introduction}
\begin{align}
    \psi_n(t) := \frac{1}{\sqrt{2^n n! \sqrt{\pi}}} \, H_n(t) \, e^{-t^2/2},
\end{align}
where 
\begin{align}
    H_{n}(t) = (-1)^n e^{t^{2}} \frac{d^{n}}{dt^{n}}e^{-t^{2}} .
\end{align}
We will rescale these functions so that they take account into account a parameter $T$, which is proportional to the variance of the pulse,
\begin{align}
    \Psi_n(t) := \frac{1}{\sqrt{T}} \, \psi_n\left(\frac{t}{T}\right).
\end{align}
An arbitrary function in this basis set has mean and variance given by 
\begin{align}
    \mu_{n} &= \int_{-\infty}^{\infty} dt\; t \Psi_{n}(t)^{2} = 0 ,\\
    \sigma_{n}^{2} &= \int_{-\infty}^{\infty} dt\; t^{2} \Psi_{n}(t)^{2} - \left(\int_{-\infty}^{\infty} dt\; t \Psi_{n}(t)^{2}\right)^{2} = \left(n+\frac{1}{2}\right)T^{2}.
\end{align}
In addition to optimization in the basis of harmonics, we also performed numerical optimization in the Hermite-Gaussian basis, and the obtained results are shown in Fig.~\eqref{fig:qfi_optimization_Hermite}. As seen here, the numerical optimization results are identical to results obtained in the basis of harmonics. That is, the optimal QFI per unit photon is $4$, and this is obtained in the large width regime $(\alpha/\sqrt{T}) \gg 1$. This confirms that we have not missed any higher frequency terms in the basis of harmonics since $\omega_{n} = n \pi /T$ and $n$ is finite for a given $T$. The populations of the optimal pulse in this basis are shown in Fig.~\eqref{fig:qfi_optimization_Hermite}~(c-e), where the blue and orange curves correspond to the population in the even and odd basis functions respectively. Note that the Hermite--Gaussian functions can be shifted by a sufficiently large positive constant $t_{0}$, \textit{i.e.}, $\Psi_{n}(t - t_{0})$, so that the pulse vanishes at $t=0$ up to numerical precision. This convention is consistent with the boundary condition used throughout the text, namely $f(0)=0$. Consequently, all relevant integrals are taken with the lower limit at $t=0$, as in Eqs.~\eqref{eq:Fp_definition}, \eqref{eq:Fz_definition}, and \eqref{eq:Fx_definition}.
\section{Positivity of \texorpdfstring{$\mathcal{F}_{p}$}{F\_p}}
\label{appendix:Fp_properties_real1}
Here we analyze the term $\mathcal{F}_{p}$ which appears in the QFI expression (Eq.~\eqref{eq:Fp_definition}). In particular, we show that $\mathcal{F}_{p}$ is always positive and the maximum value it can achieve is $4 \alpha^{2}$. By definition,
\begin{align}
    \mathcal{F}_{p}(t) &= \frac{2}{\Gamma} \int_{0}^{t} d\tau\, f(\tau) \left(\int_{0}^\tau d\tau' \, e^{-\Gamma(\tau - \tau')/2} f(\tau') \right),
\end{align}
where the pulse $f(t)$ is assumed to be real. Let 
\begin{align}
    p(t) & =\sqrt{\Gamma} \int_{0}^t d\tau' \, e^{-\Gamma (t - \tau')/2} f(\tau'), 
\end{align}
which evolves under the following ODE
\begin{align}
    p'(t) +\frac{\Gamma}{2}p(t) &= \sqrt{\Gamma} f(t) .
\end{align}
Using the above, $\mathcal{F}_{p}$ can be written as
\begin{align}
    \Gamma^{2} \mathcal{F}_{p}(t) &= 2\sqrt{\Gamma}  \int_{0}^{t} d\tau\, f(\tau) p(\tau) =  \Gamma \int_{0}^{t} d\tau\,p(\tau)^{2} + \int_{0}^{t} d\tau\, \partial_{\tau} p(\tau)^{2} \\
    &= p(t)^{2} + \Gamma \int_{0}^{t} d\tau\,p(\tau)^{2}  \geq 0
\end{align}
Also note that $\mathcal{F}_{p}$ is proportional to $\alpha^{2}$ since $f(t) \propto \alpha$, and therefore $\mathcal{F}_{p}/\alpha^{2}$ is independent of $\alpha$.
\section{Perturbation theory for real pulses}
\label{appendix:perturbation_theory_real}
In the subsection, we provide a perturbative solution to the QFI for real pulses in the limit $\varepsilon \equiv \alpha/\sqrt{T} \ll 1$ where the input pulse is given by $f(t) = \varepsilon \phi(t/T)$. Using the regular perturbation theory ansatz, $u(t) = u^{(0)}(t)+\varepsilon u^{(1)}(t)+\varepsilon^{2} u^{(2)}(t)+... $ where $u \in \{x,z,\xi,\mathcal{F}\}$ in the ODEs given in Eq.~\eqref{eq:xdot} to Eq.~\eqref{eq:Fdot}. The zeroth-order solution is given by $x^{(0)}(t) =0 $, $y^{(0)}(t)=0$ and $z^{(0)}(t) =-1 $, which correspond to the initial conditions. The first-order solution is given by
\begin{align}
x^{(1)}(t) &= -2\sqrt{\Gamma} \int_{0}^{t} d\tau \; e^{\frac{\Gamma (\tau-t)}{2}}f(\tau) = -2 p(t), \\
z^{(1)}(t) &= 0, \\
\xi^{(1)}(t) &= \frac{1} {2\sqrt{\Gamma}} \int_{0}^{t} d\tau\; e^{\frac{\Gamma (\tau-t)}{2}}f(\tau) +  \frac{1} {4}\int_{0}^{t}  d\tau\; e^{\frac{\Gamma (\tau-t)}{2}}x_{1}(\tau) =\frac{1}{2\Gamma}p(t) -  \frac{1} {2}\int_{0}^{t}  d\tau\; e^{\frac{\Gamma (\tau-t)}{2}}p(\tau),\\
\mathcal{F}^{(1)}(t) &= 0,
\end{align}
and the second-order solution is given by
\begin{align}
x^{(2)}(t) &= 0, \\
z^{(2)}(t) &= 4 \Gamma e^{-\Gamma t}\int_{0}^{t} d\tau \int_{0}^{\tau} d\tau'\; e^{\Gamma(\tau'+\tau)/2} f(\tau') f(\tau) = 2 \left(\sqrt{\Gamma}\int_{0}^{t} ds \; e^{(s-t)/2}\,f(s) \right)^{2} =2 \left(p(t)\right)^{2},\\
\xi^{(2)}(t) &= 0, \\
\mathcal{F}^{(2)}(t) &=  \frac{1}{\Gamma} \int_{0}^{t} d\tau\; p(\tau)^{2} + \frac{2}{\Gamma^{3/2}} \int_{0}^{t}
 d\tau \; f(\tau) p(\tau) - \frac{2}{\sqrt{\Gamma}} \int_{0}^{t}  d\tau \; f(\tau) \int_{0}^{\tau} d\tau' \;  e^{\frac{\Gamma (\tau'-\tau)}{2}} p(\tau') ,
\end{align}
where 
\begin{align}
    p(t) &\equiv  \sqrt{\Gamma} \int_{0}^{t} d\tau \; e^{\frac{\Gamma (\tau-t)}{2}}f(\tau)
\end{align}
is proportional to the probability amplitude of the atom to be excited up to second order. This can be seen by noting that $\frac{(1 + z(t))}{2}=p(t)^{2}$. The Fisher information is then given by $F^{(2)}(t)$ since both the zeroth and the first order corrections are zero. That is, 
\begin{align}
    \mathcal{F}_{\text{long}}(t) &=  \frac{1}{\Gamma} \int_{0}^{t} d\tau\; p(\tau)^{2} + \frac{2}{\Gamma^{3/2}} \int_{0}^{t}
 d\tau \; f(\tau) p(\tau) +4 \int_{0}^{t}  d\tau \; f(\tau) q(\tau),
\end{align}
where 
\begin{align}
    q(t) &\equiv \frac{\partial}{\partial \Gamma} \left(\frac{p(t)}{\sqrt{\Gamma}}\right) =\int_{0}^{t} d\tau \left(\frac{\tau -t}{2}\right) e^{\Gamma\left(\frac{\tau-t}{2}\right)} f(\tau) = -\frac{1}{2\sqrt{\Gamma}}  \int_{0}^{t} d\tau' \;  e^{\frac{\Gamma (\tau'-t)}{2}} p(\tau') ,
\end{align}
We use the subscript ``long" to highlight that this is the QFI in the long pulse width limit. Equivalently, the second order solution for the QFI of a coherent state in this limit, can be expressed as the solution to the following coupled ODEs
\begin{align}
\frac{\partial p(t)}{\partial t} &= -\frac{\Gamma}{2} p(t) + \sqrt{\Gamma}f(t), \\
\frac{\partial q(t)}{\partial t} &= -\frac{\Gamma}{2} q(t) - \frac{p(t)}{2\sqrt{\Gamma}}, \\
\frac{\partial \mathcal{F}_{\text{long}}}{\partial t}   &= \frac{1}{\Gamma}p(t)^2 + \frac{2}{\Gamma^{3/2}} f(t) p(t) + 4 f(t) q(t).
\end{align}
Note that both the functions $p(t)$ and $q(t)$ are $\Gamma$ dependent, but their dependence is not shown explicitly for the sake of compactness. We can then express $\mathcal{F}_{\text{long}}(t)$ in the long-time limit only as a function of $q(t)$,
\begin{align}
    \mathcal{F}_{\text{long}}(t \rightarrow \infty) \equiv \mathcal{F}_{\text{long}}^{\infty}    &= \int_{0}^{\infty} dt \left[ 8 \left(\dot{q}(t)\right)^{2} -8 \ddot{q}(t)q(t) \right] \\
    \label{eq:QFI_q_form}
    &=   16\int_{0}^{\infty} dt \left(\dot{q}(t)\right)^{2},
\end{align}
where we used the fact that $p(0)=p(\infty)=q(0)=q(\infty)=0$. It is straightforward to see that $p(0)=q(0)=0$ from their respective definitions. The term $p(\infty)=0$ can be seen from the fact probability of the atom to be in the excited state eventually goes to zero, while the term $q(\infty)=0$ can be noted from the fact that the driving term $p(t)$ goes to zero at large times.
\section{The QFI of Single-Photon Pulses}
\label{appendix:single_photon_pulses}
In this section, we express the QFI of single-photon pulses $\mathcal{F}_{\text{single}}$ given in~\cite{albarelli2023fundamental} for a real pulse in a form that allows us to compare with the QFI of a coherent state in the perturbative limit. In the long-time limit after the atom has spontaneously decayed to the ground state, the QFI is given by (see Appendix C in~\cite{albarelli2023fundamental}) 
\begin{align}
    \mathcal{F}_{\text{single}}(t) &= 4\int_{0}^t  d \tau \left( \int_{0}^\tau dt' e^{-\frac{\Gamma}{2}(\tau-t')} f(t') \right)^2 + 4\Gamma^2 \int_{0}^t  d \tau \left( \int_{0}^\tau dt'\frac{(t'-\tau)}{2} e^{-\frac{\Gamma }{2}(\tau-t')} f(t') \right)^2 \nonumber \\ 
    & + 8 \Gamma \int_{0}^t d \tau \left( \int_{0}^\tau dt' e^{-\frac{\Gamma}{2}(\tau-t')} f(t') \right)\left( \int_{0}^\tau dt'\frac{(t'-\tau)}{2} e^{-\frac{\Gamma}{2}(\tau-t')} f(t') \right).
\end{align}
Using the definitions of $p(t)$ and $q(t)$ given in  Eq.\eqref{eq:pt_definition} and Eq.~\eqref{eq:qt_definition} respectively, the above can be expressed as the solution to the following coupled ODEs
\begin{align}
\frac{\partial p(t)}{\partial t} &= -\frac{\Gamma p(t)}{2}  + \sqrt{\Gamma} f(t) ,\\
\frac{\partial q(t)}{\partial t} &= - \frac{\Gamma  q(t)}{2} -\frac{p(t)}{2\sqrt{\Gamma}},   \\
\frac{\partial \mathcal{F}_{\text{single}}}{\partial t}   &= 16 \left(\frac{\partial q(t)}{\partial t}\right)^{2},
\end{align}
where we use the subscript ``single" on $\mathcal{F}$ to highlight that this is the expression corresponding to the single-photon pulse. In the context of a single-photon problem, $p(t)$ is the same as the probability amplitude of the atom to be in the excited state up to a minus sign. The QFI for a single-photon pulse can then be alternatively expressed as
\begin{align}
        \mathcal{F}_{\text{single}}(t\to\infty)
        &= 16 \int_{0}^{\infty}d\tau \;
             \bigl(\partial_{\tau}q(\tau)\bigr)^{2},
\end{align}
which is identical to the QFI expression up to second order in $\varepsilon = \alpha/\sqrt{\Gamma T}$ for the coherent state as shown in Eq.~\eqref{eq:QFI_q_form}.
\section{Diagonalization of QFI up to second order for real pulses}
\label{appendix:diagonalizing_qfi_long_time_real}
The QFI in the long pulse width limit is given by 
\begin{align}
    \mathcal{F}_{\text{long}}(t) &=  \mathcal{F}_{1}(t) + \mathcal{F}_{2}(t)+ \mathcal{F}_{3}(t)
\end{align}
where 
\begin{align}
    \mathcal{F}_{1}(t) &= \frac{2}{\Gamma^{3/2}} \int_{0}^{t}
    d\tau \; f(\tau) p(\tau), \\
    \mathcal{F}_{2}(t) &=\frac{1}{\Gamma} \int_{0}^{t} d\tau\; p(\tau)^{2}, \\
 \mathcal{F}_{3}(t) &=4 \int_{0}^{t}  d\tau \; f(\tau) q(\tau).
\end{align}
We expand $f(t)$ in the basis of harmonics, $f(t) = \sum_{n}c_{n} \xi_{n}(t)=\sqrt{2/T}\sum_{n}c_{n} \sin\left(n\pi t/T\right)$, and express all the three terms in the bilinear form. We first evaluate the expressions for $p(t)$ and $q(t)$ in this basis, 
\begin{align}
   p(t) &=  \sqrt{\Gamma} \int_{0}^{t} d\tau \; e^{\frac{\Gamma (\tau-t)}{2}}f(\tau)\equiv \sum_{n} c_{n} \phi_{n}(t), \\
   q(t) &=  \sum_{n} c_{n} \frac{\partial}{\partial \Gamma} \left(\frac{\phi_{n}(t)}{\sqrt{\Gamma}}\right) = \sum_{n}c_{n} \eta_{n}(t),
\end{align}
where $\sum_{n} c_{n}^{2}=\alpha^{2}$,
\begin{align}
\label{eq:phi_definition}
\phi_{n}(t) &= \begin{cases} 
\frac{\sqrt{8\Gamma T}}{4n^2\pi^2 + \Gamma^2 T^2}\left[2n\pi e^{-\tfrac{t\Gamma}{2}}-2n\pi\cos\left(\frac{n\pi t}{T}\right) + \Gamma T\sin\left(\frac{n\pi t}{T}\right)\right] & 0\leq t\leq T \\
\frac{2\,n\pi\sqrt{8\Gamma T}}{4n^{2}\pi^{2} + T^{2}\Gamma^{2}}\left[e^{-\tfrac{t\Gamma}{2}} - (-1)^{n}e^{-\tfrac{(t - T)\Gamma}{2}}\right] & t\geq T
\end{cases},
\end{align}
and 
\begin{align}
\eta_{n}(t) &= \frac{\sqrt{8\,T}}{\left(4 n^{2} \pi^{2} + T^{2} \Gamma^{2} \right)^{2}} 
\left[ 4 n \pi T^{2} \Gamma \cos\left( \frac{n \pi t}{T} \right) + T \left( 4 n^{2} \pi^{2} - T^{2} \Gamma^{2} \right) \sin\left( \frac{n \pi t}{T} \right) - n \pi e^{-\frac{t \Gamma}{2}} \left( 4 n^{2} \pi^{2} t + T^{2} \Gamma \left( 4 + t \Gamma \right) \right) \right],
\end{align}
which is defined for $t\leq T$, and its value for $t>T$ is not required for our proof. 
\subsection{Evaluation of \texorpdfstring{$\mathcal{F}_{1}$}{F\_1}}
\label{appendix:Fp_properties_real2}
\begin{figure*}[t]
\includegraphics[width=\linewidth]{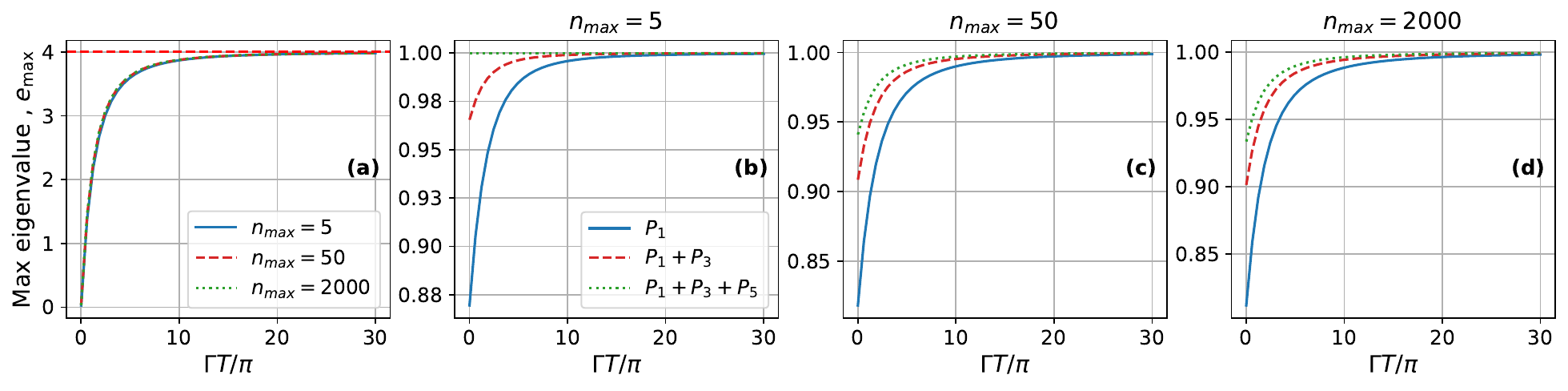}
\caption{\textbf{(a)} Maximum eigenvalue as a function of the width of the pulse $\Gamma T = r \pi$ for $n_{\text{max}}=5$, $n_{\text{max}}=50$ and $n_{\text{max}}=2000$ where $n_{\text{max}}$ is the maximum number of basis functions considered. As seen, the maximum eigenvalue is $4$ for all $r$. \textbf{(b-d)} The population of the eigenvector corresponding to the maximum eigenvalue in the first three odd harmonics is shown as a function of $r$. Note that $P_{n}$ is the population in the $n^{th}$ harmonic. For larger values of $r$, the first harmonic is the eigenvector corresponding to the maximum eigenvalue $4$.}
\label{fig:qfi1}
\end{figure*}
Note that $\mathcal{F}_{1}$ is identical to $\mathcal{F}_{p}$, which is one of the terms that appears in the QFI expression without any approximations as shown in Eq.~\eqref{eq:QFI_expression}. This term can be rewritten in terms of harmonic functions as
\begin{align}
    \mathcal{F}_{1}(t \rightarrow \infty)=\mathcal{F}_{1}^{\infty}  &= \sum_{m,n} c_{m}  \left[\frac{2}{\Gamma} \int_{0}^{T}d\tau \int_{0}^{\tau}d\tau' e^{-\frac{\Gamma(\tau-\tau')}{2}}\xi_{m}(\tau)\xi_{n}(\tau')\right]c_{n}.
\end{align}
Setting $\Gamma T = r \pi$, we have
\begin{align}
    \mathcal{F}_{1}^{\infty} &= \frac{2}{\Gamma^{3/2}} \int_{0}^{T} d\tau\, f(\tau) p(\tau) = \sum_{m, n=1}^{\infty} c_{m} \left[K_{1}(r) \right]_{m,n} c_{n}, \\
    \left[K_1 (r) \right]_{m,n} &= \begin{cases} 
      \frac{16 m n r }{\Gamma^{2} \pi (4 m^2 + r^2)} \left( \frac{4}{4 n^2 + r^2}\left(1 - (-1)^{n} e^{- \frac{\pi r}{2}} \right) + \frac{1-(-1)^{m + n}}{m^2 - n^2} \right) & m\neq n \\
     \frac{4r}{\Gamma^{2} (4n^2 + r^2)^2} \left(r^3 + \frac{4n^2}{\pi} \left(4 - 4 (-1)^n e^{- \frac{\pi r}{2}} + r \pi \right)\right) & m = n
   \end{cases}.
\end{align}
It is now sufficient to diagonalize the $K_{1}(r)$ matrix in order to identify the pulse function that maximizes $\mathcal{F}_{2} = \mathcal{F}_{p}$. It should be noted that even though the matrix $K_{1}(r)$ is not symmetric, the antisymmetric part of the matrix does not contribute, so we can analyze just the symmetric part. That is, $K_{1}(r) =  \frac{1}{2}(K_{1}(r)+K_{1}^{T}(r)) + \frac{1}{2}(K_{1}(r)-K_{1}^{T}(r)) \equiv K_{1}^{S}(r) + K_{1}^{A}(r) $,
\begin{align}
    \sum_{m,n}c_{m} \left[K_{1}^{A}(r)\right]_{m,n} c_{n} &= -\sum_{m,n}c_{m}\left[ K_{1}^{A}(r)\right]_{m,n} c_{n}. 
\end{align}
Therefore, we have
\begin{align}
     \mathcal{F}_{1}^{\infty} &= \sum_{m,n} c_{m} \left[K_{1}^{S}(r)\right]_{m, n} c_{n} \\
     \label{eq:F2_evaluation_real_pulses}
     &= \begin{cases} 
    \frac{32 m n r}{\Gamma^{2} \pi (4 m^2 + r^2)(4 n^2 + r^2) } \left[\left(1 + (-1)^{m + n} \right) -e^{- \frac{r \pi}{2}}((-1)^{m} + (-1)^{n}) \right] & m\neq n \\
    \frac{4r}{(4n^2 + r^2)^2} \left[r^3 + \frac{4n^2}{\pi} \left(4 - 4 (-1)^n e^{- \frac{r \pi}{2}} + r \pi \right)\right] & m = n
   \end{cases}.
\end{align}
The matrix $K^{S}(r)$ commutes with the parity operator defined by $P_{n,m} = \delta_{n,m} (-1)^{n}$.
We analyze the above expression assuming that we have a finite number of basis functions and truncate the sum to $n_{\text{max}}$. The maximum eigenvalue of this matrix obtained numerically as a function of $r$ is shown in Fig.~\eqref{fig:qfi1} where it can be seen that it approaches $4$ as $r$ is made larger. Also, the eigenvector corresponding to the maximum eigenvalue has a major contribution from the first harmonic, with small contributions from the other odd harmonics. Likewise, the eigenvector corresponding to the second largest eigenvalue has most of its contribution from the second harmonic with small contributions from the other even harmonics. This separation between the odd and the even harmonics results from the presence of parity symmetry mentioned above. In the long-time limit, the eigenvalues of the above matrix can be obtained analytically. Since $r \gg 1$, we ignore the terms that are exponentially suppressed in $r$ in both the diagonal and the non-diagonal part of the matrix and introducing $\tilde{n}_{1} \equiv n_{1}/r$ and $\tilde{n}_{2} \equiv n_{2}/r$  as done in the previous subsection, we have
\begin{align}
     \left[K^{S}_{1}(r)\right]_{m, n} &=
     \begin{cases}
         \frac{32 \tilde{n}_1 \tilde{n}_2}{r(4 \tilde{n}_1^2 + 1)(4 \tilde{n}_2^2 + 1) \pi} \left(1 + (-1)^{m + n} \right) & m \neq n \\
        \frac{4}{(4\tilde{n}^2 + 1)}  +  \frac{64}{r\pi} \left[\frac{\tilde{n}^2}{(4\tilde{n}^2 + 1)^2} \right] & m=n 
     \end{cases}.
\end{align}
In the above, all the non-diagonal elements and some diagonal elements in the above matrix can be neglected
because $0 \leq x/(4x^2+1) \leq 0.25$ for all $x \geq 0$, so these terms are suppressed by $r$ leading to a diagonal matrix,
\begin{align}
    \mathcal{F}_{1}^{\infty}(r \gg 1) &=\sum_{m,n}c_{m}\left( \frac{4\delta_{m,n}}{4 \tilde{n}^{2}+1} \right) c_{n}.
\end{align}
The above represents a monotonically decreasing function with $\tilde{n}$ that has values between $4$ and $0$. The maximum eigenvalue is obtained when $\tilde{n}=1/r$, and the corresponding eigenvector is the first harmonic. The maximum eigenvalue approaches $4$ as $r$, which is proportional to the width of the pulse, becomes larger.
\subsection{Evaluation of \texorpdfstring{$\mathcal{F}_{2}$}{F\_2}}
Note that $\mathcal{F}_{z}$ in the expression for the QFI becomes $\mathcal{F}_{2}$ in the long pulse width limit. Setting the width of the pulse as $\Gamma T = r\pi$, $\mathcal{F}_{2}$ in the long-time limit is given by
\begin{align}
\label{eq:F1_expression_real_pulse}
\mathcal{F}_{2}(t \rightarrow \infty)=\mathcal{F}_{2}^{\infty} &\equiv \frac{1}{\Gamma}\sum_{m,n} c_{m} 
    \Bigl(\int_{0}^{\infty} d\tau\;\phi_{m}(\tau)\phi_{n}(\tau)\Bigr) c_{n} \\
    &\equiv \sum_{m,n} c_{m}\left[K_{2}(r) \right]_{m,n}c_{n}.
\end{align}
where 
\begin{align}
\left[K_{2}(r)\right]_{m,n}&= 
\begin{cases}
    \frac{32 m n r }{\Gamma^{2} \pi (4m^{2} + r^{2}) (4n^{2} + r^{2})} \left[ \left(1-e^{-\pi r} \right) + \left( (-1)^{m} -e^{-\frac{\pi r}{2}} \right)
\left((-1)^{n} -e^{-\frac{\pi r}{2}} \right) \right]& m \neq n \\
\frac{4 r\Bigl(\pi r^{3} + 4 n^{2}\bigl(-2e^{-\pi r}  + (2 + \pi r)\bigr)\Bigr)}{\Gamma^{2}\pi (4n^{2} + r^{2})^{2}} + 
\frac{32 \bigl((-1)^{n}-e^{-\frac{\pi r}{2}}\bigr)^{2}n^{2}r}
{\Gamma^{2} \pi (4n^{2} + r^{2})^{2}} & m = n
\end{cases},
\end{align}
Ignoring terms that are exponentially decreasing in $r$, and introducing variables $\tilde{m} \equiv \frac{m}{r}$ and $\tilde{n} \equiv \frac{n}{r}$, we have 
\begin{align}
\left[K_{2}(r)\right]_{m,n} &= 
\begin{cases}
\frac{32}{\Gamma^{2} r\pi}\left(\frac{\tilde{m}}{4\tilde{m}^2 + 1} \right) \left(\frac{\tilde{n}}{4\tilde{n}^2+1}\right)\left[1+(-1)^{m+n}\right], 
& m \neq n \\
\frac{4}{\Gamma^{2}}\left[\frac{1}{4\tilde{n}^2 + 1} +\frac{1}{\pi r} \frac{8\tilde{n}^{2}}{(4\tilde{n}^{2}+1)^{2}}\right]+ \frac{32}{\Gamma^{2} r\pi}\left(\frac{\tilde{n}}{4\tilde{n}^2 + 1} \right)^{2}, 
& m = n
\end{cases},
\end{align}
where $\tilde{m}$ and $\tilde{n}$ can be $\{1/r,2/r,...,n/r,...\}$. Note that $0 \leq x/(4x^2+1) \leq 0.25$ for any positive $x$, so all the non-diagonal elements and some diagonal elements are suppressed by $r$ and can be ignored for large $r$. Hence, 
\begin{align}
    \Gamma^{2} \mathcal{F}_{2}^{\infty} &=\sum_{m,n}c_{m}\left( \frac{4\delta_{m,n}}{4 \tilde{n}^{2}+1} \right) c_{n}.
\end{align}
\subsection{Evaluation of \texorpdfstring{$\mathcal{F}_{3}$}{F\_3}}
In the long-time limit with $\Gamma T=r\pi$, the final term $\mathcal{F}_{3}$ is given by  
\begin{align}
     \mathcal{F}_{3}(t \rightarrow \infty) \equiv \mathcal{F}_{3}^{\infty}  &=4 \int_{0}^{\infty}  d\tau \; f(\tau) q(\tau)\\
     &= \sum_{m,n}c_{m} \left(4 \int_{0}^{T} d\tau \xi_{m}(\tau) \eta_{n}(\tau)\right) c_{n}\\
     &\equiv
\sum_{m,n} c_{m}\left[K_{3}(r) \right]_{m,n} c_{n},
\end{align}
where 
\begin{align}
\left[K_{3}(r) \right]_{m,n} &=
\begin{cases}
\frac{64  mnr^{2} \left[
r\left((4n^{2}+r^{2})^{2} -(-1)^{m+n}(4m^{2}+r^{2})^{2} \right) + (-1)^{m}e^{-\frac{\pi r}{2}}(m^2-n^2)\bigl(4n^{2}r(4+\pi r) + r^{3}(8+\pi r) + 4m^{2}(4n^{2}\pi + r(4+\pi r))\bigr) \right]}
{\Gamma^{2} \pi(m^2-n^2)(4m^{2}+r^{2})^{2}(4n^{2}+r^{2})^{2}}  & m \neq n \\
\frac{8 r^2} {\Gamma^2 \pi(4n^2 + r^2)^3} \left[ 8 e^{-\frac{\pi r}{2}} (-1)^{n}n^2\bigl(4n^2\pi + r(8 + \pi r)\bigr) +\bigl(16n^4\pi - 64n^2r - \pi r^4\bigr) \right] & m=n
\end{cases}.
\end{align}
Focusing on the non-diagonal terms, taking the symmetric part of the matrix because the non-symmetric part does not contribute, 
\begin{align}
\left[K_{3}^{S}(r) \right]_{m \neq n} &=\frac{32 \tilde{m} \tilde{n}}
{\Gamma^{2} \pi (4\tilde{m}^{2}+1)^{2} (4\tilde{n}^{2}+1)^{2} }\biggl[\left(4(\tilde{n}^{2}+\tilde{m}^{2})\pi+\pi+16\tilde{m}^{2}\tilde{n}^{2}\pi + \frac{8}{r}+\frac{16(\tilde{n}^{2}+\tilde{m}^{2}) }{r}  \right)\left((-1)^{m}+ (-1)^{n}\right) e^{-\frac{\pi r}{2}} \biggr] \nonumber \\
& -\frac{256}{\Gamma^{2} r \pi} \left(\frac{ \tilde{m} \tilde{n} \bigl(2\tilde{m}^{2}+2\tilde{n}^{2}+1\bigr)}
{ (4\tilde{m}^{2}+1)^{2} (4\tilde{n}^{2}+1)^{2} }\right)\left(1+(-1)^{m+n}\right).
\end{align}
Ignoring terms that are exponentially suppressed, we have
\begin{align}
\left[K_{3}^{S}(r) \right]_{m \neq n} &= -\frac{256}{\Gamma^{2} r \pi} \left(\frac{ \tilde{m} \tilde{n} \bigl(2\tilde{m}^{2}+2\tilde{n}^{2}+1\bigr)}
{ (4\tilde{m}^{2}+1)^{2} (4\tilde{n}^{2}+1)^{2} }\right)\left(1+(-1)^{m+n}\right).
\end{align}
Since
\begin{align}
    0 \leq \left(\frac{ \tilde{m} \tilde{n} \bigl(2\tilde{m}^{2}+2\tilde{n}^{2}+1\bigr)}
{ (4\tilde{m}^{2}+1)^{2} (4\tilde{n}^{2}+1)^{2} }\right) \leq \frac{1}{27},
\end{align}
the non-diagonal term is suppressed by $r$, and therefore the matrix $K_{3}^{S}$ is diagonal. The diagonal part after ignoring the exponentially suppressed terms is given by 
\begin{align}
    \left[K_{3}^{S}(r) \right]_{m \neq n} &= \frac{8  \left(16\tilde{n}^4 - 1 \right)} {\Gamma^2 (4\tilde{n}^2 + 1)^3} -  \frac{512} {\Gamma^2 r \pi}  \frac{\tilde{n}^2} {(4\tilde{n}^2 + 1)^3} .
\end{align}
Note that $0 \leq x^2/(4x^2+1)^3 \leq  1/27$, so ignoring the second term because it is suppressed by $r$. In the end, we have
\begin{align}
    \left[K_{3}^{S}(r) \right]_{m \neq n} &= \frac{8  \left(4\tilde{n}^2 - 1 \right)} {\Gamma^2 (4\tilde{n}^2 + 1)^2} .
\end{align}
Hence
\begin{align}
    \Gamma^{2}\mathcal{F}_{3}^{\infty} &=\sum_{m,n}c_{m}\left(  \frac{8  \left(4\tilde{n}^2 - 1 \right) \delta_{m,n}} {(4\tilde{n}^2 + 1)^2} \right) c_{n}.
\end{align}
\subsection{Maximum eigenvalue}
In summary, we have 
\begin{align}
    \frac{\Gamma^{2} \mathcal{F}_{\text{long}}^{\infty}}{\alpha^{2}} &= \frac{\Gamma^{2} \mathcal{F}_{1}^{\infty}}{\alpha^{2}}+\frac{\Gamma^{2} \mathcal{F}_{2}^{\infty}}{\alpha^{2}}+\frac{\Gamma^{2} \mathcal{F}_{3}^{\infty}}{\alpha^{2}}\\
    &=\frac{1}{\alpha^{2}} \sum_{m,n}c_{m}\left(  \frac{8  \left(4\tilde{n}^2 - 1 \right) \delta_{m,n}} {(4\tilde{n}^2 + 1)^2} +\frac{8\delta_{m,n}}{4 \tilde{n}^{2}+1} \right) c_{n} = \sum_{m,n}\tilde{c}_{m}\left(  \frac{64 \tilde{n}^{2} \delta_{m,n}} {(4\tilde{n}^2 + 1)^2} \right) \tilde{c}_{n}.
\end{align}
 Note that the coefficients $\{\tilde{c}_{m}\}$ are normalized, and the eigenvalues are then given by $ \frac{64 \tilde{n}^{2}} {(4\tilde{n}^2 + 1)^2}$ which range between $0$ and $4$.
The maximum eigenvalue is obtained at $\tilde{n} = 1/2$ or equivalently at 
\begin{align}
    \omega= \frac{\Gamma}{2}.
\end{align}
In other words, the maximum QFI per unit photon in this limit is given by $4$, and $\sqrt{2\alpha/T} \sin(\Gamma t/2)$ is the pulse that optimizes it.
\section{Complex Pulses}
\subsection{Perturbation theory for complex pulses}
The equations of motion with complex pulses and nonzero detuning given in Eqs.~\eqref{eq:complex_ODE_x1_noncompact}-\eqref{eq:complex_ODE_F_noncompact} can be written in a more compact form,
\begin{align}
    r_1'(t) &= -k \, r_1(t)
            + 2\sqrt{\Gamma}\,f^*(t)\,z_{1}(t),\\
    z_1'(t) &= -\Gamma\,(1+z_1(t))-2\sqrt{\Gamma}\Re\Bigl(f(t) r_1(t)\Bigr),\\
    w_1'(t) &= \frac{1}{2\sqrt{\Gamma}}\;\Im\bigl(f(t)\,r_1(t)\bigr),\\
    r_2'(t) &= -k\,r_2(t)
            + 2\sqrt{\Gamma}\,f^*(t)\,z_2(t)
            + \frac{i}{4}\,r_1(t)
            + \frac{i}{2\sqrt{\Gamma}}\,f^*(t),\\
    z_2'(t) &= -\Gamma\left(z_2(t) + w_{1}(t)\right)-2\sqrt{\Gamma}\Re\Bigl(f(t)\,r_2(t)\Bigr) ,\\
    w_2'(t) &= \frac{1}{2\sqrt{\Gamma}}\;\Im\bigl(f(t)\,r_2(t)\bigr), \\
    \mathcal{F}'(t) &= \frac{1}{2\Gamma} \left( 1 + z_1(t) \right)  +4 \frac{d}{dt}\left(2 w_{2}(t)-w_{1}(t)^{2} \right).
\end{align}
where $r_{m}(t) = x_{m}(t) + i y_{m}(t)$ where $m=\{1,2\}$ and $k = \frac{\Gamma}{2} - i\delta$. We introduce the following two variables, 
\begin{align}
    P(t) &\equiv e^{-i \delta t}\sqrt{\Gamma} \int_{0}^{t} d\tau \; e^{k (\tau-t)}  f^{*}(\tau) =\sqrt{\Gamma} \int_{0}^{t} d\tau \; e^{\frac{\Gamma}{2} (\tau-t)} e^{-i\delta \tau} f^{*}(\tau) ,\\
    Q(t) &\equiv \frac{\partial}{\partial \Gamma}\left(\frac{P(t)}{\sqrt{\Gamma}}\right) = e^{-i \delta t}\int_{0}^{t} d\tau \;\left(\frac{\tau-t}{2}\right) e^{k (\tau-t)} f^{*}(\tau)= -\frac{1}{2\sqrt{\Gamma}}\int_{0}^{t} d\tau \; e^{\frac{\Gamma}{2}(\tau-t)}P(\tau).
\end{align}
Note that the above variables are just the generalization of $p(t)$ and $q(t)$ defined in Eq.~\eqref{eq:pt_definition} and Eq.~\eqref{eq:qt_definition} respectively in the case of real pulses with zero detuning. As before, we use the regular perturbation theory ansatz, $u(t) = u^{(0)}(t)+\varepsilon u^{(1)}(t)+\varepsilon^{2} u^{(2)}(t)+... $ where $u \in \{r_{1},z_{1},w_{1},r_{2}, z_{2},w_{2},\mathcal{F}\}$ in the ODEs shown above where $f(t) = \varepsilon \phi(t/T)$ and $\varepsilon \equiv \alpha/\sqrt{T} \ll 1$. The zeroth-order solution is given by $r_{1}^{(0)}(t) =0 $, $z_{1}^{(0)}(t) =-1 $, $w_{1}^{(0)}(t) =0 $, $r_{2}^{(0)}(t)=0$, $w_{2}^{(0)}(t) =0 $ and $z_{2}^{(0)}(t)=0$, which correspond to the initial conditions. 
\subsubsection{First-order solution}
The first order ODEs are given by
\begin{align}
    \dot{r}^{(1)}_1(t) &= -k \, r_1^{(1)}(t)
            - 2\sqrt{\Gamma}\,f^*(t),\\
    \dot{z}^{(1)}_1(t) &= -\Gamma\,z^{(1)}_{1}(t),\\
    \dot{w}_{1}^{(1)}(t) &= 0,\\
    \dot{r}_{2}^{(1)}(t) &= -k\,r_2^{(1)}(t)
            + \frac{i}{4}\,r_{1}^{(1)}(t)
            + \frac{i}{2\sqrt{\Gamma}}\,f^*(t),\\
    \dot{z}^{(1)}_{2}(t) &= -\Gamma\left(z_{2}^{(1)}(t) + w_{1}^{(1)}(t)\right) ,\\
    \dot{w}^{(1)}_2(t) &= 0,\\
    \mathcal{F}^{(1)}(t) &= \frac{1}{2\Gamma}z_{1}^{(1)}(t),
\end{align}
with $ r_1^{(1)}(0)=z^{(1)}_{1}(0)=w^{(1)}_{1}(0)= r_2^{(1)}(0)=z^{(1)}_{2}(0)=w^{(1)}_{2}(0)=\mathcal{F}^{(1)}(0)=0$, and the solution to the above ODEs is given by
\begin{align}
    r_{1}^{(1)}(t) &= -2 e^{i\delta t}P(t), \\
    z_{1}^{(1)}(t) &= 0 ,\\
    w_{1}^{(1)}(t) &=0 ,\\
    r_{2}^{(1)}(t) &= \frac{i e^{i\delta t}}{2 \Gamma} P(t) + i e^{i\delta t}\sqrt{\Gamma} Q(t),\\
    z_{2}^{(1)}(t) &= 0, \\
    \mathcal{F}^{(1)}(t) &=0 .
\end{align}
\subsubsection{Second-order ODEs}
The second-order ODEs are given by 
\begin{align}
    \dot{r}^{(2)}_1(t) &= -k \, r_1^{(2)}(t),\\
    \dot{z}_1^{(2)}(t) &= -\Gamma\,z_1^{(2)}(t)-2\sqrt{\Gamma}\Re\Bigl[f(t) r_1^{(1)}(t)\Bigr],\\
    \dot{w}_1^{(2)}(t) &= \frac{1}{2\sqrt{\Gamma}}\;\Im\bigl[f(t)\,r_1^{(1)}(t)\bigr],\\
    \dot{r}_2^{(2)}(t) &= -k\,r_2^{(2)}(t)
            + \frac{i}{4}\,r_1^{(2)}(t), \\
    \dot{z}_2^{(2)}(t) &= -\Gamma\left(z_2^{(2)}(t) + w_{1}^{(2)}(t)\right)-2\sqrt{\Gamma}\Re\left[f(t)\,r^{(1)}_2(t)\right] ,\\
    \dot{w}_2^{(2)}(t) &= \frac{1}{2\sqrt{\Gamma}}\;\Im\left[f(t)\,r^{(1)}_2(t)\right],\\
    \dot{\mathcal{F}}^{(2)}(t) &= \frac{z_{1}^{(2)}(t)}{2\Gamma} + 8 \dot{w}_{2}^{(2)}(t),
\end{align}
with $ r_1^{(2)}(0)=z^{(2)}_{1}(0)=w^{(2)}_{1}(0)= r_2^{(2)}(0)=z^{(2)}_{2}(0)=w^{(2)}_{2}(0)=\mathcal{F}^{(2)}(0)=0$, and the solution to the above ODEs is given by
\begin{align}
    r_{1}^{(2)}(t) &=0, \\
    z_{1}^{(2)}(t) &=4\sqrt{\Gamma} \Re\left[\int_{0}^{t}d\tau\; e^{\Gamma(\tau-t)} e^{i\delta \tau}f(\tau)P(\tau)\right] =  2 |P(t)|^{2}, \\
    w_1^{(2)}(t) &= - \frac{1}{\sqrt{\Gamma}}\int_{0}^{t} d\tau \; \Im\bigl(e^{i\delta \tau}f(\tau)P(\tau)\bigr),\\
    r_{2}^{(2)}(t) &= 0,\\
    z_{2}^{(2)}(t) &= - \Gamma\int_{0}^{t} d\tau \, e^{\Gamma(\tau- t)}w_{1}^{(2)}(\tau) - 2\sqrt{\Gamma}\int_{0}^{t} d\tau \; e^{\Gamma (\tau -t)}\Re\left[f(\tau)r_{2}^{(1)}(\tau)\right], \\ 
    w_2^{(2)}(t) &= \frac{1}{4 \Gamma^{3/2}} \int_{0}^{t} d\tau \; \Re\left[e^{i\delta \tau}f(\tau) P(\tau)\right]+\frac{1}{2} \int_{0}^{t} d\tau\Re\left[e^{i\delta \tau}f(\tau)Q(\tau))\right], \\
    \mathcal{F}^{(2)}(t) &= \frac{1}{\Gamma}\int_{0}^{t} d\tau\; |P(\tau)|^{2} +\frac{2}{\Gamma^{3/2}}\int_{0}^{t} d\tau\;  \Re\left[e^{i\delta \tau}f(\tau) P(\tau)\right] +4 \int_{0}^{t} d\tau\; \Re\left[e^{i\delta \tau}f(\tau)Q(\tau))\right] ,
\end{align}
where the final expression for $z_{1}^{(2)}$ is obtained by substituting the definition of $P(t)$, writing the real part as the sum of the argument and its conjugate, and then swapping the integrals in the double integral on one of the terms. Note that $1+z_{1} = 2|P(t)|^{2}$ up to second order in $\varepsilon$ implies that $P(t)$ is the proportional to the probability amplitude of the atom to be in the excited state (up to a complex conjugate) in this long pulse width limit. In the case of real pulses with zero detuning, the above expression for the QFI reduces to Eq.~\eqref{eq:QFI_long} as expected.
\subsubsection{Second order solution in terms of coupled ODEs}
We now express the QFI expression up to second-order in $\varepsilon$ as the exact solution to the following coupled ODEs in order to relate it to the QFI of single-photon pulses
\begin{align}
    \dot{P}(t) &= - \frac{\Gamma}{2} P(t) + \sqrt{\Gamma} e^{-i \delta t} f^{*}(t),  \\
    \dot{Q}(t) &= - \frac{\Gamma}{2} Q(t)  -\frac{1}{2\sqrt{\Gamma}} P(t), \\
    \label{eq:QFI_long_complex}\dot{\mathcal{F}}_{\text{long}}(t) &= \frac{1}{\Gamma} |P(t)|^{2} + \frac{2}{\Gamma^{3/2}} \Re \left[e^{i\delta t}f(t) P(t)\right] + 4 \Re \left[e^{i\delta t}f(t)Q(t)\right],
\end{align}
where we added a subscript ``long" on the QFI to emphasize that it is the QFI expression in the long pulse width limit. We now express different terms in the ODEs for $\mathcal{F}_{\text{long}}$ completely in terms of $Q(t)$,
\begin{align}
    \dot{\mathcal{F}}_{\text{long}}(t) &= \underbrace{8|\dot{Q}|^{2} + 2\Gamma^{2}|Q(t)|^{2} +  8\Gamma \Re\left[\dot{Q}(t)Q^{*}(t)\right] +\frac{1}{\Gamma^{2}}\partial_{t}\left(P(t)P^{*}(t) \right)}_{\frac{1}{\Gamma} |P(t)|^{2} + \frac{2}{\Gamma^{3/2}} \Re \left[e^{i\delta t}f(t) P(t)\right]} \underbrace{-8 \Re\left[\ddot{Q}^{*}(t)Q(t)\right]-8 \Gamma \Re\left[\dot{Q}(t) Q^{*}(t)\right] - 2 \Gamma^{2}|Q(t)|^{2}}_{ 4 \Re \left[e^{i\delta t}f(t)Q(t)\right]}.
\end{align}
Note that a couple of terms in the above can be written as full derivatives of time of $Q(t)$, 
\begin{align}
    \Re\left[\dot{Q}^{*}(t)Q(t)\right] &= \partial_{t}\left(Q(t) Q^{*}(t) \right), \\
    \Re\left[\ddot{Q}^{*}(t)\dot{Q}(t)\right] &= \partial_{t}\left(\dot{Q}(t) \dot{Q}^{*}(t) \right),
\end{align}
so these terms vanish along with $\partial_{t}\left(P(t)P^{*}(t) \right)$ under the integral if we are interested in the long-time QFI since $P(t)$, $Q(t)$ and their derivatives vanish at the boundaries. Hence, the expression for the QFI in the long-time limit simplifies to
\begin{align}
   \mathcal{F}_{\text{long}}(t\rightarrow \infty) \equiv  \mathcal{F}_{\text{long}}^{\infty}&=  \int_{0}^{\infty}d\tau \left(8|\dot{Q}(\tau)|^{2}   -8 \Re\left[\ddot{Q}^{*}(\tau)Q(\tau)\right] \right).
\end{align}
Using the fact that $ \int_{0}^{\infty} dt\; \Re\left[\ddot{Q}^{*}(t)Q(t)\right] = -\int_{0}^{\infty} dt\; |\dot{Q}(t)|^{2}$ under integration by parts, we have
\begin{align}
\label{eq:QFI_complex_long_width_limit}
   \mathcal{F}_{\text{long}}^{\infty} &= 16  \int_{0}^{\infty}d\tau |\dot{Q}(\tau)|^{2}, 
\end{align}
which is the analog of the result in Eq.~\eqref{eq:QFI_q_form} for complex pulses with nonzero detuning.
\subsection{QFI of complex single-photon pulse with nonzero detuning}
\label{appendix:single_photon_complex_pulses}
The QFI for a single-photon pulse with complex pulse shape $f(t)$ in the long-time limit is provided in \cite{darsheshdar2024role},
\begin{align}\label{eq:qfigamma1}
    \mathcal{F}_{\text{single}}^{\infty}&= 
    \frac{1}{\Gamma}\int_{-\infty }^{\infty }{d}\omega {{\left| \tilde{f} (\omega ) \right|}^{2}}  \left | g(\omega) + 2\Gamma \partial_\Gamma g(\omega) \right |^2  -\left|\frac{1}{\sqrt{\Gamma}}\int_{-\infty }^{\infty }{d}\omega {{\left| \tilde{f} (\omega ) \right|}^{2}}\left( 1-\sqrt{ \Gamma}g^*(\omega) \right) \left( g(\omega) + 2\Gamma \partial_\Gamma g(\omega) \right)\right|^{2},
\end{align}
where
\begin{align}
    \tilde{f}(\omega) &= \frac{1}{\sqrt{2 \pi}} \int_{-\infty}^{\infty} d\tau f(\tau) e^{i\omega \tau}, \\
    g(\omega) &= \frac{\sqrt{{\Gamma} }}{ \frac{\Gamma}{2}+i(\delta-\omega)},
\end{align}
assuming that the pulse is zero at the boundaries $f(-\infty)=f(\infty)=0$ (as opposed to $f(0)=f(\infty)=0$ in our work). In this subsection, the integrals are extended from $-\infty$ to $\infty$ under the implicit assumption that $f(t)=0$ for $t \leq 0$, thereby facilitating the use of Fourier transforms. We now rewrite the above expression in the time-domain and put it in a manner that allows us to compare it with the perturbation theory expression obtained for coherent states in the long pulse width limit. Substituting $g(\omega)$, we have 
\begin{align}
    \mathcal{F}_{\text{single}}^{\infty}&= 64  \int_{-\infty}^{\infty} d\omega\; (\delta - \omega)^{2} \frac{|\tilde{f}(\omega)|^{2}}{\left(\Gamma^{2}+4(\delta - \omega)^{2}\right)^{2}} - 64 \left|\int_{-\infty}^{\infty}d\omega\; (\delta - \omega) \frac{|\tilde{f}(\omega)|^{2}}{\Gamma^{2}+4(\delta - \omega)^{2}}\right|^{2}.
\end{align}
We now use the variables $P(t)$ and $Q(t)$ defined in the previous subsection. Note that $P(t)$ is proportional to the probability amplitude of the atom to be excited by the single-photon state (up to complex conjugate), $P(t) = - e^{-i \delta t}\psi_{e}^{*}(t)$. 
\begin{align}
    P(t) &\equiv \sqrt{\Gamma} \int_{-\infty}^{t} d\tau \; e^{\frac{\Gamma}{2} (\tau-t)} e^{-i\delta \tau} f^{*}(\tau) =  \sqrt{\Gamma} \int_{-\infty}^{\infty} d\tau \; \Theta(t-\tau) e^{-\frac{\Gamma}{2} (t-\tau)} e^{-i\delta \tau} f^{*}(\tau) =(a*b)(t),
\end{align}
where $ a(t) = \sqrt{\Gamma} \Theta(t) e^{-\frac{\Gamma}{2} t}$ and $ b(t) = e^{-i\delta t} f^{*}(t)$ with their Fourier transforms given by $   \mathscr{F}\left[a(t)\right] =\frac{1}{\sqrt{2 \pi}} \frac{2\sqrt{\Gamma}}{\Gamma -2i \omega}$ and $\mathscr{F}\left[b(t)\right] =\tilde{f}^{*}(\delta-\omega)$. With our definition of the Fourier transform, $\mathscr{F}\left[(a*b)(t)\right] =\sqrt{2 \pi}\mathscr{F}\left[a(t)\right]\mathscr{F}\left[b(t)\right] $. Hence the Fourier transforms of $P(t)$ and $Q(t)$ are given by
\begin{align}
    \mathscr{F}[P(t)](\omega) &\equiv  \tilde{P}(\omega) = \frac{2\sqrt{\Gamma} \tilde{f}^{*}(\delta-\omega)}{\Gamma -2i \omega}, \\
    \mathscr{F}[Q(t)](\omega) &\equiv  \tilde{Q}(\omega) = -\frac{2 \tilde{f}^{*}(\delta-\omega)}{(\Gamma-2i\omega)^{2}}.
\end{align}
We know that $\mathscr{F}[\partial_{t}Q(t)] = - i\omega \mathscr{F}[Q(t)]=- i\omega \tilde{Q}(\omega)$,   $|\mathscr{F}[\partial_{t}Q(t)]|^{2} = \omega^{2} |\tilde{Q}(\omega)|^{2}$ and $\mathscr{F}[\partial_{t}P(t)]\left(\mathscr{F}[P(t)]\right)^{*} = -i\omega |\tilde{P}(\omega)|^{2}$. Using these relationships, we can re-express the QFI as follows
\begin{align}
    \mathcal{F}_{\text{single}}^{\infty} &= 16 \int_{-\infty}^{\infty} d\omega\; (\delta-\omega)^{2} |\tilde{Q}(\delta-\omega)|^{2} - \frac{4}{\Gamma^2}\left|\int_{-\infty}^{\infty} d\omega \;(\delta-\omega) |\tilde{P}(\delta- \omega)|^{2}\right|^{2} \\
    &= 16 \int_{-\infty}^{\infty} d\omega \; \omega^{2} |\tilde{Q}(\omega)|^{2} - \frac{4}{\Gamma^{2}}\left|\int_{-\infty}^{\infty}d\omega \; \omega|\tilde{P}(\omega)|^{2}  \right|^{2} \\
    &= 16 \int_{-\infty}^{\infty} d\omega \; |\mathscr{F}[\partial_{t}Q(t)]|^{2}  -\frac{4}{\Gamma^{2}} \left|\int_{-\infty}^{\infty} d\omega \; \mathscr{F}[\partial_{t}P(t)](\mathscr{F}[P(t)])^{*} \right|^{2}.
\end{align}
Using the Parseval's theorem, which is given by
\begin{align}
    \int_{-\infty}^{\infty} dt \; k(t) m^{*}(t) &= \int_{-\infty}^{\infty} d\omega \; \tilde{k}(\omega) \tilde{m}(\omega)^{*},
\end{align}
we have
\begin{align}
    \mathcal{F}_{\text{single}}^{\infty} = 16 \int_{-\infty}^{\infty} dt \; |\partial_{t}Q(t)|^{2}  - \frac{4}{\Gamma^2} \left|\int_{-\infty}^{\infty} dt \; \left(\partial_{t}P(t)\right) P(t)^{*} \right|^{2}.
\end{align}
Since $P(t=0)=Q(t=0)=0$, the above expression reduces to
\begin{align}
\label{eq:single_photon_QFI_complex}
    \mathcal{F}_{\text{single}}^{\infty} = 16 \int_{0}^{\infty} dt \; |\partial_{t}Q(t)|^{2}  - \frac{4}{\Gamma^2} \left|\int_{0}^{\infty} dt \; \left(\partial_{t}P(t)\right) P(t)^{*} \right|^{2}.
\end{align}
Note that when the pulse is real with zero detuning, $\left(\partial_{t}P(t)\right) P(t)^{*} = \frac{1}{2}\partial_{t}\left(P(t)^{2}\right)$, and $P(t)$ is zero at the boundaries. Therefore, the above expression reduces to the integral of just $16 (\partial_{t}Q(t))^{2}$, which is the expression we obtained for the real pulses. The above can be expressed as the long-time limit solution to the following coupled ODEs
\begin{align}
    \frac{\partial P(t)}{\partial t} &= -\frac{\Gamma P(t)}{2}  + \sqrt{\Gamma} e^{-i\delta t}f^{*}(t), \\
    \frac{\partial Q(t)}{\partial t} &= - \frac{\Gamma  Q(t)}{2} -\frac{P(t)}{2\sqrt{\Gamma}},   \\
    \frac{\partial R(t)}{\partial t}   &=  P(t)^{*} \partial_{t}P(t),\\
    \frac{\partial \mathcal{F}_{\text{single}}(t)}{\partial t}   &=  16 |\partial_{t}Q(t)|^{2} - \frac{4}{\Gamma^{2}}\partial_{t}|R(t)|^{2}.
\end{align} 
Note that the QFI expression for a complex single-photon pulse has an additional term that is fourth order in the pulse shape $f(t)$. Even though the second order term from the perturbation theory (see Eq.~\eqref{eq:QFI_complex_long_width_limit}) matches with the term $16 |\partial_{t}Q(t)|^{2}$ for a single photon state, the fourth-order term from the perturbation theory for the coherent states obtained through numerical simulations does not match with the second term in Eq.~\eqref{eq:single_photon_QFI_complex}. In summary, the QFI per unit photon of a real coherent state is identical to the single-photon QFI for a small amplitude coherent state, while the QFI per unit photon of complex coherent state is different from a single-photon complex pulse's QFI in the same limit. 
\subsection{Diagonalizing QFI up to second order with the closed boundary conditions}
\label{appendix:diagonalize_open_boundary_conditions}
The QFI in the long pulse width limit (Eq.~\eqref{eq:QFI_long_complex}) is given by
\begin{align}
    \mathcal{F}_{\text{long}}(t) &= \frac{2}{\Gamma^{3/2}}\int_{0}^{t} d\tau\;  \Re\left[e^{i\delta \tau}f(\tau) P(\tau)\right] +\frac{1}{\Gamma}\int_{0}^{t} d\tau\; |P(\tau)|^{2} +4 \int_{0}^{t} d\tau\; \Re\left[e^{i\delta \tau}f(\tau)Q(\tau))\right] \\
    &\equiv \mathcal{F}_{1}(t) + \mathcal{F}_{2}(t)+\mathcal{F}_{3}(t) 
\end{align}
where 
\begin{align}
    P(t) &=\sqrt{\Gamma} \int_{0}^{t} d\tau \; e^{\frac{\Gamma}{2} (\tau-t)} e^{-i\delta \tau} f^{*}(\tau), \\
    Q(t) &= \frac{\partial}{\partial \Gamma}\left(\frac{P(t)}{\sqrt{\Gamma}}\right).
\end{align}
Since the above expression is quadratic in $f(t)$, the optimum value can be found analytically in a straightforward manner. We express QFI in the bilinear form, $\mathcal{F}_{\text{long}}(t) = \sum_{m,n}c_{m}^{*}\left[K \right]_{mn}c_{n}$ where $K$ is a Hermitian matrix, and the optimum value of the QFI in this limit corresponds to the maximum eigenvalue of the $K$ matrix. For this, we seek a basis that allows us to decompose all the functions that go to zero at the boundaries. That is, $f(t=0)=f(t=T)=0$, and one such basis is given by 
$f(t) = \sum_{n}c_{n}\xi_{n}(t) = \sum_{n}c_{n} \left(\sqrt{(2/T)}\sin{\omega_{n}t}\right)$ where $\omega_{n}T=n\pi$ for $n =\{1,2,...\}$, $c_{n}$ can be complex and $\sum_{n} |c_{n}|^{2}=\alpha^{2}$. We modify the basis functions so that $f(t)=\sum_{n}c_{n}\tilde{\xi}_{n}(t)  = \sum_{n}c_{n}\left(\sqrt{(2/T)} e^{-i\delta t}\sin{\omega_{n}t}\right)$, and the new basis functions still obey the usual orthonormality condition. Alternatively, we can move to a different frame of reference where the effect of detuning manifests as an overall phase factor in the pulse shape as explained in the main text. Then we have,
\begin{align}
   P(t)     &\equiv \sum_{n} c_{n}^{*} \phi_{n}(t)=\sum_{n} c_{n}^{*} \left(\sqrt{\Gamma} \int_{0}^{t} d\tau \; e^{\frac{\Gamma}{2} (\tau-t)} e^{-i\delta \tau} \tilde{\xi}_{n}^{*}(\tau) \right) ,
\end{align}
where $\phi_{n}$ is real and given by
\begin{align}
\phi_{n}(t) &= \begin{cases} 
\frac{\sqrt{8\Gamma T}}{4n^2\pi^2 + \Gamma^2 T^2}\left[2n\pi e^{-\tfrac{t\Gamma}{2}}-2n\pi\cos\left(\frac{n\pi t}{T}\right) + \Gamma T\sin\left(\frac{n\pi t}{T}\right)\right] & 0\leq t\leq T \\
\frac{n\pi\sqrt{32\Gamma T}}{4n^{2}\pi^{2} + T^{2}\Gamma^{2}}\;\;\left[e^{-\tfrac{t\Gamma}{2}} - (-1)^{n}e^{-\tfrac{(t - T)\Gamma}{2}}\right] & t\geq T
\end{cases},
\end{align}
Note that the above definitions of $\phi_{n}$ are identical to the ones defined in Eq.~\eqref{eq:phi_definition} for the real pulse case.
\subsubsection{Evaluation of \texorpdfstring{$\mathcal{F}_{1}$}{F\_1}}
With $\Gamma T = r \pi$, the second term is given by 
\begin{align}
    \mathcal{F}_{1}(t\rightarrow \infty)&\equiv \mathcal{F}_{1}^{\infty} = \frac{2}{\Gamma^{3/2}}\int_{0}^{T} d\tau\;  \Re\left[e^{i\delta \tau}f(\tau) P(\tau)\right]= \sum_{m, n=1}^{\infty} c_{m}^{*} \left(\frac{\left[K_{1}(r) \right]_{m,n} + \left[K_{1}(r) \right]_{n,m}}{2}\right) c_{n} ,
\end{align}
where 
\begin{align}
    \left[K_{1}(r) \right]_{m,n} &= \frac{2}{\Gamma^{3/2}}\int_{0}^{T} d\tau \; e^{i\delta \tau}\phi_{m}(\tau) \tilde{\xi}_{n}(\tau) \\
    &= \begin{cases}
        \frac{16 m r}{\Gamma^{2}\pi \left(4 m^{2} + r^{2}\right)} \left[ \frac{\left(-1 +(-1)^{m+n}\right) n}{n^{2}-m^{2}} + \frac{\left(4 - 4 (-1)^{n} e^{-\frac{\pi r}{2}}\right) n}{4 n^{2} + r^{2}}
\right] & m \neq n\\
\frac{4 r }{\Gamma^{2} \left(4 n^{2} + r^{2}\right)^{2} }\left[ r^{3} + \frac{4 n^{2}}{\pi} \left( 4 - 4 (-1)^{n} e^{-\frac{\pi r}{2}} + \pi r \right) \right]
& m= n
    \end{cases},
\end{align}
and 
\begin{align}
\left[K_{1}^{(S)}(r)\right]_{m,n} \equiv \frac{\left[K_{1}(r) \right]_{m,n} + \left[K_{1}(r) \right]_{n,m}}{2}  &= \begin{cases}
    \frac{ 32 m n r }{\Gamma^{2}\pi \left(4 m^{2} + r^{2}\right) \left(4 n^{2} + r^{2}\right) } \left[\left(1 +(-1)^{m+n}\right)   -e^{-\frac{\pi r}{2}}((-1)^{m} + (-1)^{n}) \right] & m\neq n\\
    \frac{4 r }{\Gamma^{2} \left(4 n^{2} + r^{2}\right)^{2} }\left[ r^{3} + \frac{4 n^{2}}{\pi} \left( 4 - 4 (-1)^{n} e^{-\frac{\pi r}{2}} + \pi r \right) \right] & m= n
\end{cases},
\end{align}
which is the same expression we obtained for the second term in the real pulse case (see Eq.~\eqref{eq:F2_evaluation_real_pulses}). Following the same arguments as in the real pulse case, we then have
for large $r$,
\begin{align}
     \frac{2}{\Gamma^{3/2}}\int_{0}^{t} d\tau\;  \Re\left[e^{i\delta \tau}f(\tau) P(\tau)\right] &=\sum_{n} c_{m}^{*}\left(\frac{4 \delta_{m,n}}{4 \tilde{n}^2+1}\right) c_{n},
\end{align}
where $\tilde{n}=n/r$, as before. 
\subsubsection{Evaluation of \texorpdfstring{$\mathcal{F}_{2}$}{F\_2}}
We set the width of the pulse as $\Gamma T = r\pi$ and evaluate $\mathcal{F}_{1}$ in the long-time limit,
\begin{align}
\mathcal{F}_{2}(t \rightarrow \infty)=\mathcal{F}_{2}^{\infty} &\equiv \frac{1}{\Gamma}\sum_{m,n} c_{m}^{*} 
    \Bigl(\int_{0}^{\infty} d\tau\;\phi_{m}(\tau)\phi_{n}(\tau)\Bigr) c_{n} \\
    &\equiv \sum_{m,n} c_{m}^{*}\left[K_{2}(r) \right]_{m,n}c_{n},
\end{align}
where 
\begin{align}
\left[K_{2}(r)\right]_{m,n} &= 
\begin{cases}
    \frac{32 m n r }{\Gamma^{2} \pi (4m^{2} + r^{2}) (4n^{2} + r^{2})} \left[ \left(1-e^{-\pi r} \right) + \left( (-1)^{m} -e^{-\frac{\pi r}{2}} \right)
\left((-1)^{n} -e^{-\frac{\pi r}{2}} \right) \right] & m \neq n \\
\frac{4 r\Bigl(\pi r^{3} + 4 n^{2}\bigl(-2e^{-\pi r}  + (2 + \pi r)\bigr)\Bigr)}{\Gamma^{2}\pi (4n^{2} + r^{2})^{2}} + 
\frac{32 \bigl((-1)^{n}-e^{-\frac{\pi r}{2}}\bigr)^{2}n^{2}r}
{\Gamma^{2} \pi (4n^{2} + r^{2})^{2}} & m = n
\end{cases},
\end{align}
Note that the expression for $\mathcal{F}_{2}^{\infty}$ is identical to its expression given in in Eq.~\eqref{eq:F1_expression_real_pulse} for the real case except the coefficient $c_{m}$ is conjugated. Following the same arguments as in the real pulse case, the QFI can be written as 
\begin{align}
    \Gamma^{2} \mathcal{F}_{2}^{\infty} &=\sum_{m,n}c_{m}^{*}\left( \frac{4\delta_{m,n}}{4 \tilde{n}^{2}+1} \right) c_{n},
\end{align}
where $\tilde{n}=n/r$.
\subsubsection{Evaluation of \texorpdfstring{$\mathcal{F}_{3}$}{F\_3}}
Evaluating third term with $\Gamma T = r \pi$, we have
\begin{align}
     \mathcal{F}_{3}(t \rightarrow \infty) \equiv \mathcal{F}_{3}^{\infty} &=4\int_{0}^{T}  d\tau \;\Re\left[e^{i\delta \tau} f(\tau) Q(\tau)\right] ,
\end{align}
where 
\begin{align}
    Q(t) &= \frac{\partial}{\partial \Gamma}\left(\frac{P(t)}{\sqrt{\Gamma}}\right) =  \sum_{n}c_{n}^{*} \frac{\partial}{\partial \Gamma}\left(\frac{\phi_{n}(t)}{\sqrt{\Gamma}}\right)=  \sum_{n}c_{n}^{*} \eta_{n}(t),
\end{align}
and 
\begin{align}
    \eta_{n}(t) &= \frac{
\sqrt{8 T}
}{\left( 4 n^{2} \pi^{2} + T^{2} \Gamma^{2} \right)^{2}}\left[
4 n \pi T^{2} \Gamma \cos\left( \frac{n \pi t}{T} \right)
+ T \left( 4 n^{2} \pi^{2} - T^{2} \Gamma^{2} \right) \sin\left( \frac{n \pi t}{T} \right)- n \pi e^{-\frac{t \Gamma}{2}} \left( 4 n^{2} \pi^{2} t + T^{2} \Gamma \left( 4 + t \Gamma \right) \right)
\right].
\end{align}
Note that we are only concerned here about the derivative of $\phi_{n}(t)$ for $t\leq T$ since the upper bound on the integral for $\mathcal{F}_{3}$ is $T$. Hence, 
\begin{align}
    \mathcal{F}_{3}^{\infty} &= \sum_{m,n} \Re\left[c_{m}^{*}\left(4  \int_{0}^{T} d\tau \; e^{i\delta \tau}\tilde{\xi}_{n}(\tau)\eta_{m}(\tau)\right)c_{n}\right] \equiv \sum_{m,n} c_{m}^{*} \left(\frac{\left[K_{3}\right]_{m,n}+\left[K_{3}\right]_{n,m}}{2}\right)c_{n}.
\end{align}
The symmetric part of the non-diagonal terms is given by
\begin{align}
\left[K_{3}^{S}\right]_{m,n} &= \frac{32 m n r^{2} \left(4\pi m^{2} r^{2}+  4\pi n^{2} r^{2}+ \pi r^{4}+  16 m^{2} n^{2} \pi+ 8r^{3}  + 16 m^{2}r+ 16 n^{2} r \right)}{\Gamma^{2}\pi \left(4 m^{2}+r^{2}\right)^{2} \left(4 n^{2}+r^{2}\right)^{2}}\left[ \left( (-1)^{m} + (-1)^{n} \right) e^{-\frac{\pi r}{2}} \right]\nonumber \\
& \qquad\qquad -\frac{256 m n r^{3}\left( 2m^{2} + 2 n^{2} + r^{2} \right) }{\Gamma^{2}\pi \left(4 m^{2}+r^{2}\right)^{2} \left(4n^{2}+r^{2}\right)^{2}} \left(1 + (-1)^{m+n}\right),
\end{align}
and the diagonal terms are given by 
\begin{align}
\left[K_{3}^{S}\right]_{n,n} &= \frac{8 r^{2}}{\Gamma^{2}\pi \left( 4 n^{2}+r^{2}\right)^{3}}\left[8 e^{-\frac{\pi r}{2}}(-1)^{n} n^{2} \left( 4 n^{2} \pi + r \left( 8 + \pi r \right) \right) + \left(16 n^{4} \pi - 64 n^{2} r - \pi r^{4} \right) \right].
\end{align}
Note that the $\mathcal{F}_{3}$ expression obtained in this case is identical to the one obtained with real pulses and zero detuning. Following the same arguments as in the case of real pulse,  we have in the limit $T \rightarrow \infty$, 
\begin{align}
    \Gamma^{2}\mathcal{F}_{3}^{\infty} &=\sum_{m,n}c_{m}^{*}\left(  \frac{8  \left(4\tilde{n}^2 - 1 \right) \delta_{m,n}} {(4\tilde{n}^2 + 1)^2} \right) c_{n}.
\end{align}
\subsubsection{Optimal QFI and optimal pulse}
As in the real pulse case, we have
\begin{align}
    \frac{\Gamma^{2} \mathcal{F}_{\text{long}}^{\infty}}{\alpha^{2}}    &=\frac{1}{\alpha^{2}} \sum_{m,n}c_{m}^{*}\left(  \frac{8  \left(4\tilde{n}^2 - 1 \right) \delta_{m,n}} {(4\tilde{n}^2 + 1)^2} +\frac{8\delta_{m,n}}{4 \tilde{n}^{2}+1} \right) c_{n} = \sum_{m,n}\tilde{c}_{m}^{*}\left(  \frac{64 \tilde{n}^{2} \delta_{m,n}} {(4\tilde{n}^2 + 1)^2} \right) \tilde{c}_{n}.
\end{align}
 Note that the coefficients $\{\tilde{c}_{m}\}$ are normalized, and the eigenvalues are then given by $ \frac{64 \tilde{n}^{2}} {(4\tilde{n}^2 + 1)^2}$ which range between $0$ and $4$ as $\tilde{n}$ varies in the range $\{1/r,2/r,...\}$.
The maximum eigenvalue is obtained at $\tilde{n} = 1/2$ or equivalently at 
\begin{align}
    \omega= \frac{\Gamma}{2}.
\end{align}
In other words, the maximum QFI per unit photon in this limit even when the coefficients in the basis are allowed to complex is $4$, and $\sqrt{2\alpha/T} \sin(\Gamma t/2)e^{-i\delta t}$ is the pulse that optimizes it.
\subsection{Diagonalizing QFI up to second order with the periodic boundary conditions}
\label{appendix:diagonalize_periodic_boundary_conditions}
In the plane-wave basis, we have $f(t) = \sum_{n}c_{n} \xi_{n}(t) = \sqrt{1/T}\sum_{n} c_{n} e^{i\omega_{n}t}$ where $\omega_{n}=2n\pi/T$ and $n=\{..., -2,-1,0,1,2,...\}$ due to periodic boundary conditions, $f(0)=f(T)$. We first assume that the detuning is zero, and the result for nonzero detuning will be obtained in a straightforward manner at the end. Expressing $P(t)$ and $Q(t)$ in the plane-wave basis,
\begin{align}
    P(t) &=\sqrt{\Gamma} \int_{0}^{t} d\tau \; e^{\frac{\Gamma}{2} (\tau-t)} f^{*}(\tau)=\sum_{n} c_{n}^{*}\left(\sqrt{\Gamma} \int_{0}^{t} d\tau \; e^{\frac{\Gamma}{2} (\tau-t)} \xi_{n}^{*}(\tau)\right) \equiv \sum_{n}c_{n}^{*}\phi_{n}(t) ,
\end{align}
where 
\begin{align}
\phi_{n}(t) &= \begin{cases} 
\frac{\sqrt{4 T \Gamma}}{4 i n \pi - T \Gamma} \left( e^{-\frac{t \Gamma}{2}} - e^{-\frac{2 i n \pi t}{T}} \right)
 & 0\leq t\leq T \\
\frac{\sqrt{4 T \Gamma}}{4 i n \pi - T \Gamma}  \left( e^{-\frac{t \Gamma}{2}} -  e^{-\frac{\Gamma(t -T)}{2}}\right)
 & t\geq T
\end{cases}.
\end{align}
Likewise, $Q(t)$ for $t \leq T$, which is the range of interest to us, is given by 
\begin{align}
        Q(t) &= \frac{\partial}{\partial \Gamma}\left(\frac{P(t)}{\sqrt{\Gamma}}\right) = \sum_{n}c_{n}^{*}\frac{\partial}{\partial \Gamma}\left(\frac{\phi_{n}(t)}{\sqrt{\Gamma}}\right)  \equiv \sum_{n}c_{n}^{*}\eta_{n}(t) ,
\end{align}
where 
\begin{align}
    \eta_{n}(t) &= \frac{ e^{ -\frac{1}{2} t \left( \frac{4 i n \pi}{T} + \Gamma \right) } \sqrt{T} }
{\left( T \Gamma - 4 i n \pi \right)^{2}}
\left[
e^{ \frac{2 i n \pi t}{T} } \left( 2 T + \Gamma t T - 4 i n \pi t \right)
- 2 e^{ \frac{t \Gamma}{2} } T
\right].
\end{align}
\subsubsection{Evaluation of \texorpdfstring{$\mathcal{F}_{1}$}{F\_1}}
With $\Gamma T = r \pi$, the second term is given by 
\begin{align}
    \mathcal{F}_{1}(t\rightarrow \infty)&\equiv \mathcal{F}_{1}^{\infty} = \frac{2}{\Gamma^{3/2}}\int_{0}^{T} d\tau\;  \Re\left[f(\tau) P(\tau)\right]= \sum_{m, n=1}^{\infty} c_{m}^{*} \left(\frac{\left[K_{1}(r) \right]_{m,n} + \left[K_{1}(r) \right]_{n,m}^{*}}{2}\right) c_{n} ,
\end{align}
where the Hermitian part of the $K_{1}$ matrix is given by
\begin{align}
    \left[K_{1}^{H}(r) \right]_{m,n} &\equiv \left(\frac{\left[K_{1}(r) \right]_{m,n} + \left[K_{1}(r) \right]_{n,m}^{*}}{2}\right)=\frac{2}{\Gamma^{3/2}}\int_{0}^{T} d\tau \; \left[\phi_{m}(\tau) \xi_{n}(\tau)+\phi_{n}^{*}(\tau) \xi_{m}^{*}(\tau)\right]\\
    &=
    \begin{cases}
        \frac{8r\left( 16 m n - r^{2} \right)}
{\Gamma^{2}\pi \left( 16 m^{2} + r^{2} \right) \left( 16 n^{2} + r^{2} \right) }\left( 1-e^{ -\frac{\pi r}{2} }  \right) 
 & m\neq n \\
  \frac{1}
{\Gamma^{2}\pi \left( 16 n^{2} + r^{2} \right)^{2} }\left[ 
4 r^{3} \left( \pi r -2+ 2e^{ -\frac{\pi r}{2} }  \right)
+ 64 n^{2} r  \left(\pi r + 2  -2e^{ -\frac{\pi r}{2} }\right)
\right]
 & m=n
    \end{cases},
\end{align}
Ignoring exponentially small terms in $r$, and relabeling the variables $\tilde{m} = m/r$ and $\tilde{n}=n/r$, we have we have
\begin{align}
    \left[K_{1}^{H}(r) \right]_{m,n} &=
    \begin{cases}
        \frac{8\left( 16 \tilde{m} \tilde{n} - 1 \right)}
{\Gamma^{2}\pi r \left( 16 \tilde{m}^{2} + 1 \right) \left( 16 \tilde{n}^{2} + 1 \right) }
 & m\neq n \\
  \frac{1}
{\Gamma^{2}\pi \left( 16 \tilde{n}^{2} + 1 \right)^{2} }\left[ 
4  \left( \pi  -\frac{2}{r} \right)
+ 64 \tilde{n}^{2}  \left(\pi  + \frac{2}{r} \right)
\right]
 & m=n
    \end{cases}.
\end{align}
Since the term $\frac{8\left( 16 \tilde{m} \tilde{n} - 1 \right)}
{\Gamma^{2}\pi \left( 16 \tilde{m}^{2} + 1 \right) \left( 16 \tilde{n}^{2} + 1 \right) }$ is bounded for any $m$ and $n$, the non-diagonal terms of the $K_{2}^{H}$ matrix can be ignored for large $r$. Ignoring terms that are suppressed by $r$, the diagonal matrix is given by
\begin{align}
    \left[K_{1}^{H}(r) \right]_{m,n}
    &= \sum_{m,n}c_{m}^{*} \left(\frac{4 \delta_{n,m}}
{\Gamma^{2} \left( 16 \tilde{n}^{2} + 1 \right)}\right)c_{n}.
\end{align}
Note that the above expression ranges between $0$ and $4$ as $\tilde{n}$ varies in the range $\tilde{n}=\{..., -\frac{2}{r},-\frac{1}{r},0,\frac{1}{r}, \frac{2}{r},...\}$.
\subsubsection{Evaluation of \texorpdfstring{$\mathcal{F}_{2}$}{F\_2}}
\begin{align}
\mathcal{F}_{2}(t \rightarrow \infty)=\mathcal{F}_{2}^{\infty} &\equiv \frac{1}{\Gamma}\sum_{m,n} c_{m}^{*} 
    \Bigl(\int_{0}^{\infty} d\tau\;\phi_{m}(\tau)\phi_{n}^{*}(\tau)\Bigr) c_{n} \\
    &\equiv \sum_{m,n} c_{m}^{*}\left[K_{2}(r) \right]_{m,n}c_{n},
\end{align}
where
\begin{align}
\left[K_{2}(r)\right]_{m,n} &= 
\begin{cases}
    \frac{2 r}{\Gamma^{2}\pi(r-4 i m ) (r+4in)}\left[ 2 - 2 e^{ -\pi r } + \frac{ 4r \left(e^{ -\frac{\pi r}{2} } -1\right)}{r+4im} + \frac{ 4 r\left(e^{ -\frac{\pi r}{2} }-1 \right)}{r-4in}\right] + \frac{4r\left(1-e^{-\frac{\pi r}{2}}\right)^{2}}{ \Gamma^{2} \pi (r-4 i m ) (r+4in)}
& m \neq n \\
\frac{1}{\Gamma^{2}\pi (16 n^{2} + r^{2})^{2}} \left[ 
4 r^{3} \left(4 e^{- \frac{\pi r}{2} }  -e^{ -\pi r }+  \pi r - 3 \right) 
+ 64 n^{2} r \left( 1 + \pi r  -e^{ -\pi r } \right) 
\right] +  \frac{ 4 
r\left( 1 - e^{- \frac{\pi r}{2} } \right)^{2}}
{ \Gamma^{2} \pi  (16 n^{2} + r^{2})}
 & m = n
\end{cases},
\end{align}
Ignoring terms that are exponentially suppressed in $r$, we have
\begin{align}
\left[K_{2}(r)\right]_{m,n} &= 
\begin{cases}
    \frac{2 r}{\Gamma^{2}\pi(r-4 i m ) (r+4in)}\left[ 2 - \frac{4r}{r+4im} - \frac{ 4 r}{r-4in}\right] + \frac{4r}{ \Gamma^{2} \pi (r-4 i m ) (r+4in)}
& m \neq n \\
\frac{1}{\Gamma^{2}\pi (16 n^{2} + r^{2})^{2}} \left[ 
4 r^{3} \left(\pi r - 3 \right)
+ 64 n^{2} r \left( 1 + \pi r   \right) 
\right] +  \frac{4r}{\Gamma^{2}\pi(16n^{2}+r^{2})}
 & m = n
\end{cases},
\end{align}
Relabeling variables as before $\tilde{m}=m/r$, $\tilde{n}=n/r$, we have
\begin{align}
\left[K_{2}(r)\right]_{m,n} &= 
\begin{cases}
    \frac{8}{\Gamma^{2}r\pi(1-4 i \tilde{m} ) (1+4i\tilde{n})}\left[ 1 - \frac{1}{1+4i\tilde{m}} - \frac{1}{1-4i\tilde{n}}\right]
& m \neq n \\
\frac{1}{\Gamma^{2}\pi(16 \tilde{n}^{2} + 1)^{2}} \left[ 
4 \left(\pi  - \frac{3}{r} \right)
+ 64 \tilde{n}^{2} \left( \frac{1}{r} + \pi  \right) 
\right] +  \frac{4}{\Gamma^{2}r\pi(16\tilde{n}^{2}+1)}
 & m = n
\end{cases},
\end{align}
Upon further simplification, we have
\begin{align}
\left[K_{2}(r)\right]_{m,n} &= 
\begin{cases}
    \frac{ 8 \left( 16 \tilde{m} \tilde{n}-1) \right) }
{ \Gamma^{2}r\pi\left( 1 + 16 \tilde{m}^{2} \right) \left( 1 + 16 \tilde{n}^{2} \right)  }
& m \neq n \\
\frac{1}{\Gamma^{2}\pi(16 \tilde{n}^{2} + 1)^{2}} \left[ 
4 \left(\pi  - \frac{3}{r} \right)
+ 64 \tilde{n}^{2} \left( \frac{1}{r} + \pi  \right) 
\right] +  \frac{4}{\Gamma^{2}r\pi(16\tilde{n}^{2}+1)}
 & m = n
\end{cases}.
\end{align}
Note that $ \frac{ 8 \left( 16 \tilde{m} \tilde{n}-1) \right) }{\pi\left( 1 + 16 \tilde{m}^{2} \right) \left( 1 + 16 \tilde{n}^{2} \right)}$ is bounded for all $m$ and $n$, so the non-diagonal term is suppressed by $r$. Hence, the matrix becomes diagonal. Ignoring terms that are suppressed by $r$, we have
\begin{align}
\left[K_{2}(r)\right]_{m,n} &= 
\sum_{m,n}c_{m}^{*}\left(\frac{4 \delta_{n,m}}{\Gamma^{2}(16 \tilde{n}^{2} + 1)}\right)c_{n},
\end{align}
which is the same as the $K_{1}(r)$ matrix.
\subsubsection{Evaluation of \texorpdfstring{$\mathcal{F}_{3}$}{F\_3}}
\begin{align}
      \mathcal{F}_{3}(t \rightarrow \infty) \equiv \mathcal{F}_{3}^{\infty} &=4\int_{0}^{T}  d\tau \;\Re\left[ f(\tau) Q(\tau)\right] 
\end{align}
\begin{align}
    \mathcal{F}_{3}^{\infty} &= \sum_{m,n} \Re\left[c_{m}^{*}\left(4  \int_{0}^{T} d\tau \; \xi_{n}(\tau)\eta_{m}(\tau)\right)c_{n}\right] \equiv \sum_{m,n} c_{m}^{*} \left(\frac{\left[K_{3}(r)\right]_{m,n}+\left[K_{3}(r)\right]_{n,m}^{*}}{2}\right)c_{n}
\end{align}
and the non-diagonal elements are given by
\begin{align}
    \left[K_{3}^{H}(r)\right]_{m \neq n} &\equiv \left(\frac{\left[K_{3}(r)\right]_{m,n}+\left[K_{3}(r)\right]_{n,m}^{*}}{2}\right)\\
    &=\frac{8 r^2 e^{-\frac{\pi r}{2}}}{\pi \Gamma^{2} \left(m - n\right)} \left[
 \frac{ 16 n^{3} \pi + n r \left( 4 + \pi r \right) }
       { \left( 16 n^{2} + r^{2} \right)^{2} }  -\frac{ m \left( 16 m^{2} \pi + r \left( 4 + \pi r \right) \right) }
       { \left( 16 m^{2} + r^{2} \right)^{2} }
\right]\\
& +\frac{32 r^{3}}{\pi \Gamma^{2} (m - n)} \left[ 
\frac{m}{\left( 16 m^{2} + r^{2} \right)^{2}}
- \frac{n}{\left( 16 n^{2} + r^{2} \right)^{2}}
\right].
\end{align}
Ignoring exponentially small terms in $r$ and relabeling terms using $\tilde{m}=m/r$ and $\tilde{n}=n/r$, we have
\begin{align}
    \left[K_{3}^{H}\right]_{m \neq n} &= \frac{32}{\pi \Gamma^{2} r(\tilde{m} - \tilde{n})} \left[ 
\frac{\tilde{m}}{\left( 16 \tilde{m}^{2} + 1 \right)^{2}}
- \frac{\tilde{n}}{\left( 16 \tilde{n}^{2} + 1 \right)^{2}}
\right],
\end{align}
which is suppressed by $r$, so it can be ignored when $r$ is very large. The diagonal element is given by 
\begin{align}
    \left[K_{3}^{H}\right]_{m= n} &= -\frac{8r^{2}}
{\pi \left( 16 n^{2} + r^{2} \right)^{3} \Gamma^{2}}\left[
-256 \left( e^{ -\frac{\pi r}{2} } + 1 \right) n^{4} \pi
+ 192 \left( -e^{ -\frac{\pi r}{2} } + 1 \right) n^{2} r
+ r^{3} \left( 4e^{ -\frac{\pi r}{2} } + \pi r e^{ -\frac{\pi r}{2} }+ \pi r -4 \right)
\right].
\end{align}
Ignoring exponentially suppressed terms and relabeling as before, $\tilde{m}=m/r$ and $\tilde{n}=n/r$, we have
\begin{align}
    \left[K_{3}^{H}\right]_{m= n} &= -\frac{8}
{\Gamma^{2} \pi\left( 16 \tilde{n}^{2} + 1\right)^{3} }\left[ \pi -256  \tilde{n}^{4} \pi +   \frac{192 \tilde{n}^{2}}{r}  -\frac{4}{r} \right] .
\end{align}
Ignoring terms suppressed by $r$, we have
\begin{align}
    \left[K_{3}^{H}\right]_{m= n} & =  \frac{8 \left(16  \tilde{n}^{2} -1\right) }
{\Gamma^{2} \left( 16 \tilde{n}^{2} + 1\right)^{2} },
\end{align}
where $\tilde{n}=\{..., -\frac{2}{r},-\frac{1}{r},0,\frac{1}{r}, \frac{2}{r},...\}$
\subsubsection{Optimal QFI and optimal pulse}
In summary, we have
\begin{align}
    \frac{\Gamma^{2} \mathcal{F}_{\text{long}}^{\infty}}{\alpha^{2}}    &=\frac{1}{\alpha^{2}} \sum_{m,n}c_{m}^{*}\left(  \frac{8  \left(16\tilde{n}^2 - 1 \right) \delta_{m,n}} {(16\tilde{n}^2 + 1)^2} +\frac{8\delta_{m,n}}{16 \tilde{n}^{2}+1} \right) c_{n} = \sum_{m,n}\tilde{c}_{m}^{*}\left(  \frac{256 \tilde{n}^{2} \delta_{m,n}} {(16\tilde{n}^2 + 1)^2} \right) \tilde{c}_{n}.
\end{align}
 Note that the coefficients $\{\tilde{c}_{m}\}$ are normalized, and the eigenvalues are then given by $ \frac{256 \tilde{n}^{2}} {(16\tilde{n}^2 + 1)^2}$ which range between $0$ and $4$ as $\tilde{n}$ varies in the range $\{0, \pm 1/r,\pm 2/r,...\}$.
The maximum eigenvalue is obtained at $\tilde{n} =\pm 1/4$ or equivalently at $    \omega= \pm \frac{\Gamma}{2}$, 
and the corresponding eigenvector is given by $\sqrt{\alpha/T} \exp(\pm \frac{i \Gamma t}{2})$. With nonzero detuning $\delta$, the optimal QFI is still $\mathcal{F}^{\infty}/\alpha^2=4$, and the optimal frequency is given by 
\begin{align}
    \omega = \delta \pm \frac{\Gamma}{2},
\end{align}
Hence, the optimal pulse is $\sqrt{\alpha/T} \exp\left[-i( \delta \pm \frac{\Gamma}{2})t\right]$. 
\end{widetext}
\end{document}